\documentclass[a4paper]{article}
\usepackage[margin=2cm]{geometry}
\usepackage[brazilian, english]{babel}
\usepackage{amsmath, amssymb}
\usepackage{graphicx, subcaption}
\usepackage{array, multirow}
\usepackage{xurl, hyperref}
\usepackage{abstract}
\usepackage{xcolor}

\usepackage[font=small]{caption}

\begin{document}
\sloppy

\title{\bf F\'isica de Part\'iculas no Ensino M\'edio\\
	Parte II: F\'isica Nuclear\\[4mm]
	\normalsize Particle Physics in High School \\  Part II: Nuclear Physics
}
\author{ \normalsize 
		{\bf Thaisa~Carneiro~da Cunha~Guio}\thanks{thaisa@fisica.ufmg.br}~,
		{\bf Gl\'auber~Carvalho~Dorsch}\thanks{glauber@fisica.ufmg.br}\\[3mm]
		\small \it Departamento de F\'isica, Universidade Federal de Minas Gerais, 31270-901, Belo Horizonte, MG\\
}
\date{}

\twocolumn[
    \maketitle
    \selectlanguage{brazilian}
    \vspace*{-5mm}
    \begin{onecolabstract}
        Apresentamos a segunda parte de uma s\'erie de artigos que prop\~oe uma nova sequ\^encia did\'atica sobre F\'isica de Part\'iculas para o ensino m\'edio. O tema do presente trabalho \'e a  F\'isica Nuclear. O objetivo principal da sequ\^encia \'e abordar o assunto de modo a fomentar a alfabetiza\c{c}\~ao cient\'ifica dos(as) estudantes, em uma perspectiva que relaciona Ci\^encia, Tecnologia, Sociedade e Meio Ambiente (CTSA). Avaliamos as potencialidades e a efetividade do material aqui proposto, aliado a uma postura dial\'ogica docente, a partir da an\'alise de indicadores de alfabetiza\c{c}\~ao cient\'ifica e de engajamento dos(as) estudantes durante as interven\c{c}\~oes realizadas em uma escola p\'ublica estadual do Esp\'irito Santo.
        \\
        
        \noindent {\bf Palavras-chave:} F\'isica de part\'iculas, ensino m\'edio, alfabetiza\c{c}\~ao cient\'ifica, CTSA, engajamento. 
    \end{onecolabstract}
    
    \selectlanguage{english}
    \begin{onecolabstract}
    We present the second part of a series of papers proposing a novel teaching sequence for Particle Physics in high school. The topic of the present work is Nuclear Physics. The goal of the sequence is to approach the subject in a way as to stimulate scientific literacy, from a perspective involving Science, Technology, Society and Environment (STSE). We evaluate the potentialities and effectiveness of the material proposed here, allied to a dialogical approach by the teacher, by analyzing the presence of scientific literacy and engagement indicators during interventions applied to a public school in Esp\'irito Santo, Brazil.\\
     
    \end{onecolabstract}

    \vspace*{1cm}
]
\saythanks

\selectlanguage{brazilian}

\section{Introdu\c{c}\~ao}

Em trabalho anterior~\cite{CarvalhoDorsch:2021lvd} argumentamos em favor da inser\c{c}\~ao da tem\'atica de F\'isica de Part\'iculas em salas de aula no ensino m\'edio e apresentamos uma sequ\^encia did\'atica sobre elementos de eletrodin\^amica qu\^antica. Partimos de conhecimentos elementares, possivelmente j\'a apropriados por estudantes desse n\'ivel, como a no\c{c}\~ao de estrutura at\^omica da mat\'eria e aspectos da intera\c{c}\~ao eletromagn\'etica, para pavimentar, pouco a pouco, um caminho rumo a uma discuss\~ao s\'olida sobre o conceito de f\'oton, e como a intera\c{c}\~ao eletromagn\'etica pode ser entendida como mediada pela troca dessas part\'iculas. Sustentamos a tese central do artigo, de que \'e poss\'ivel introduzir a tem\'atica de F\'isica de Part\'iculas em salas de aula do ensino m\'edio de forma sistem\'atica e com engajamento dos estudantes, a partir dos resultados da an\'alise das interven\c{c}\~oes que ministramos com base na sequ\^encia proposta.

Neste artigo damos continuidade \`a proposta desta s\'erie, desta vez com a aten\c{c}\~ao voltada \`a din\^amica do n\'ucleo at\^omico. 

Na se\c{c}\~ao~\ref{sec:momentos} iniciamos a segunda parte da sequ\^encia did\'atica com uma discuss\~ao sobre o experimento de Rutherford-Geiger-Marsden, que deu origem ao conceito de n\'ucleo at\^omico. Trata-se de um t\'opico que n\~ao foge \`a ementa tradicional do ensino m\'edio, portanto adequado como ponto de partida da segunda parte da sequ\^encia. 
Discutiremos a necessidade de se postular uma nova forma de intera\c{c}\~ao, chamada intera\c{c}\~ao nuclear forte, para explicar a estabilidade do n\'ucleo. Apresentamos diversas consequ\^encias fenomenol\'ogicas que podem ser discutidas em sala, tais como: rea\c{c}\~oes de fiss\~ao e energia nuclear; a fus\~ao nuclear como fonte de energia do Sol e demais estrelas; a fus\~ao como rea\c{c}\~ao geradora de todos os elementos na Natureza, justificando a famosa frase de Carl Sagan de que ``somos poeira das estrelas''~\cite{CarlSagan}; e decaimentos nucleares e aplica\c{c}\~oes tecnol\'ogicas. Enfatizaremos como cada um desses subtemas t\^em rico potencial de discuss\~ao sob uma abordagem que interliga Ci\^encia, Tecnologia, Sociedade e Ambiente (CTSA)~\cite{Bybee1987, LopezCerezo1996, Krasilchik1987, SantosMortimer2001, SantosMortimer2002, SasseronCarvalho2011, Waks1990}, bem ao esp\'irito desta s\'erie de artigos. 

Cabe destacar, desde j\'a, que a f\'isica nuclear \'e intrinsicamente convidativa \`a contemporaneidade e \`a interdisciplinaridade, constituindo-se, assim, em uma plataforma ideal para uma abordagem CTSA e de alfabetiza\c{c}\~ao cient\'ifica em sala de aula. Exatamente por isso, trata-se tamb\'em de uma tem\'atica que atrai a aten\c{c}\~ao de diversos(as) estudantes, sejam apaixonados(as) pela f\'isica ou n\~ao. O intuito da prepara\c{c}\~ao desta parte da sequ\^encia foi tornar expl\'icito as potencialidades de se tratar dessa tem\'atica sob essa abordagem pedag\'ogica. Enfatizamos como tal abordagem possibilita um campo extremamente frut\'ifero para, ainda, incluir atividades que desenvolvem diversas habilidades norteadas pela Base Nacional Curricular Comum (BNCC)~\cite{BNCC} na \'area de Ci\^encias da Natureza e suas tecnologias no ensino m\'edio.

Ap\'os a apresenta\c{c}\~ao da sequ\^encia, discutimos a din\^amica em sala de aula na se\c{c}\~ao~\ref{sec:analise}, quando da aplica\c{c}\~ao desta sequ\^encia a uma turma mista de estudantes de 1$^{\rm a}$, 2$^{\rm a}$ e 3$^{\rm a}$ s\'eries do ensino m\'edio de um Centro Estadual de Ensino M\'edio em Tempo Integral (CEEMTI) no munic\'ipio de Vila Velha, estado do Esp\'irito Santo. Usando indicadores presentes na literatura, avaliamos ind\'icios de alfabetiza\c{c}\~ao cient\'ifica e de engajamento da turma durante a aplica\c{c}\~ao da sequ\^encia.

As conclus\~oes s\~ao apresentadas na se\c{c}\~ao~\ref{sec:conclusoes}. No ap\^endice~\ref{sec:atividades} sugerimos atividades que podem ser discutidas em sala de aula, al\'em das mencionadas no corpo principal do texto.

\section{Propostas de momentos did\'aticos}
\label{sec:momentos}

Como no trabalho anterior~\cite{CarvalhoDorsch:2021lvd}, destacamos que os momentos did\'aticos apresentados abaixo n\~ao t\^em a pretens\~ao de constituir um material formulaico, um receitu\'ario j\'a pronto para ser seguido acriticamente pelo(a) docente, com um roteiro fixo das aulas. Preferimos elaborar um material que sirva como mat\'eria-prima para que o(a) docente prepare sua aula como julgar mais adequada em seu contexto, um material de car\'ater sugestivo e male\'avel, adapt\'avel \`as aspira\c{c}\~oes, necessidades e circunst\^ancias de cada docente e de cada turma. 

Entretanto, o conjunto de momentos did\'aticos abaixo foi elaborado de modo a iniciar com um conte\'udo j\'a familiar a discentes e docentes, progredindo paulatinamente de modo a se construir a estrutura e a din\^amica do n\'ucleo at\^omico de forma intuitiva, acess\'ivel \`a apreens\~ao e sempre suscet\'ivel a debates entre os estudantes. 

\subsection{O experimento de Rutherford-Geiger-Marsden}
\label{sec:Rutherford}

O ponto de partida natural para uma discuss\~ao sobre a din\^amica do n\'ucleo at\^omico \'e a pr\'opria descoberta da exist\^encia do n\'ucleo, ou seja, um estudo sobre o experimento de Geiger e Marsden, bem como sobre a proposta de modelo at\^omico de Rutherford visando explicar os dados observados. Al\'em de essa abordagem explicitar o desenvolvimento hist\'orico de nosso conhecimento sobre o \'atomo, o experimento da folha de ouro de Geiger-Marsden e o modelo de Rutherford s\~ao assuntos j\'a comumente discutidos no ensino m\'edio, e tamb\'em por isso constituem um ponto de partida ideal para o aprofundamento da discuss\~ao rumo a outras tem\'aticas de F\'isica Nuclear e de Part\'iculas nesse n\'ivel de ensino.

\subsubsection{O modelo at\^omico de J.~J.~Thomson}
\label{sec:Thomson}

\begin{figure}
	\centering
	\includegraphics[width=.17\textwidth]{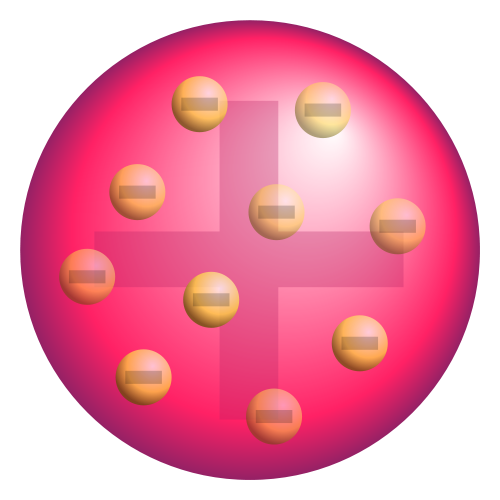}\qquad
	\includegraphics[width=.25\textwidth]{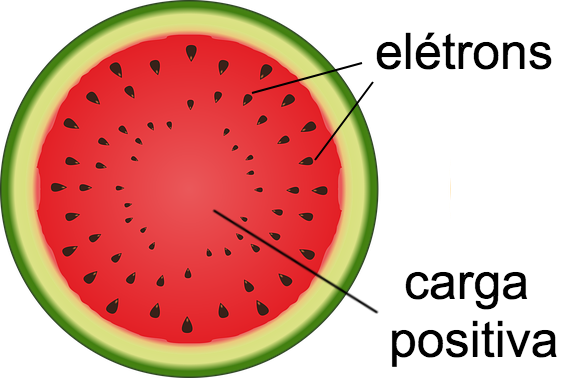}
	\caption{(Esquerda) Ilustra\c{c}\~ao do modelo at\^omico de Thomson. Os el\'etrons est\~ao imersos em uma esfera uniforme de carga positiva, de modo que a carga total se anula. 
	(Direita) Uma analogia \'e uma melancia, em que as sementes s\~ao el\'etrons e a polpa \'e a subst\^ancia de carga positiva. 
	Entretanto, cabe ressaltar que, no modelo proposto por Thomson, os el\'etrons {\bf n\~ao} est\~ao est\'aticos no interior da esfera, mas giram em circunfer\^encias de raios diversos, cada uma com uma capacidade m\'axima de el\'etrons, determinada por condi\c{c}\~oes de estabilidade diante da m\'utua intera\c{c}\~ao eletromagn\'etica. Assim, o modelo de Thomson prev\^e um comportamento peri\'odico para os \'atomos, dado pelo preenchimento de consecutivas camadas eletr\^onicas. Fonte: Dom\'inio P\'ublico.	}
	\label{fig:thomson}
\end{figure}

Em seu experimento de 1897, J.~J.~Thomson demonstrou a exist\^encia de part\'iculas de massa menor do que a do \'atomo mais leve\footnote{Para uma abordagem desse experimento em sala de aula no ensino m\'edio, ver~\cite{CarvalhoDorsch:2021lvd}.}, e argumentou que tais corp\'usculos devem ser constituintes fundamentais de todos os elementos qu\'imicos, pois as propriedades dessas part\'iculas independiam do material do eletrodo de onde elas eram emitidos. Imediatamente se imp\^os a quest\~ao de como esses corp\'usculos --- os el\'etrons --- estariam organizados no interior dos \'atomos, para a qual Thomson ofereceu uma proposta de solu\c{c}\~ao em artigo datado de 1904~\cite{Thomson:1904bjw}. No chamado ``modelo at\^omico de Thomson'', os el\'etrons, que possuem carga el\'etrica negativa, est\~ao imersos em uma esfera uniforme de carga positiva, de modo que a carga total do \'atomo se anula, como ilustrado na figura~\ref{fig:thomson}.
O raio da esfera positiva \'e o raio at\^omico, $r\sim 10^{-10}$~m, valor j\'a conhecido na \'epoca\footnote{O raio at\^omico pode ser estimado a partir do n\'umero de Avogadro, da densidade e da massa molar de uma determinada subst\^ancia. Por exemplo, a densidade do carbono \'e $\approx 2.26~{\rm g/cm}^3$, e a massa molar \'e $12$~g, portanto 1 mol de carbono ocupa um volume de $5.31~{\rm cm}^3$. Como 1 mol possui $\approx 6.02\times 10^{23}$ \'atomos, cada \'atomo ocupa um volume de $\sim 10^{-23}~{\rm cm}^3$, o que corresponde a uma esfera de raio $\sim 10^{-10}$~m.}. Importante ressaltar que, embora usualmente o modelo de Thomson seja imaginado e apresentado como est\'atico, o artigo de 1904 considerava os el\'etrons girando em circunfer\^encias de diversos raios, cada uma com uma capacidade m\'axima de el\'etrons ditada pelas condi\c{c}\~oes de estabilidade eletromagn\'etica.   Assim, o modelo de Thomson j\'a fazia algumas previs\~oes a respeito de um comportamento peri\'odico de propriedades at\^omicas, dado pelo preenchimento de consecutivas camadas eletr\^onicas. 

\begin{figure}
	\centering
	\includegraphics[scale=.45]{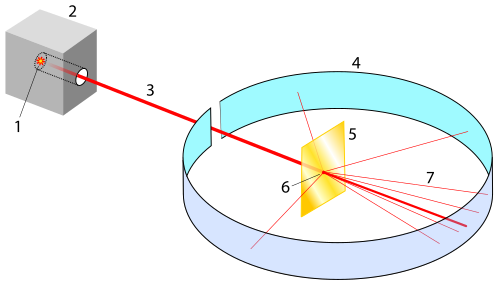}
	\caption{Esquematiza\c{c}\~ao do experimento de Rutherford-Geiger-Marsden. (1) Fonte radioativa emissora de part\'iculas $\alpha$. (2) Inv\'olucro de chumbo com um pequeno orif\'icio para colimar o feixe. (3) Feixe de part\'iculas $\alpha$. (4) Tela de sulfeto de zinco (ZnS) que fluoresce quando atingida por uma part\'icula $\alpha$. (5) Folha de ouro. (6) Ponto de impacto do feixe na folha de ouro. (7) Feixes de part\'iculas $\alpha$ espalhadas. Fonte: Wikimedia Commons/Dom\'inio P\'ublico.}
	\label{fig:GeigerMarsden}
\end{figure}

Entre 1908 e 1913~\cite{Geiger:1910, Geiger:1913}, visando testar o modelo de Thomson para o \'atomo, Rutherford sugeriu a Geiger e Marsden que realizassem uma s\'erie de experimentos bombardeando uma fina folha de ouro com part\'iculas $\alpha$ altamente energ\'eticas, oriundas do decaimento radioativo do Pol\^onio-214\footnote{\`A \'epoca, a maioria desses elementos radioativos n\~ao havia sido nomeada.  Eles eram referenciados segundo a posi\c{c}\~ao que ocupavam em certa cadeia de decaimento radioativo. Assim, os elementos oriundos de sucessivos decaimentos a partir do R\'adio-226 eram chamados R\'adio A, B, C, etc. Nos trabalhos originais de Rutherford, Geiger e Marsden, o elemento emissor das part\'iculas $\alpha$ usadas no experimento \'e denominado R\'adio C, que, hoje, sabemos tratar-se do $^{214}$Po, emissor de part\'iculas $\alpha$ com energias de 7.7~MeV.}. O experimento est\'a ilustrado na figura~\ref{fig:GeigerMarsden}. A fonte radioativa emite part\'iculas $\alpha$, que s\~ao n\'ucleos de H\'elio, com carga $+2e$ e massa de aproximadamente quatro vezes a massa de um pr\'oton, e 8000 vezes a massa de um el\'etron. Essas part\'iculas incidem sobre a folha de ouro e s\~ao espalhadas, repelidas pela carga positiva do \'atomo. 

Modelos diferentes para a estrutura at\^omica fazem previs\~oes distintas sobre a probabilidade de as part\'iculas incidentes serem defletidas a um determinado \^angulo. Assim, realizando-se uma contagem do n\'umero de part\'iculas espalhadas como fra\c{c}\~ao do n\'umero de part\'iculas incidentes no alvo, \'e poss\'ivel testar diferentes modelos at\^omicos. O primeiro passo, portanto, \'e investigar a previs\~ao de cada modelo quanto \`a fra\c{c}\~ao de part\'iculas desviadas a um dado \^angulo.

\begin{figure}
	\centering
	\includegraphics[trim=0 0 40 0 , clip, scale=.4]{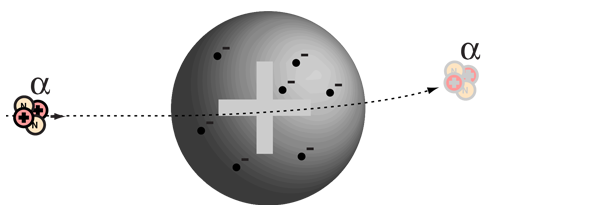}
	\caption{Incid\^encia frontal de uma part\'icula $\alpha$ em uma esfera uniformemente carregada, como um \'atomo de Thomson. Fonte: Hyperphysics \textcopyright~Rod Nave~\cite{Hyperphysics}, adaptado e publicado com autoriza\c{c}\~ao do autor.}
	\label{fig:frontal_collision}
\end{figure}

\subsubsection{Deflex\~ao das part\'iculas $\alpha$ no modelo de Thomson}
\label{sec:Thomson_deflection}

Qual \'e a predi\c{c}\~ao para o espalhamento do feixe incidente segundo o modelo at\^omico de Thomson? Nesse modelo a carga positiva do \'atomo est\'a distribu\'ida por todo seu volume, ocupando uma esfera de raio $r\sim 10^{-10}$~m. Pelo fato de essa carga estar assim ``dilu\'ida'', o efeito repulsivo \'e baixo e a part\'icula $\alpha$ incidente \'e muito pouco desviada. Esse resultado pode ser obtido mediante uma aplica\c{c}\~ao simples da  conserva\c{c}\~ao da energia, e oferece a possibilidade de se discutir esse princ\'ipio fundamental em uma situa\c{c}\~ao diversa dos sistemas mec\^anicos simples em que esse assunto \'e usualmente introduzido, como blocos em rampas ou em queda livre. A figura~\ref{fig:frontal_collision} ilustra a incid\^encia frontal de uma part\'icula $\alpha$ em um \'atomo de Thomson, i.e. uma esfera uniformemente carregada com carga $+Ze$, e contendo em seu interior $Z$ el\'etrons de carga $-e$ cada, de maneira que a carga total \'e nula. Enquanto a part\'icula $\alpha$ est\'a longe do \'atomo, a for\c{c}a repulsiva n\~ao atua, pois o \'atomo \'e efetivamente neutro. Precisamos apenas considerar o que ocorre quando a part\'icula $\alpha$ come\c{c}a a adentrar a esfera.
Como os el\'etrons s\~ao muito mais leves do que a part\'icula $\alpha$, suas contribui\c{c}\~oes ao desvio ou frenagem da part\'icula $\alpha$ s\~ao desprez\'iveis\footnote{Essa conclus\~ao pode ser facilmente atingida usando conserva\c{c}\~ao de momento linear, ou apelando-se para a intui\c{c}\~ao usando como analogia a colis\~ao entre uma bola de basquete e uma bolinha de gude, situa\c{c}\~ao em que \'e f\'acil ver que apenas a bolinha de gude \'e desviada significativamente.}, e podemos restringir a an\'alise \`a intera\c{c}\~ao com a carga positiva uniformemente distribu\'ida no \'atomo. Nesse caso, \`a medida que a part\'icula $\alpha$ adentra o \'atomo, a carga positiva que a repele \'e efetivamente menor, at\'e ser igual a zero no centro da esfera. Portanto a part\'icula $\alpha$ est\'a submetida a uma maior for\c{c}a de repuls\~ao quando ela come\c{c}a a adentrar a esfera, igual a 
\begin{equation}
    F_\text{max} = -k\frac{Ze^2}{r^2}.
\end{equation}
Portanto o trabalho realizado pela for\c{c}a repulsiva quando a part\'icula atravessa da extremidade ao centro da esfera, percorrendo uma dist\^ancia $r$, certamente \'e \emph{inferior} \`a $F_\text{max}\times r$, de modo que a varia\c{c}\~ao da energia cin\'etica da part\'icula $\alpha$ \'e
\begin{equation}
	\Delta K_\alpha \lesssim - k\dfrac{Ze^2}{r}\sim -\dfrac{10^{-13}~{\rm m}}{r}~{\rm MeV},
	\label{eq:DeltaK}
\end{equation}
onde substitu\'imos os valores num\'ericos $Z=79$ (correspondendo ao elemento ouro), $e\approx 1.6\times 10^{-19}$~C a carga el\'etrica elementar, $k\approx 9\times 10^9~{\rm J\cdot m/C}^2$ a constante de Coulomb, e convertemos $1~{\rm J}=(1.6\times 10^{-19})^{-1}$~eV. 

A energia cin\'etica de uma part\'icula $\alpha$ emitida por $^{214}$Po \'e 7.7~MeV. 

Ou seja, para $r\sim 10^{-10}$~m, a part\'icula $\alpha$ perde apenas cerca de 0.1\% de sua energia cin\'etica ao atravessar o \'atomo, uma varia\c{c}\~ao desprez\'ivel, de modo que o \'atomo de Thomson \'e essencialmente transparente \`as part\'iculas $\alpha$ altamente energ\'eticas.

\begin{figure}
	\centering
	\includegraphics[scale=.17]{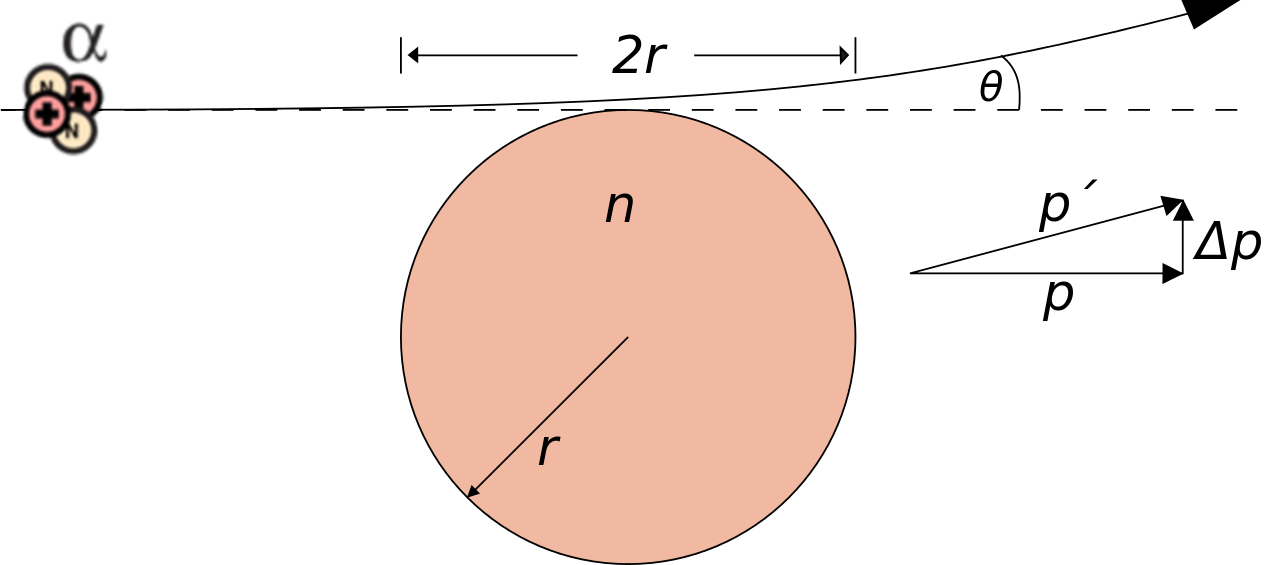}
	\caption{Part\'icula $\alpha$ colidindo tangencialmente com uma esfera uniforme. Fonte: Wikimedia Commons, sob licen\c{c}a Creative Commons BY-SA 4.0~\cite{BY-SA40}.}
	\label{fig:thomson_tangencial}
\end{figure}
O que ocorre no caso de uma colis\~ao n\~ao-frontal? O maior desvio se d\'a quando a colis\~ao \'e tangencial, como ilustrada na figura~\ref{fig:thomson_tangencial}. Isso porque uma part\'icula $\alpha$ que penetra a esfera, chegando a uma dist\^ancia $d<r$ de seu centro, interage somente com a carga contida no interior da esfera de raio $d$ (pela lei de Gauss), portanto inferior \`a carga total do \'atomo, o que reduz o efeito repulsivo. Sendo $v_\alpha$ a velocidade inicial da part\'icula $\alpha$, o intervalo de tempo necess\'ario para atravessar a esfera de di\^ametro $2r$ \'e
\begin{equation}
	\Delta t \simeq \dfrac{2r}{v_\alpha}.
	\label{eq:Deltat}
\end{equation}
Durante esse intervalo de tempo, a part\'icula est\'a submetida \`a repuls\~ao coulombiana, que causa uma varia\c{c}\~ao $\Delta p_y$ da componente transversal da velocidade da part\'icula, dada por 
\begin{equation}
	F_\text{Coulomb} = m\dfrac{\Delta v_y}{\Delta t} = k\dfrac{2Ze^2}{r^2}.
\end{equation}
Da figura~\ref{fig:thomson_tangencial} v\^e-se que o \^angulo $\theta$ de desvio \'e tal que
\begin{equation}
	\sin\theta = \dfrac{\Delta v_y}{v^\prime} \simeq \dfrac{\Delta v_y}{v_\alpha},
	\label{eq:theta}
\end{equation}
onde usamos o fato de que $v^\prime \simeq v_\alpha$, como j\'a esperado devido \`a an\'alise energ\'etica discutida acima. Dessa mesma an\'alise segue-se que o \^angulo de desvio \'e pequeno, $\theta\ll 1$ e portanto $\sin\theta\approx \theta$. Logo, das eqs.~(\ref{eq:Deltat}) e (\ref{eq:theta}), vem que
\begin{equation}
	\theta \simeq  k\dfrac{4Ze^2}{mv_\alpha^2 r} \simeq k\dfrac{2Ze^2}{K_\alpha r} \simeq 0.02^{\rm o}.
\end{equation}

Em suma, o desvio \emph{m\'aximo} da trajet\'oria da part\'icula $\alpha$, ap\'os interagir com um \'atomo de Thomson, \'e $0.02^{\rm o}$. No experimento de Geiger-Marsden a folha de ouro tinha espessura da ordem de $1~\mu$m, o que significa que o n\'umero de camadas de \'atomos da folha \'e $\mathcal{N}=10^{-6}~{\rm m}/10^{-10}~{\rm m}=10^4$. Na colis\~ao aleat\'oria com muitos \'atomos, alguns desvios ocorrer\~ao em uma dire\c{c}\~ao, e outros na dire\c{c}\~ao oposta. Nessa situa\c{c}\~ao, o valor m\'edio da deflex\~ao total \'e $\sqrt{\mathcal{N}}\theta \simeq 2^{\rm o}$, e \'e necess\'ario uma combina\c{c}\~ao extremamente improv\'avel de deflex\~oes numa mesma dire\c{c}\~ao para que se totalize um desvio maior que $90^{\rm o}$~\footnote{O c\'alculo dessa probabilidade no caso em quest\~ao \'e demasiadamente complexo para ser apresentado no ensino m\'edio, mas a no\c{c}\~ao de qu\~ao improv\'avel \'e essa situa\c{c}\~ao pode ser explicada da seguinte maneira. Suponha que, em cada colis\~ao, a deflex\~ao s\'o pode ser a m\'axima, $0.02^{\rm o}$, em uma dire\c{c}\~ao ou na oposta (``para cima'' ou ``para baixo'', caso a part\'icula $\alpha$ passe acima ou abaixo do \'atomo, vide figura~\ref{fig:thomson_tangencial}). Para totalizar uma deflex\~ao de $90^{\rm o}$ \'e necess\'ario que haja $90^{\rm o}/0.02^{\rm o}=4500$ deflex\~oes a mais em uma dire\c{c}\~ao do que em outra. Ou seja, \'e como se lan\c{c}\'assemos 10000 moedas e requer\^essemos ao menos 7250 ``coroas''. A probabilidade disso ocorrer \'e 
	\[ \dfrac{10000!}{7250!\ 2750!} \left(\dfrac{1}{2}\right)^{10000} \approx 10^{-450}.\]
	Uma aproxima\c{c}\~ao ainda melhor seria sup\^or que a part\'icula pode ser desviada em quatro dire\c{c}\~oes poss\'iveis: ``para cima'', ``para baixo'', ``para a esquerda'' e ``para a direita''. Nesse caso o fator 1/2 acima deve ser substitu\'ido por 1/4, resultando em uma probabilidade $\approx 10^{-3450}$.}. De fato, a probabilidade de que isso ocorra \'e da ordem de $10^{-3500}$~\cite{Eisberg}, ou seja, efetivamente zero. 

\subsubsection{O modelo de Rutherford}
\label{sec:Rutherford_model}

Ap\'os realizarem o experimento, Geiger e Marsden notaram que cerca de 1 em cada 8000 part\'iculas $\alpha$ eram defletidas de um \^angulo maior que $90^{\rm o}$, em flagrante desacordo com a expectativa do modelo de Thomson. Rutherford assim expressou seu espanto diante de tal resultado:
\begin{flushright}
	\parbox{.4\textwidth}{\it ``Foi a coisa mais incr\'ivel que j\'a vi em minha vida. Foi quase t\~ao incr\'ivel quanto se lan\c{c}\'assemos um m\'issil contra um len\c{c}o de papel, e ele rebatesse e voltasse para nos atingir.''}\\[2mm]
	E.~Rutherford 
\end{flushright}

O resultado do experimento demandava uma reformula\c{c}\~ao da estrutura at\^omica, que Rutherford completou em 1911~\cite{Rutherford:1911}. Sua solu\c{c}\~ao pode ser compreendida a partir da equa\c{c}\~ao~(\ref{eq:DeltaK}): para que a repuls\~ao e\-le\-tros\-t\'a\-ti\-ca seja suficiente para frear completamente uma part\'icula $\alpha$ frontalmente incidente e faz\^e-la retornar no sentido oposto, \'e necess\'ario que a varia\c{c}\~ao da energia cin\'etica seja da ordem da energia cin\'etica incidente, $\Delta K_\alpha \sim 7.7$~MeV, o que demanda que a carga positiva do \'atomo esteja concentrada em uma regi\~ao de raio $r\lesssim 10^{-14}$~m.
\begin{figure}
	\centering
	\includegraphics[scale=.09]{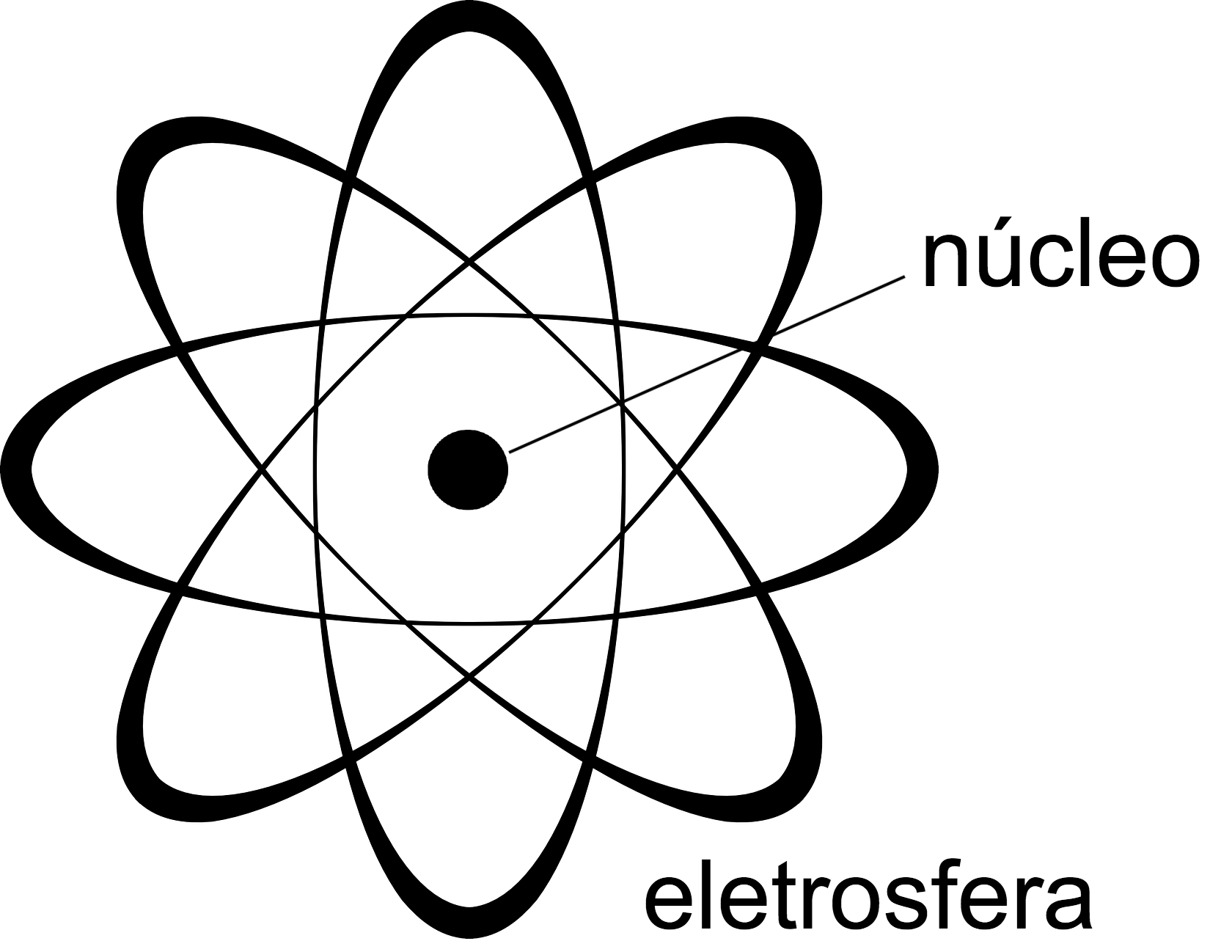}
	\caption{Esquematiza\c{c}\~ao do modelo de Rutherford para o \'atomo, com a carga positiva concentrada no n\'ucleo ao centro, circundada pelas \'orbitas dos el\'etrons, que constituem a regi\~ao denominada eletrosfera. Fonte: Dom\'inio P\'ublico. 	}
	\label{fig:atomo_Rutherford}
\end{figure}

Rutherford prop\^os, ent\~ao, que toda a carga positiva do \'atomo estivesse contida em um \'unico ponto, em torno do qual os el\'etrons orbitam a uma dist\^ancia do centro da ordem do raio at\^omico, $r\sim 10^{-10}$~m. Essa regi\~ao da carga positiva do \'atomo foi posteriormente denominada \emph{``n\'ucleo at\^omico''}. Um esquema do modelo do \'atomo proposto por Rutherford est\'a ilustrado na figura~\ref{fig:atomo_Rutherford}.

\begin{figure}
	\centering
	\includegraphics[scale=.4]{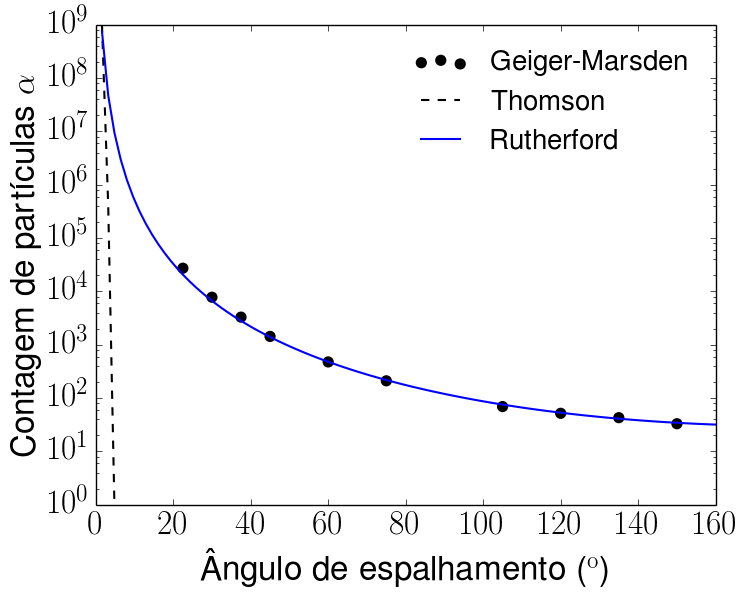}
	\caption{Contagem de part\'iculas $\alpha$ incidentes sobre a placa de ZnS em fun\c{c}\~ao do \^angulo de espalhamento. S\~ao mostradas as curvas te\'oricas para o modelo de Thomson (tracejada) e o modelo de Rutherford (linha s\'olida azul), bem como os pontos experimentais obtidos por Geiger e Marsden. O acordo entre o experimento e o modelo de Rutherford \'e impressionante, corroborando a hip\'otese de que a carga positiva do \'atomo se concentra em um pequeno n\'ucleo.}
	\label{fig:rutherford_data}
\end{figure}

A partir desse modelo, \'e poss\'ivel computar o n\'umero de part\'iculas incidentes na placa de ZnS como fun\c{c}\~ao do \^angulo $\theta$ de espalhamento, que \'e proporcional a
\begin{equation}
	N(\theta) \propto \dfrac{1}{K_\alpha^2\sin^4\dfrac{\theta}{2}}.
	\label{eq:Rutherford_scattering}
\end{equation}
A figura~\ref{fig:rutherford_data} ilustra os dados experimentais obtidos por Geiger-Marsden, a curva te\'orica predita pelo modelo de Rutherford, dada pela equa\c{c}\~ao~(\ref{eq:Rutherford_scattering}), bem como a curva esperada para o modelo de Thomson. O acordo entre o modelo de Rutherford e os dados experimentais \'e surpreendente, indicando a validade da hip\'otese do n\'ucleo at\^omico.

\subsubsection{Discuss\~ao: como se constr\'oi conhecimento cient\'ifico?}
\label{sec:ciencia}

Os momentos did\'aticos sugeridos acima culminam em uma excelente oportunidade de se discutir, em sala, sobre a natureza da ci\^encia e o ``m\'etodo cient\'ifico''. Nota-se, no processo descrito acima, as seguintes etapas: (i) constru\c{c}\~ao de um modelo te\'orico para representa\c{c}\~ao de um objeto f\'isico, como o modelo de Thomson para o \'atomo; (ii) elabora\c{c}\~ao de um experimento capaz de testar as previs\~oes te\'oricas do modelo, como o experimento de Geiger-Marsden no caso presente; note que, para isso, \'e essencial que o modelo seja capaz de prever o resultado de algum experimento, e que saibamos determinar essa previs\~ao, como fizemos acima para o \'atomo de Thomson; (iii) realiza-se o experimento e, se o resultado contradiz a expectativa do modelo, ele deve ser modificado; (iv) um novo modelo \'e formulado de modo a explicar o resultado dos experimentos; com isso, retorna-se ao item (i) acima.

Esse processo, embora extremamente simplificado, ilustra bem a dial\'etica entre teoria e experimento, fundamental para o desenvolvimento cient\'ifico. Assim como n\~ao basta formular uma teoria que n\~ao se sustente frente a resultados experimentais, \'e igualmente ilus\'orio imaginar que a ci\^encia possa progredir pela simples observa\c{c}\~ao \emph{passiva} de fen\^omenos, sem uma estrutura te\'orica pr\'evia que sirva de orienta\c{c}\~ao para o desenvolvimento de experimentos voltados a test\'a-la. Ressalta-se, a esse respeito, que a ideia do experimento de Geiger-Marsden surgiu ap\'os a formula\c{c}\~ao do modelo at\^omico de Thomson. Antes disso, o \'atomo era visto como uma esfera neutra, e n\~ao se esperaria nenhum desvio das part\'iculas $\alpha$ por for\c{c}as eletrost\'aticas, portanto o experimento jamais teria sido elaborado.

O(a) docente pode propor que os(as) estudantes debatam em sala as seguintes quest\~oes:
\begin{itemize}
	\item \'E necess\'ario enxergarmos o \'atomo para entendermos sua estrutura?
	\item Qual \'e o papel dos experimentos no desenvolvimento das ci\^encias? 
	\item E qual \'e o papel dos modelos te\'oricos?
	\item Por que \'e importante que uma teoria fa\c{c}a previs\~oes concretas sobre resultados experimentais?
	\item Qual o papel da predi\c{c}\~ao te\'orica no desenvolvimento cient\'ifico?
	\item O que \'e uma teoria cient\'ifica?
	\item O que \'e ci\^encia?
\end{itemize}

Enfim, essa discuss\~ao tamb\'em demonstra a import\^ancia da matem\'atica  para o desenvolvimento da F\'isica. \'E somente atrav\'es desses c\'alculos que podemos determinar o \^angulo de desvio esperado para o \'atomo de Thomson e a probabilidade de se observar um desvio total maior do que 90$^{\rm o}$. E \'e somente devido \`a discrep\^ancia radical entre a previs\~ao te\'orica e o resultado experimental que podemos considerar o modelo de Thomson como insatisfat\'orio. Aos estudantes deve restar claro, nesse caso, que a matem\'atica n\~ao aparece como mero exerc\'icio abstrato para a obten\c{c}\~ao de n\'umeros que servem como resposta de uma quest\~ao de prova. Ela exerce um papel essencial na falsifica\c{c}\~ao de um modelo e no consequente desenvolvimento da F\'isica, para um melhor entendimento dos fen\^omenos da Natureza".

\subsection{Estrutura do n\'ucleo at\^omico: pr\'otons e n\^eutrons}
\label{sec:protons_neutrons}

Em 1919, continuando suas investiga\c{c}\~oes a respeito da intera\c{c}\~ao entre part\'iculas $\alpha$ e a mat\'eria, Rutherford notou que a passagem dessas part\'iculas pelo ar produzia um fluxo de n\'ucleos de hidrog\^enio detectados na chapa de ZnS~\cite{Rutherford:1919}. O efeito desaparecia quando o ar era substituindo por g\'as oxig\^enio, O$_2$, ou por di\'oxido de carbono, CO$_2$, mas era ainda mais not\'avel para g\'as nitrog\^enio, N$_2$. Hoje, \'e sabido que Rutherford observara a rea\c{c}\~ao nuclear 
\begin{center}
	$^{14}$N + $^4$He $\to\, ^{17}$O + $^{1}$H.
\end{center}
A partir desses resultados, Rutherford postulou que o n\'ucleo de hidrog\^enio --- o mais leve dentro todos os elementos --- \'e um constituinte fundamental de todos os n\'ucleos at\^omicos~\cite{Rutherford:1920}, e que, no processo acima, um deles \'e ejetado do nitrog\^enio pela colis\~ao com a part\'icula $\alpha$. Com isso, Rutherford reavivou a tese proposta um s\'eculo antes por William Prout, que notara que as massas molares de v\'arios elementos eram m\'ultiplos inteiros da massa molar do hidrog\^enio, e sup\^os ser esse o elemento primordial do qual todos s\~ao constitu\'idos\footnote{A tese de Prout foi abandonada poucos anos depois de proposta, ao se constatar a exist\^encia de v\'arios elementos cujas massas molares {\bf n\~ao} s\~ao m\'ultiplos inteiros da massa molar do H, como o cloro (massa molar 35.45~g). Hoje se sabe que essa diverg\^encia \'e devido \`a exist\^encia de is\'otopos, i.e. variantes de um mesmo elemento com massas diferentes. A massa de cada is\'otopo obedece, em boa aproxima\c{c}\~ao, a rela\c{c}\~ao notada por Prout, mas a massa molar do elemento \'e a m\'edia das massas dos diferentes is\'otopos, ponderada por suas abund\^ancias, podendo portanto resultar em valores fracion\'arios de massa. No entanto, a exist\^encia de is\'otopos de um mesmo elemento era desconhecida no s\'eculo XIX, e a hip\'otese de Prout acabou n\~ao vingando, at\'e ser revivida por Rutherford.}. Prout designou de \emph{pr\'otil} esse elemento primordial, enquanto Rutherford chamou de \emph{pr\'oton} o n\'ucleo de H (ambas palavras advindas do grego antigo \emph{protos}, que significa \emph{primeiro}).

A partir dessa descoberta de Rutherford, o \'atomo passou a ser visto como constitu\'ido de pr\'otons e el\'etrons. Como a massa do el\'etron \'e $\approx$ 1800 vezes menor que a do pr\'oton, sua contribui\c{c}\~ao \`a massa at\^omica \'e desprez\'ivel, de modo que se supunha que toda a massa fosse devida ao n\'umero de pr\'otons no n\'ucleo. Assim, um n\'ucleo com n\'umero de massa $A$ (i.e. cuja massa \'e $A$ vezes a massa do H) era suposto composto de $A$ pr\'otons. Mas, como explicar a observa\c{c}\~ao de que a carga de um n\'ucleo \'e quase sempre menor do que a carga de $A$ pr\'otons? Para isso, supunha-se que, al\'em dos el\'etrons da eletrosfera, havia tamb\'em el\'etrons no interior do n\'ucleo, tal que um n\'ucleo de massa $A$ e carga $Z$ (em unidades da massa e carga do n\'ucleo de H) era visto como composto de $A$ pr\'otons e $A-Z$ el\'etrons. Por um lado, a exist\^encia de el\'etrons nucleares era uma hip\'otese bem-vinda, pois ajudava a explicar decaimentos radioativos que eram acompanhados pela emiss\~ao de el\'etrons --- chamados decaimentos do tipo $\beta$, discutidos abaixo na se\c{c}\~ao~\ref{sec:beta}. Por outro lado, a exist\^encia de el\'etrons no interior do n\'ucleo at\^omico engendra v\'arias contradi\c{c}\~oes com resultados experimentais. Por exemplo, o princ\'ipio da incerteza de Heisenberg assegura que um el\'etron confinado em uma regi\~ao de raio $r\lesssim 10^{-14}$~m teria uma energia cin\'etica da ordem de $20$~MeV, muito acima da energia t\'ipica dos el\'etrons emitidos em decaimentos $\beta$, da ordem de 1 a 5 MeV. Ademais, medi\c{c}\~oes do momento angular e momento magn\'etico do n\'ucleo de nitrog\^enio-14 indicam que ele deve ser constitu\'ido de um n\'umero \emph{par} de part\'iculas\footnote{Experimentos apontam que o $^{14}$N tem momento angular inteiro (em unidades de $\hbar$), ao passo que o momento angular intr\'inseco de pr\'otons e el\'etrons \'e semi-inteiro, igual a $\hbar/2$. O soma de um n\'umero \'impar de momentos angulares semi-inteiros \'e, tamb\'em, semi-inteiro. Portanto, o $^{14}$N deve ser constitu\'ido de um n\'umero par de pr\'otons e el\'etrons.}, em contradi\c{c}\~ao com a expectativa de 14 pr\'otons + 7 el\'etrons.

Essas contradi\c{c}\~oes foram resolvidas em 1932, quando James Chadwick demonstrou experimentalmente a exist\^encia do \emph{n\^eutron}: part\'icula eletricamente neutra com massa muito semelhante \`a do pr\'oton. Assim, estabeleceu-se que um n\'ucleo de carga $Z$ e massa $A$ \'e composto de $Z$ pr\'otons e $A-Z$ n\^eutrons. Um n\'ucleo de nitrog\^enio-14 \'e, portanto, constitu\'ido de 7 pr\'otons + 7 n\^eutrons, totalizando um n\'umero par de constituintes, como requerido pelos experimentos. No ano seguinte, em 1933, Enrico Fermi explicou o decaimento $\beta$ sob essa nova perspectiva: o el\'etron emitido n\~ao existia anteriormente no n\'ucleo, mas surgiu do decaimento de um n\^eutron em um pr\'oton e um el\'etron --- uma perspectiva revolucion\'aria de transmuta\c{c}\~ao de part\'iculas que inaugurou uma nova era na F\'isica de Altas Energias.

A descoberta dos n\^eutrons tamb\'em possibilitou uma explica\c{c}\~ao simples \`a exist\^encia de is\'otopos. A natureza de um elemento \'e definida pelo n\'umero de pr\'otons que existem no n\'ucleo --- i.e. pela sua carga el\'etrica ---, pois \'e isso que define a quantidade de el\'etrons na eletrosfera, e portanto determina as suas possibilidades de liga\c{c}\~oes qu\'imicas, que nada mais s\~ao do que reminisc\^encias da intera\c{c}\~ao eletromagn\'etica. Por outro lado, a massa do elemento \'e determinada pela quantidade de pr\'otons e n\^eutrons no n\'ucleo. Portanto \'e poss\'ivel que um mesmo elemento possua variantes com diferentes massas, dada pela diferen\c{c}a no n\'umero de n\^eutrons no n\'ucleo.

Pr\'otons e n\^eutrons s\~ao coletivamente denominados {\bf\it n\'ucleons}, i.e. part\'iculas do n\'ucleo.

\subsection{O raio nuclear}
\label{sec:raio}

Qual \'e o tamanho do n\'ucleo at\^omico? Os resultados do experimento de Geiger-Marsden, aliados \`a an\'alise energ\'etica que resultou na equa\c{c}\~ao~(\ref{eq:DeltaK}) acima, j\'a indicam que pr\'otons e n\^eutrons devem estar confinados em uma regi\~ao de dimens\~oes $\lesssim 10^{-14}$~m. Mas essa estimativa \'e apenas um limite superior. Uma determina\c{c}\~ao mais precisa do raio nuclear pode ser obtida atrav\'es de experimentos de espalhamento de el\'etrons altamente energ\'eticos incidentes sobre o n\'ucleo --- semelhantes ao experimento de Geiger-Marsden, mas com el\'etrons ao inv\'es de part\'iculas $\alpha$.

Uma discuss\~ao desses experimentos em sala apresenta uma excelente oportunidade de se abordar temas como a \emph{dualidade onda-part\'icula}~\cite{CarvalhoDorsch:2021lvd} (veja a ref.~\cite{Elcio} para uma sequ\^encia did\'atica investigativa sobre esse t\'opico) e a \emph{difra\c{c}\~ao de ondas} ao atravessar um obst\'aculo.

\begin{figure*}
	\centering
	\includegraphics[trim=20 0 10 15, clip, scale=.8]{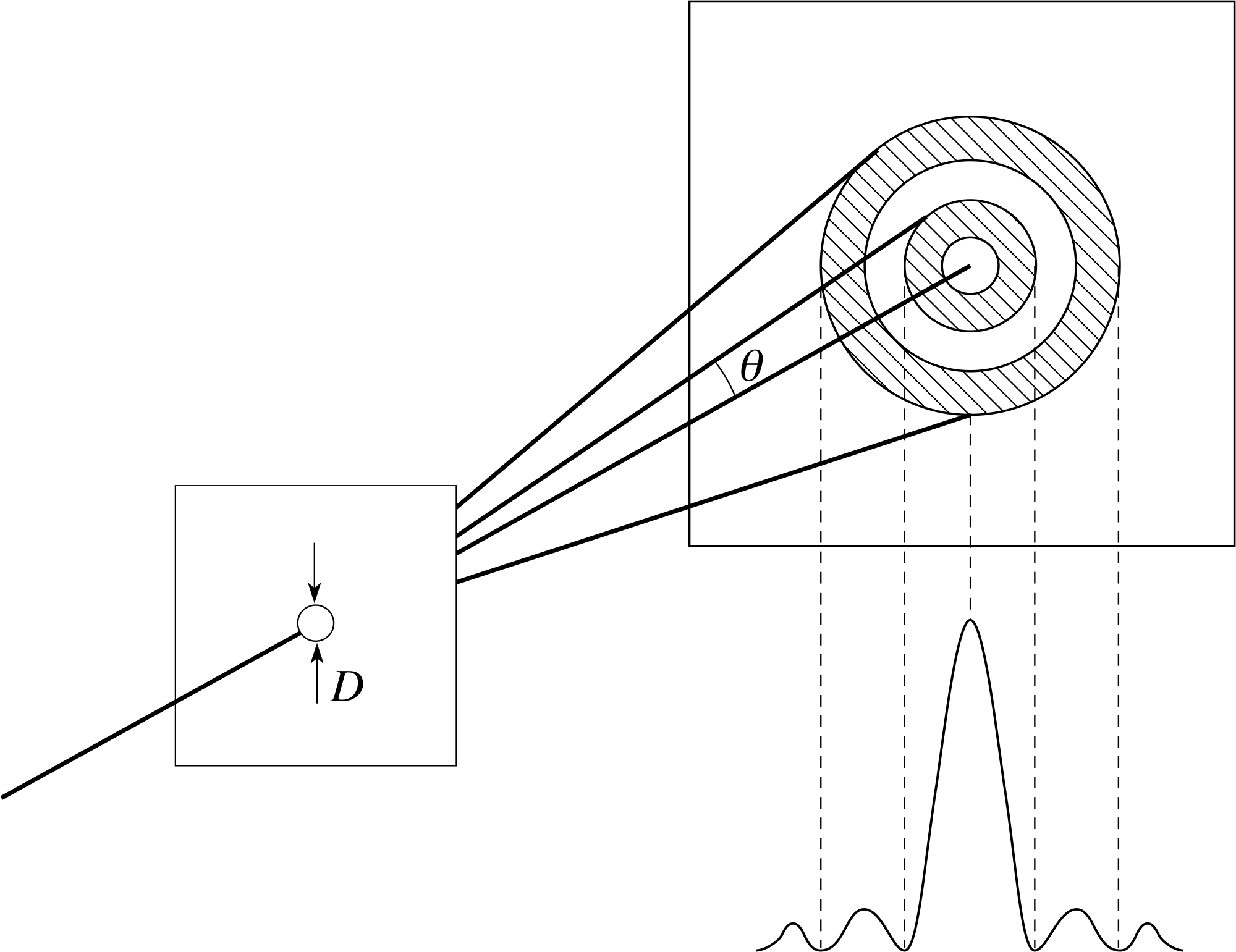}\qquad\qquad
	\includegraphics[trim=25 0 17 20 , clip, scale=1.7]{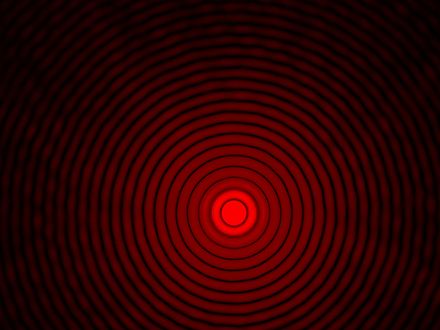}
	\caption{(Esquerda) Ilustra\c{c}\~ao do padr\~ao de difra\c{c}\~ao produzido por uma onda de comprimento $\lambda$ ao atravessar um orif\'icio de di\^ametro $D$. O disco luminoso no centro da imagem \'e circundado de an\'eis conc\^entricos, alternando regi\~oes sombreadas e luminosas. O gr\'afico ilustra a intensidade da luz nessas regi\~oes, em padr\~ao caracter\'istico da difra\c{c}\~ao por um orif\'icio. O \^angulo $\theta$ referente \`a localiza\c{c}\~ao do primeiro anel sombreado \'e da ordem de $\theta\sim \lambda/D$. Fonte: PPlato/University of Reading~\cite{PPlato}. (Direita) Padr\~ao de difra\c{c}\~ao obtido pela passagem de um laser por um orif\'icio. Fonte: Wikimedia Commons/Dom\'inio P\'ublico.}
	\label{fig:difracao}
\end{figure*}

De fato, devido \`a dualidade onda-part\'icula, o feixe de el\'etrons incidentes no alvo comporta-se como uma onda de comprimento
\begin{equation}
	\lambda = \dfrac{h}{p},
\end{equation}
onde $h$ \'e a constante de Planck --- uma constante fundamental da Natureza, de valor $h\approx 4.136\times 10^{-21}~{\rm MeV\cdot s}$ ---, e $p$ o momento linear dos el\'etrons. Ao incidir sobre o n\'ucleo at\^omico, essa onda sofre difra\c{c}\~ao, tal como a luz difrata ao atravessar um orif\'icio circular, vide figura~\ref{fig:difracao}. Se o el\'etron for suficientemente energ\'etico, seu comprimento de onda associado ser\'a suficientemente pequeno para que se observe um m\'inimo de difra\c{c}\~ao no padr\~ao de espalhamento. A posi\c{c}\~ao $\theta$ desse m\'inimo permite determinar a dimens\~ao $D$ do obst\'aculo a partir da rela\c{c}\~ao 
\begin{equation}
	\theta\sim \dfrac{\lambda}{D}.
	\label{eq:difr}
\end{equation}

\begin{figure}
	\centering
	\includegraphics[scale=.45]{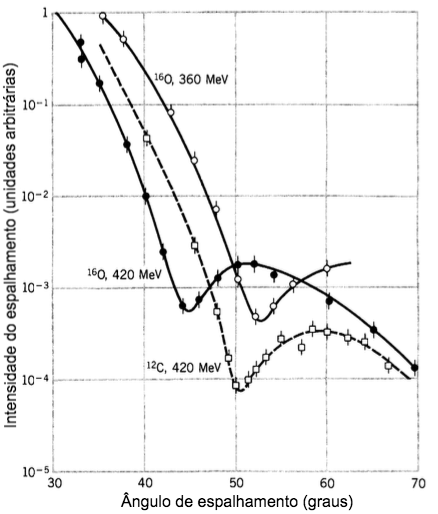}
	\caption{No experimento de Ehrenberg et al.~\cite{Ehrenberg}, feixes de el\'etrons com energias $360$~MeV e $420$~MeV s\~ao espalhados por n\'ucleos de carbono e oxig\^enio. O gr\'afico acima~\cite{Krane}, compilado a partir dos resultados de~\cite{Ehrenberg}, mostra a intensidade do feixe incidente no detector em fun\c{c}\~ao do \^angulo de deflex\~ao. A exist\^encia de um m\'inimo se deve ao fen\^omeno de difra\c{c}\~ao dos el\'etrons ao atravessarem os n\'ucleos, similar aos an\'eis sombreados no padr\~ao de difra\c{c}\~ao da luz ao passar por um obst\'aculo circular.}
	\label{fig:Krane}
\end{figure}

A figura~\ref{fig:Krane} demonstra a exist\^encia de um tal m\'inimo na intensidade do feixe resultante ap\'os o espalhamento de el\'etrons altamente energ\'eticos sobre alvos de $^{12}$C e $^{16}$O. Nesse caso, os n\'ucleos agem como o orif\'icio na figura~\ref{fig:difracao}, e o feixe eletr\^onico tamb\'em sofre difra\c{c}\~ao. Por exemplo, para o $^{12}$C o m\'inimo ocorre em $\theta\approx 50^\circ\approx0.873$~rad quando se incide el\'etrons de energia cin\'etica $E\approx 420$~MeV. Note que essa energia \'e muito maior que a energia de repouso dos el\'etrons, $m_e c^2\approx 0.511$~MeV, portanto esses el\'etrons se movem a uma velocidade pr\'oxima \`a da luz, $c\approx 3\times 10^8$~m/s, e vale a rela\c{c}\~ao $E=pc$. O comprimento de onda \'e ent\~ao $\lambda = hc/E = 2.95\times 10^{-15}$~m, e, da equa\c{c}\~ao~(\ref{eq:difr}), vem que
\begin{equation}
	D \sim 3~{\rm fm},
\end{equation}
com 1~fm = $10^{-15}$~m. O valor tabelado para o raio do n\'ucleo de $^{12}$C \'e $\approx 2.47$~fm~\cite{Angeli}, portanto seu di\^ametro \'e $\sim 5$~fm, n\~ao muito distante da estimativa acima.

Ao contr\'ario do padr\~ao de difra\c{c}\~ao da luz por um orif\'icio circular, a intensidade do m\'inimo na figura~\ref{fig:Krane} n\~ao \'e exatamente zero porque o n\'ucleo n\~ao possui bordas bem definidas. Ademais, o c\'alculo acima \'e apenas uma estimativa da dimens\~ao do n\'ucleo, que \'e, em geral, da ordem de 1~fm. Para a obten\c{c}\~ao de um valor mais preciso, \'e necess\'ario uma modelagem mais detalhada da forma do n\'ucleo, que foge do escopo de uma discuss\~ao no ensino m\'edio. 

\subsection{A for\c{c}a nuclear}

\subsubsection{Problema da estabilidade do n\'ucleo}
\label{sec:estabilidade}

Como \'e poss\'ivel que o n\'ucleo at\^omico seja est\'avel, se \'e composto somente de part\'iculas neutras ou com carga positiva, todas confinadas a uma dist\^ancia m\'utua da ordem de $10^{-15}$~m e, portanto, com os pr\'otons constantemente sujeitos a intensas for\c{c}as eletrost\'aticas repulsivas?

Tal quest\~ao pode ser proposta \`a turma como ponto de partida de uma discuss\~ao que desemboque na conceitua\c{c}\~ao da \emph{for\c{c}a nuclear}. 

Em um primeiro momento, os estudantes usualmente levantam duas propostas para explicar a estabilidade nuclear:
\begin{itemize}
	\item A {\bf for\c{c}a gravitacional} \'e atrativa e talvez possa manter o n\'ucleo est\'avel diante da repuls\~ao eletrost\'atica. No entanto --- e aqui entra novamente o valor da matem\'atica para corroborar ou excluir hip\'oteses f\'isicas ---, como j\'a discutido na sequ\^encia anterior~\cite{CarvalhoDorsch:2021lvd}, a intera\c{c}\~ao gravitacional \'e \emph{muito} mais fraca do que a eletromagn\'etica, e jamais seria suficiente para equilibrar a repuls\~ao dos pr\'otons. De fato, em um n\'ucleo composto de $Z$ pr\'otons e $A-Z$ n\^eutrons, cada pr\'oton \'e submetido a uma for\c{c}a repulsiva $F_\text{el}$ devido \`as outras $Z-1$ cargas positivas, e a uma atra\c{c}\~ao gravitacional $F_\text{grav}$ devido a $A-1$ pr\'otons e n\^eutrons, dadas por
	\begin{equation}
		F_\text{el} \sim k\dfrac{(Z-1)\,e^2}{r^2},\qquad 
		F_\text{grav} \sim G\dfrac{(A-1)\, m_p^2}{r^2},
	\end{equation}
	onde supomos que as massas dos pr\'otons e n\^eutrons s\~ao iguais --- o que \'e v\'alido em boa aproxima\c{c}\~ao ---, e que a dist\^ancia m\'edia de afastamento entre cada part\'icula \'e da mesma ordem de magnitude, $r\sim 10^{-15}$~m.
	
	Substituindo valores num\'ericos,
	\begin{equation*}\begin{split} 
		k\approx &~8.98\times 10^9~{\rm N\cdot m^2/C^2},\\
		G\approx &~6.67\times 10^{-11}~{\rm N\cdot m^2/kg^2},\\
		 e&\approx 1.6\times 10^{-19}~{\rm C},\\
		 m_p&\approx 1.67\times 10^{-27}~{\rm kg},
	\end{split}\end{equation*}
	vem que a condi\c{c}\~ao de equil\'ibrio entre essas for\c{c}as requer
	\begin{equation}
		A-1 \sim (Z-1) \dfrac{ke^2}{Gm_p^2} \implies A \sim 10^{36} Z,
	\end{equation}
	ou seja, seria necess\'ario um n\'umero imensamente maior de n\^eutrons do que de pr\'otons, ao passo que o n\'ucleo mais pesado observado na Natureza --- i.e. n\~ao produzido artificialmente --- \'e o $^{238}_{\ 92}$U, para o qual $A\approx 2.6~Z$.
	\item Os {\bf el\'etrons na eletrosfera} atraem os pr\'otons no n\'ucleo. No entanto, mesmo que houvesse uma tal for\c{c}a efetiva\footnote{Como os el\'etrons est\~ao em \'orbita em torno do n\'ucleo, em m\'edia essa for\c{c}a se anula.}, ela tenderia a desestabilizar o n\'ucleo ainda mais, pois atrai os pr\'otons em dire\c{c}\~ao \`a eletrosfera, para fora do n\'ucleo. 
\end{itemize}

\subsubsection{Discuss\~ao e levantamento de hip\'oteses}
\label{sec:discussoes_nuclearforce}
A conclus\~ao que a turma deve ser capaz de alcan\c{c}ar, nesse momento, \'e que, com as intera\c{c}\~oes conhecidas at\'e ent\~ao, n\~ao \'e poss\'ivel explicar a estabilidade do n\'ucleo at\^omico. O(a) docente pode, ent\~ao, estimular os(as) estudantes a discutir solu\c{c}\~oes para esse impasse, em um \'otimo exerc\'icio de levantamento de hip\'oteses.

Note que, \emph{nesse exerc\'icio, n\~ao h\'a respostas erradas}. No desenvolvimento hist\'orico-cient\'ifico, em tais situa\c{c}\~oes de impasse, \'e comum se propor hip\'oteses que, embora hoje descartadas, ou at\'e consideradas absurdas, levaram a importantes desenvolvimentos te\'oricos e experimentais posteriores\footnote{Exemplos abundam. Em 1917, ao notar que sua teoria da gravita\c{c}\~ao previa um Universo em constante contra\c{c}\~ao, portanto din\^amico, Einstein alterou suas equa\c{c}\~oes \emph{ad hoc}, introduzindo um termo chamado \emph{constante cosmol\'ogica} que levasse \`a predi\c{c}\~ao de um Universo est\'atico. Alguns anos depois, Hubble mostrou que o Universo est\'a em expans\~ao --- e o pr\'oprio Einstein afirmou, ent\~ao, que introduzir esse termo em suas equa\c{c}\~oes foi o maior erro de sua vida profissional. Hoje se sabe, contudo, que essa constante cosmol\'ogica, embora proposta a partir de uma suposi\c{c}\~ao incorreta, \'e considerada essencial para explicar o fato (descoberto posteriormente) de a expans\~ao ser \emph{acelerada}. Outro exemplo \'e a relut\^ancia de Einstein em aceitar a interpreta\c{c}\~ao de Copenhagen da mec\^anica qu\^antica, levando-o a postular a exist\^encia de ``vari\'aveis ocultas'', hip\'otese que deu origem ao importante teorema de Bell e a v\'arios experimentos que visavam test\'a-lo. Ainda digno de men\c{c}\~ao \'e o impasse relativo \`a energia dos el\'etrons emitidos em decaimentos $\beta$: a conserva\c{c}\~ao de energia-momento previa que os el\'etrons fossem emitidos com determinada energia cin\'etica, mas a observa\c{c}\~ao n\~ao condizia com a previs\~ao. Niels Bohr chegou a supor que a pr\'opria conserva\c{c}\~ao de energia fosse violada em n\'ivel qu\^antico --- uma hip\'otese hoje tida como ousada, at\'e absurda ---, enquanto Wolfgang Pauli prop\^os a exist\^encia de uma nova part\'icula que n\~ao era detectada no decaimento, e que carregava parte da energia. A hip\'otese de Pauli foi posteriormente verificada experimentalmente, salvaguardando o princ\'ipio da conserva\c{c}\~ao da energia. E nos problemas descritos acima, quanto \`a exist\^encia de el\'etrons no interior do n\'ucleo de $^{14}$N, tampouco faltavam propostas ex\'oticas antes da descoberta dos n\^eutrons, como a suposi\c{c}\~ao de que el\'etrons no interior do n\'ucleo n\~ao contribuem ao momento angular e momento magn\'etico total do n\'ucleo, por motivos inc\'ognitos.}. O importante, nessa atividade, \'e que os(as) discentes exercitem sua criatividade e capacidade argumentativa na elabora\c{c}\~ao de suas teses e, indo ainda al\'em, que tamb\'em se empenhem na elabora\c{c}\~ao de poss\'iveis maneiras de as testarem experimentalmente. 

Por exemplo, um(a) estudante pode propor que a intera\c{c}\~ao eletromagn\'etica \'e modificada quando as part\'iculas est\~ao a dist\^ancias muito pequenas, da ordem do raio do n\'ucleo at\^omico, de modo a abrandar a repuls\~ao. Tal hip\'otese poderia ser testada submetendo-se part\'iculas carregadas \`a intera\c{c}\~ao com pr\'otons no interior do n\'ucleo, como ocorre no espalhamento de el\'etrons altamente energ\'eticos, discutido na se\c{c}\~ao~\ref{sec:raio} acima. O fato de os resultados se conformarem ao esperado para el\'etrons submetidos \`a intera\c{c}\~ao coulombiana e ao fen\^omeno de difra\c{c}\~ao --- devido ao car\'ater ondulat\'orio do feixe de el\'etrons --- atesta contra essa hip\'otese. Para outros testes da validade da Lei de Coulomb vide ref.~\cite{Tu:2004}.

Ainda outra hip\'otese possivelmente levantada seria uma modifica\c{c}\~ao da gravita\c{c}\~ao nessas escalas de dist\^ancia. Entretanto, \'e dif\'icil testar a gravita\c{c}\~ao nessas situa\c{c}\~oes.

Outra possibilidade \'e se postular a exist\^encia de uma outra intera\c{c}\~ao, que pode ser denominada de \textbf{for\c{c}a nuclear}. 

\subsubsection{A for\c{c}a nuclear}
\label{sec:nuclearforce}

O(a) docente pode, ent\~ao, estimular discuss\~oes sobre as propriedades que essa nova intera\c{c}\~ao deve possuir para que solucione o problema da estabilidade nuclear sem engendrar outras contradi\c{c}\~oes. 

Em primeiro lugar, essa for\c{c}a deve possuir uma componente atrativa, e deve ser mais intensa do que a intera\c{c}\~ao eletromagn\'etica, para superar a repuls\~ao e estabilizar o n\'ucleo. 

Al\'em disso, n\~ao deve afetar os el\'etrons na eletrosfera, pois o \'atomo \'e bem descrito com base apenas na intera\c{c}\~ao eletromagn\'etica. O experimento de espalhamento de el\'etrons por pr\'otons, discutido na se\c{c}\~ao~\ref{sec:raio}, tamb\'em \'e muito bem descrito assumindo-se que a intera\c{c}\~ao el\'etron-pr\'oton \'e inteiramente de natureza eletromagn\'etica. Em outras palavras, n\~ao h\'a evid\^encia de que el\'etrons interajam com pr\'otons ou n\^eutrons por meio dessa for\c{c}a nuclear.

Ademais, a atua\c{c}\~ao dessa for\c{c}a se d\'a apenas a dist\^ancias da ordem do raio nuclear, $\sim 1$~fm. Por exemplo, as for\c{c}as de liga\c{c}\~ao intermoleculares s\~ao exclusivamente de natureza eletromagn\'etica, sendo desprez\'ivel a contribui\c{c}\~ao de outra forma de intera\c{c}\~ao entre os n\'ucleos dos \'atomos constituintes\footnote{No terceiro artigo desta sequ\^encia discutiremos como esse curto alcance da intera\c{c}\~ao nuclear se deve ao fato de a part\'icula mediadora da intera\c{c}\~ao (o an\'alogo do f\'oton na intera\c{c}\~ao eletromagn\'etica) ser \emph{massiva}.}.

\subsection{Rea\c{c}\~oes nucleares}
\label{sec:reacoes}

Consolidando-se entre os(as) discentes o conceito de for\c{c}a nuclear, pode-se enveredar em discuss\~oes sobre diversos fen\^omenos e aplica\c{c}\~oes associadas \`a F\'isica Nuclear. Trata-se de um f\'ertil contexto para explora\c{c}\~ao interdisciplinar em uma abordagem CTSA (Ci\^encia, Tecnologia, Sociedade e Ambiente) com tem\'aticas como:
\begin{itemize}
	\item fiss\~ao nuclear e suas aplica\c{c}\~oes b\'elicas e pac\'ificas;
	\item fus\~ao nuclear como fonte de energia do Sol, e sua import\^ancia para o desenvolvimento da vida na Terra;
	\item a forma\c{c}\~ao da maioria dos elementos em processos de fus\~ao no interior das estrelas, e por que ``somos poeira das estrelas'';
	\item a matriz energ\'etica no Brasil e no mundo: diversas fontes de energia e seus impactos ambientais;
	\item decaimentos radioativos e aplica\c{c}\~oes.
\end{itemize}

\subsubsection{Energia de liga\c{c}\~ao nuclear}
\label{sec:binding}

\begin{figure}
	\centering
    \includegraphics[scale=.45]{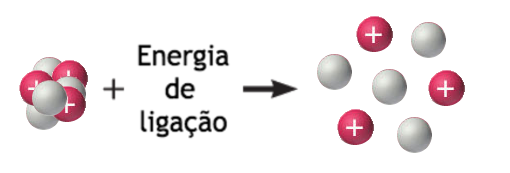}
	\caption{A energia que deve ser fornecida ao n\'ucleo para desintegr\'a-lo em seus constituintes \'e chamada \emph{energia de liga\c{c}\~ao}.}
	\label{fig:desintegrar}
\end{figure}

Como existe uma for\c{c}a nuclear atrativa que tende a manter o n\'ucleo coeso, qualquer tentativa de desintegr\'a-lo, separando os pr\'otons e n\^eutrons uns dos outros, requer que se \emph{forne\c{c}a} energia ao sistema, como mostra a figura~\ref{fig:desintegrar}. A energia necess\'aria para isso \'e denominada \emph{energia de liga\c{c}\~ao} nuclear.

\begin{figure}
	\centering
	\includegraphics[scale=.18]{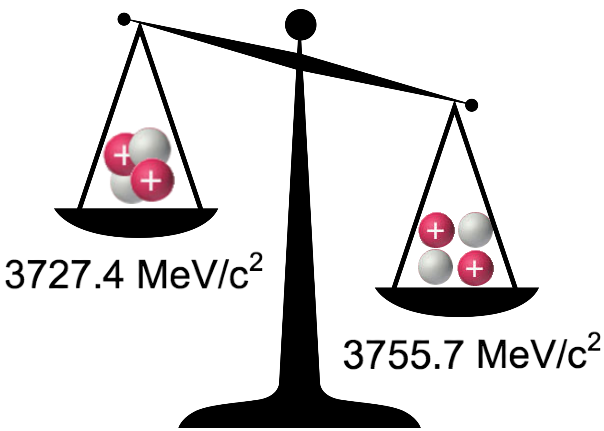}\quad
	\includegraphics[scale=.28]{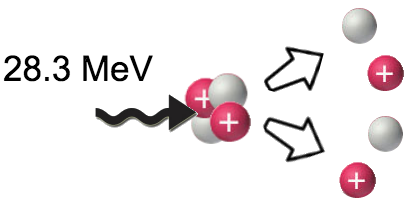}
	\caption{Ilustra\c{c}\~ao de duas maneiras de se compreender a energia de liga\c{c}\~ao de um n\'ucleo de h\'elio, i.e. uma part\'icula $\alpha$: (esquerda) d\'eficit de massa e (direita) energia necess\'aria para desintegrar o n\'ucleo em seus constituintes.}
	\label{fig:mass_deficit}
\end{figure}

A famosa rela\c{c}\~ao relativ\'istica de equival\^encia entre massa e energia, $E=mc^2$, garante que \emph{a massa de um nucl\'ideo\footnote{A palavra ``n\'ucleo'' sempre requer um predicado --- n\'ucleo {\it do qu\^e?} ---, de modo que seu uso j\'a pressup\~oe a exist\^encia de um outro objeto maior, como o \'atomo, do qual esse ``n\'ucleo'' constitui apenas a parte central. No entanto, em F\'isica Nuclear, lidamos com esses entes --- pr\'otons e n\^eutrons ligados por uma for\c{c}a nuclear --- \`a parte de qualquer contexto at\^omico, em uma situa\c{c}\~ao em que sequer est\~ao circundados por uma eletrosfera. Nesse caso, \'e comum referir-se a esses objetos como \emph{nucl\'ideos}.}  \'e menor do que a soma das massas dos constituintes isolados}, como mostra a figura~\ref{fig:mass_deficit}. De fato, como \'e necess\'ario fornecer uma quantidade de energia $E_\text{lig}$
para desintegrar um nucl\'ideo $^A_Z\,X$ em $Z$ pr\'otons e $N=A-Z$ n\^eutrons, o princ\'ipio de conserva\c{c}\~ao de energia garante que
\begin{equation}
	m_\text{nuc}c^2 + E_\text{lig} = Z\,m_pc^2 + N\, m_nc^2,
	\label{eq:Ebinding}
\end{equation}
ou seja,
\begin{equation}
	E_\text{lig} = \underbrace{(Z\,m_p + N\,m_n - m_\text{nuc})}_\text{d\'eficit de massa}\,c^2.
	\label{eq:mass_deficit}
\end{equation}
A diferen\c{c}a entre a massa de um nucl\'ideo coeso e a de seus constituintes \'e denominada \emph{d\'eficit de massa}.

\begin{figure}
	\centering
	\includegraphics[scale=.46]{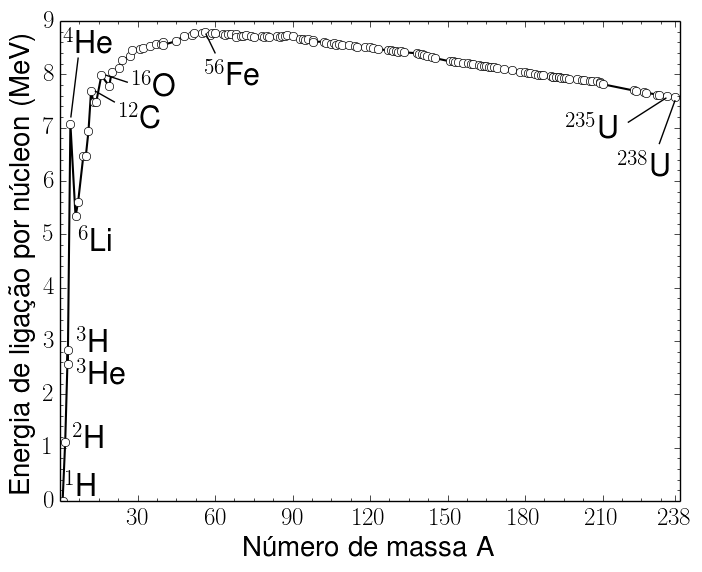}
	\caption{Energia de liga\c{c}\~ao dividida pelo n\'umero de n\'ucleons (pr\'otons e n\^eutrons).}
	\label{fig:binding_energy}
\end{figure}

A figura~\ref{fig:binding_energy} mostra a energia de liga\c{c}\~ao m\'edia por n\'ucleon --- a raz\~ao da energia de liga\c{c}\~ao pelo n\'umero de massa $A$ --- para v\'arios nucl\'ideos. Trata-se de uma medida do quanto cada n\'ucleon est\'a ligado a seus vizinhos, em m\'edia. Muita f\'isica pode ser discutida a partir dessa figura, que promove uma excelente oportunidade de desenvolvimento de interpreta\c{c}\~ao de gr\'aficos e da intui\c{c}\~ao f\'isica dos(as) discentes (ver ap\^endice~\ref{sec:atividades}).

De imediato, nota-se que a escala de energia associada \`a f\'isica nuclear --- por exemplo, a energia para dissociar um nucl\'ideo --- \'e da ordem de MeV, que \'e 1 milh\~ao de vezes maior do que as escalas de energia da f\'isica at\^omica e molecular --- a energia de ioniza\c{c}\~ao dos \'atomos ou de dissocia\c{c}\~ao de mol\'eculas \'e $\sim 1-10$~eV. Da\'i j\'a se pode depreender por que armas nucleares s\~ao t\~ao mais destrutivas do que explosivos usuais, baseados em rea\c{c}\~oes qu\'imicas, e por que a gera\c{c}\~ao de energia nuclear \'e t\~ao eficiente, capaz de produzir energia suficiente para abastecer metr\'opoles por meses ou anos a partir de alguns quilogramas de mat\'eria. Por exemplo, a ogiva detonada em Hiroshima, havendo consumido apenas $\sim 1$~kg de $^{235}$U, liberou energia equivalente \`a detona\c{c}\~ao de 20.000 toneladas de TNT~\cite{Krane}.

\subsubsection{Satura\c{c}\~ao da for\c{c}a nuclear}

J\'a foi discutido acima (se\c{c}\~ao~\ref{sec:nuclearforce}) que a for\c{c}a nuclear deve ter curto alcance, da ordem de $1$~fm. A figura~\ref{fig:binding_energy} oferece outra evid\^encia a favor dessa afirma\c{c}\~ao. 

Para entender por que, considere, primeiro, a energia associada \`a intera\c{c}\~ao eletrost\'atica entre $Z$ pr\'otons em um nucl\'ideo. Cada pr\'oton, de carga $e$, \'e repelido por $Z-1$ outros pr\'otons, distantes um do outro de $r\sim 1$~fm. A contribui\c{c}\~ao eletrost\'atica \`a energia \emph{por pr\'oton} \'e, ent\~ao, 
\begin{equation}
	\dfrac{E_\text{Coulomb}}{Z} \sim -(Z-1)\,k\dfrac{e^2}{r},
	\label{eq:Coulomb_binding}
\end{equation}
ou seja, varia linearmente com o n\'umero de pr\'otons no n\'ucleo. O sinal negativo se deve ao fato de a for\c{c}a ser repulsiva, provocando uma \emph{redu\c{c}\~ao} na energia de liga\c{c}\~ao por constituir uma tend\^encia natural \`a desintegra\c{c}\~ao do nucl\'ideo.

Analogamente, para um nucl\'ideo com $A$ n\'ucleons existem $A(A-1)/2$ maneiras de se arranj\'a-los em pares, de modo que, se supusermos que cada n\'ucleon interage com todos os demais via for\c{c}a nuclear, esperar\'iamos 
\begin{equation}
	E_\text{nuclear} \propto A(A-1),
	\label{eq:AA-1}
\end{equation}
ou seja, $E_\text{nuclear}/A$ cresceria linearmente com o n\'umero de n\'ucleons. Analogamente \`a equa\c{c}\~ao~(\ref{eq:Coulomb_binding}), o coeficiente de proporcionalidade \'e indicativo da intensidade da intera\c{c}\~ao nuclear.

Se olharmos, agora, para a figura~\ref{fig:binding_energy}, veremos que esse crescimento linear \'e de fato observado para nucl\'ideos leves, com $A\leq 4$. Mas a taxa de crescimento --- dada pelo coeficiente de proporcionalidade na equa\c{c}\~ao~(\ref{eq:AA-1}) --- diminui gradativamente, at\'e que, para nucl\'ideos mais pesados, com $A\gtrsim 16$, a curva tende a se aplainar em um patamar $E_\text{liga\c{c}\~ao}/A \sim 8$~MeV, atingindo um m\'aximo para o $^{56}$Fe. Ou seja, a hip\'otese de que cada n\'ucleon interage com todos seus vizinhos deixa de ser obedecida. Dito de outra forma, a intensidade m\'edia da intera\c{c}\~ao nuclear atuante sobre cada n\'ucleon diminui com o aumento do nucl\'ideo: outra evid\^encia sustentando que a intera\c{c}\~ao nuclear \'e de curto alcance. Disso podemos extrair, ainda, uma estimativa do alcance dessa intera\c{c}\~ao, que deve ser da ordem de magnitude das dimens\~oes nucleares, $r\sim 1$~fm, uma vez que, mesmo para n\'ucleons contidos no interior de um nucl\'ideo, essa intera\c{c}\~ao j\'a deixa de ser eficaz.

Podemos aprofundar ainda mais a an\'alise da figura~\ref{fig:binding_energy} para extrair mais informa\c{c}\~oes sobre a intera\c{c}\~ao nuclear. Partindo de um \'unico pr\'oton, e acrescentando-lhe n\^eutrons e outros pr\'otons, nota-se que a energia de liga\c{c}\~ao por n\'ucleon cresce drasticamente\footnote{Interessante notar que ambos $^3$H e $^3$He possuem tr\^es n\'ucleons, mas a energia de liga\c{c}\~ao do $^3$He \'e ligeiramente inferior, por causa da repuls\~ao eletrost\'atica entre os dois pr\'otons ali presentes. Ou seja, requer-se menos energia para desintegrar o $^3$He, porque j\'a existe uma for\c{c}a repulsiva contribuindo para isso no interior desse nucl\'ideo. No entanto, a diferen\c{c}a \'e de apenas $\sim 0.25$~MeV, ou 10\% do valor total da energia de liga\c{c}\~ao desses nucl\'ideos, indicando que a intera\c{c}\~ao nuclear \'e muito mais forte do que a eletromagn\'etica --- i.e. a maior parte da energia de liga\c{c}\~ao adv\'em da for\c{c}a nuclear entre pr\'otons e n\^eutrons, e apenas uma fra\c{c}\~ao bem inferior vem da repuls\~ao eletrost\'atica entre pr\'otons.} at\'e atingir um m\'aximo local no $^4$He, que \'e um nucl\'ideo excepcionalmente est\'avel\footnote{N\~ao \'e necess\'ario elaborar, em sala, sobre o motivo dessa estabilidade, mas pensamos ser conveniente discuti-la brevemente aqui, caso surjam perguntas por parte dos(as) estudantes a esse respeito. H\'a duas maneiras quase equivalentes de se entender essa propriedade do $^4$He. Assim como os el\'etrons na eletrosfera, os n\'ucleons tamb\'em se arranjam em ``camadas'', que, quando totalmente preenchidas, produzem nucl\'ideos altamente est\'aveis --- similar ao mecanismo subjacente \`a estabilidade qu\'imica dos gases nobres. A primeira camada nuclear comporta 2 part\'iculas, mas pr\'otons e n\^eutrons ocupam camadas distintas. Portanto o $^4$He \'e o primeiro nucl\'ideo para o qual h\'a o preenchimento completo de uma camada tanto por parte dos pr\'otons quanto dos n\^eutrons. Alternativamente, pode-se alegar que a intera\c{c}\~ao nuclear \'e mais intensa quando: (i) tanto pr\'otons quanto n\^eutrons est\~ao emparelhados; (ii) h\'a simetria entre n\'umero de pr\'otons e n\^eutrons; (iii) o nucl\'ideo cont\'em poucos n\'ucleons, que portanto est\~ao a uma curta dist\^ancia e interagem fortemente entre si.}.

Se acrescentamos ao $^4$He um pr\'oton e um n\^eutron, formando $^6$Li, a energia de liga\c{c}\~ao total aumenta --- porque um maior n\'umero de n\'ucleons implica em mais possibilidades de m\'utuas intera\c{c}\~oes ---, mas a energia m\'edia \emph{por n\'ucleon} diminui. \'E como se alguns dos n\'ucleons de $^6$Li se aglomerassem em uma combina\c{c}\~ao altamente interagente de 2 pr\'otons e 2 n\^eutrons, enquanto o pr\'oton e o n\^eutron adicionais, desemparelhados, interagissem com menor intensidade, reduzindo a energia de liga\c{c}\~ao m\'edia. Se seguirmos acrescentando n\'ucleons, os pr\'otons e n\^eutrons extras v\~ao se emparelhando, provocando um aumento de $E_\text{liga\c{c}\~ao}/A$. Mas esse crescimento ocorre a uma taxa inferior \`aquela entre $^1$H e $^4$He --- a m\'edia da m\'utua intera\c{c}\~ao entre os n\'ucleons j\'a \'e menor, devido \`a maior dist\^ancia entre eles. Essa segunda reta de crescimento na figura~\ref{fig:binding_energy} culmina no $^{12}$C, um outro nucl\'ideo excepcionalmente est\'avel por ser composto de um aglomerado\footnote{O $^8$Be, composto de 2\,$^4$He, \'e inst\'avel, decaindo via $^8\text{Be}\to 2\,\alpha$. O motivo \'e que, precisamente porque o $^4$He \'e t\~ao est\'avel, \'e mais energeticamente favor\'avel manter dois $^4$He isolados do que ligados, enfrentando a repuls\~ao eletrost\'atica entre ambos. A instabilidade do $^8$Be desfavorece o decaimento $^{12}\text{C}\to\ ^8\text{Be} + \alpha$, e o decaimento em $3\alpha$ simultaneamente \'e altamente improv\'avel, o que faz com que o $^{12}$C seja altamente est\'avel.} de 3\,$^4$He. Analogamente, o $^{16}$O pode ser visto como um aglomerado de 4\,$^4$He e possui, tamb\'em, grande energia de liga\c{c}\~ao por n\'ucleon\footnote{O $^{16}$O tamb\'em corresponde ao nucl\'ideo em que os pr\'otons e os n\^eutrons preenchem a segunda camada nuclear, e tamb\'em por isso \'e bastante est\'avel.}.

Como j\'a discutido nesta mesma subse\c{c}\~ao, \`a medida que o nucl\'ideo aumenta de tamanho, n\'ucleons diametralmente opostos deixam de interagir um com o outro pela for\c{c}a nuclear --- devido ao curto alcance dessa intera\c{c}\~ao ---, e o acr\'escimo de novos n\'ucleons tende a aumentar cada vez menos a energia de liga\c{c}\~ao m\'edia de cada n\'ucleon. Fala-se, ent\~ao, da \emph{\bf satura\c{c}\~ao da for\c{c}a nuclear}, que \'e outra forma de expressar o curto alcance dessa intera\c{c}\~ao.

Acontece que, quanto maior o nucl\'ideo, maior \'e seu n\'umero de pr\'otons\footnote{Um nucl\'ideo est\'avel tende a ter aproximadamente o mesmo n\'umero de n\^eutrons que de pr\'otons, mais especificamente $N \approx 1.0-1.6$~Z.}, e embora a intera\c{c}\~ao nuclear tenda a saturar, a repuls\~ao coulombiana continua atuante entre todos os pr\'otons, contribuindo negativamente \`a energia de liga\c{c}\~ao, conforme a equa\c{c}\~ao~(\ref{eq:Coulomb_binding}). \'E por isso que, a partir do ferro-56, a curva da figura~\ref{fig:binding_energy} apresenta uma inflex\~ao, tornando-se decrescente. 

\begin{figure}
	\centering
	\includegraphics[scale=.18]{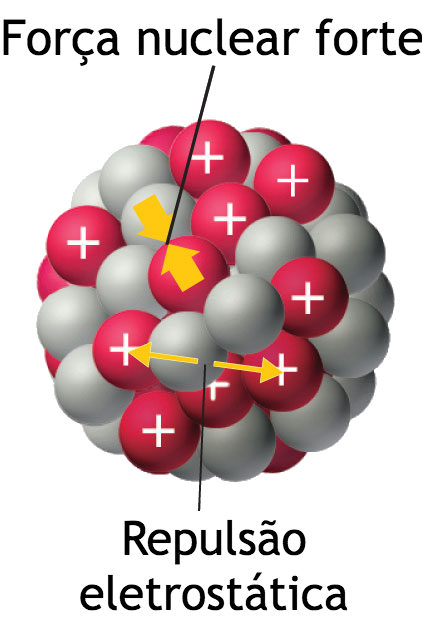}
	\caption{As for\c{c}as atuantes sobre os n\'ucleons. Fonte: Chemistry LibreTexts~\cite{ChemLibre}, sob licen\c{c}a Creative Commons BY-NC-SA 3.0~\cite{CCBYSA30}.}
	\label{fig:nucleus_forces}
\end{figure}

\subsection{Fiss\~ao nuclear}
\label{sec:fissao}

Devido ao comportamento decrescente da curva da figura~\ref{fig:binding_energy} para nucl\'ideos pesados, processos que ocasionam uma redu\c{c}\~ao do tamanho do nucl\'ideo s\~ao energeticamente favorecidos.

\begin{figure}
	\centering
	\includegraphics[scale=.15]{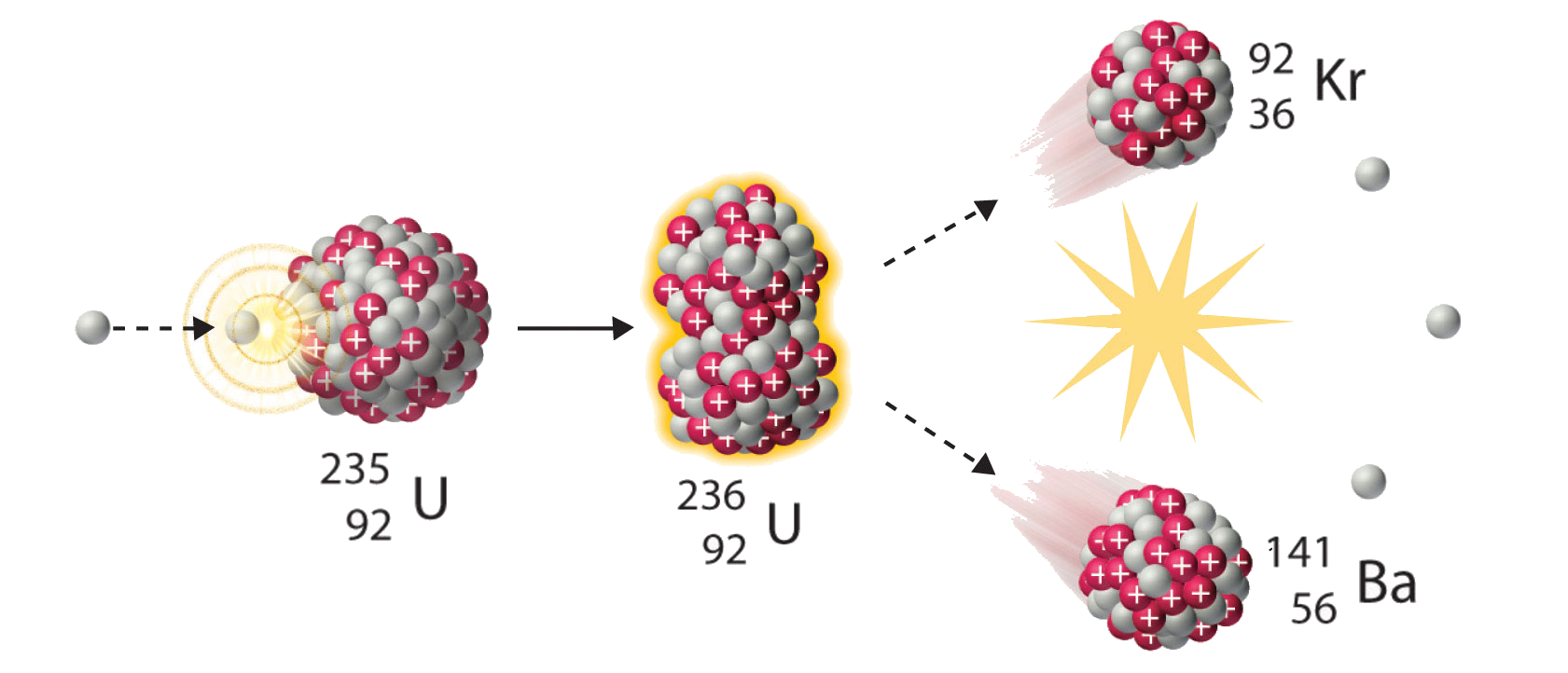}
	\caption{Fiss\~ao de $^{235}$U induzida pela captura de um n\^eutron termal. O n\^eutron incide com baixa energia cin\'etica, $\sim 0.02$~eV, e \'e absorvido, resultando em um nucl\'ideo de U$^{236}$, que \'e inst\'avel e se fissiona em dois nucl\'ideos mais leves. Neste caso o produto da rea\c{c}\~ao inclui, ainda, tr\^es n\^eutrons. Fonte: Chemistry LibreTexts~\cite{ChemLibre}, sob licen\c{c}a Creative Commons BY-NC-SA 3.0~\cite{CCBYSA30}.}
	\label{fig:fission}
\end{figure}

Considere, por exemplo, um processo em que um nucl\'ideo pesado, como o ur\^anio-$235$, se divide em dois nucl\'ideos mais leves ap\'os absorver um n\^eutron, como ilustrado na figura~\ref{fig:fission}. Esse tipo de processo \'e denominado \emph{fiss\~ao nuclear}. Como os produtos da rea\c{c}\~ao s\~ao nucl\'ideos menores --- o que aumenta a intera\c{c}\~ao nuclear entre n\'ucleons --- e possuem menos pr\'otons --- reduzindo a repuls\~ao coulombiana ---, eles s\~ao mais est\'aveis, com maior energia de liga\c{c}\~ao m\'edia por n\'ucleon, como se nota pela figura~\ref{fig:binding_energy}. Por ocorrer uma varia\c{c}\~ao da energia potencial dos n\'ucleons na rea\c{c}\~ao, deve haver uma correspondente libera\c{c}\~ao de energia durante o processo, para que a energia total se conserve. Outra maneira de se entender isso \'e notando que, quanto maior a energia de liga\c{c}\~ao, menor \'e a massa total do nucl\'ideo, de acordo com a equa\c{c}\~ao~(\ref{eq:mass_deficit}). Portanto, a forma\c{c}\~ao de nucl\'ideos mais est\'aveis est\'a associada \`a redu\c{c}\~ao da energia do sistema em forma de massa\footnote{Em uma rea\c{c}\~ao de fiss\~ao, como a ilustrada na figura~\ref{fig:fission}, o n\'umero de pr\'otons e n\^eutrons se conserva, e a diferen\c{c}a entre a massa dos reagentes e dos produtos \'e inteiramente devida \`a diferen\c{c}a das energias de liga\c{c}\~ao, \[ (m_\text{reagentes}-m_\text{produtos})c^2 = E(^{89}\text{Kr}) + E(^{144}\text{Ba})-E(^{235}\text{U})>0.\] Como a energia total se conserva, deve-se ter \[ m_\text{reagentes}c^2 = m_\text{produto} c^2 + E_\text{liberada}. \]}, e a diferen\c{c}a constitui-se em energia liberada em outras formas, como radia\c{c}\~ao ou energia cin\'etica dos fragmentos (ou seja, calor). Este \'e o princ\'ipio fundamental subjacente \`a gera\c{c}\~ao de energia por fiss\~ao nuclear. 

\begin{figure}
	\centering
	\includegraphics[scale=.3]{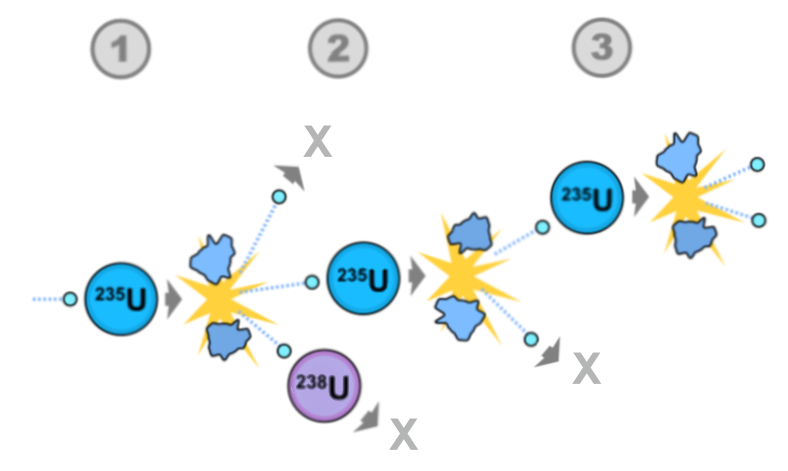}
	\caption{Rea\c{c}\~ao autossustent\'avel de fiss\~ao controlada. Em cada est\'agio, ao menos um n\^eutron ejetado se encontra com outro nucl\'ideo de $^{235}$U, originando outra fiss\~ao e garantindo a continuidade do processo. Fonte: Wikimedia Commons/Dom\'inio P\'ublico.}
	\label{fig:chain_controlled}
\end{figure}

Mas h\'a um detalhe a mais que merece ser discutido. Em rea\c{c}\~oes de fiss\~ao induzidas por capturas de n\^eutrons, \'e comum que haja libera\c{c}\~ao de dois ou mais n\^eutrons no produto final, como ilustra a figura~\ref{fig:fission}. Esses novos n\^eutrons podem colidir com outros nucl\'ideos de $^{235}$U e induzir outras rea\c{c}\~oes, liberando mais n\^eutrons, e assim sucessivamente, resultando em uma rea\c{c}\~ao autossustent\'avel, como ilustrado na figura~\ref{fig:chain_controlled}. 

Ocorre que 99.3\% do ur\^anio encontrado na natureza est\'a na forma do is\'otopo $^{238}$U, que n\~ao \'e facilmente fission\'avel, e apenas cerca de 0.7\% \'e $^{235}$U. Para otimizar a quantidade de energia liberada e garantir a autossustentabilidade da cadeia de rea\c{c}\~oes, \'e desej\'avel que o material tenha uma concentra\c{c}\~ao muito maior de $^{235}$U, de modo a aumentar a probabilidade de que os n\^eutrons emergentes em cada etapa de fiss\~ao encontrem outros nucl\'ideos desse is\'otopo. A esses processos que visam aumentar a concentra\c{c}\~ao de $^{235}$U d\'a-se o nome de \emph{enriquecimento de ur\^anio}.

Diferentes prop\'ositos requerem diferentes n\'iveis de enriquecimento. Para a extra\c{c}\~ao de energia nuclear, uma concentra\c{c}\~ao de 3\% de $^{235}$U j\'a \'e suficiente~\cite{Krane}. Isso porque, nesse caso, n\~ao se \'e sequer desej\'avel que todos os n\^eutrons resultantes de uma rea\c{c}\~ao se encontrem com outro nucl\'ideo de $^{235}$U e provoquem outras fiss\~oes, pois isso provocaria um aumento exponencial e descontrolado da energia liberada. Ao contr\'ario, uma parte essencial no desenho de um reator nuclear s\~ao as chamadas \emph{barras de controle}, constitu\'idas de materiais altamente eficientes na absor\c{c}\~ao de n\^eutrons --- como boro ou ligas de metais de terra rara ---, justamente para que os n\^eutrons em excesso sejam capturados, como representado pelos ``X'' na figura~\ref{fig:chain_controlled}, permitindo que o operador da usina mantenha controle sobre a quantidade de energia extra\'ida. Ainda outro elemento importante em um reator \'e a presen\c{c}a de um material moderador --- como \'agua ou carbono na forma de grafite ---, que reduz a energia cin\'etica dos n\^eutros secund\'arios de $\sim 1$~MeV para $\sim 0.02$~eV, o que aumenta a probabilidade de fiss\~ao do ur\^anio-235\footnote{H\'a uma categoria de reatores, chamados \emph{Fast Neutron Reactors} (FNRs) ou reatores de n\^eutrons r\'apidos, em que a fiss\~ao \'e promovida por n\^eutrons r\'apidos, e n\~ao h\'a, portanto, material moderador. O combust\'ivel mais comum desse tipo de reator \'e plut\^onio-239, que passa por fiss\~ao mais facilmente quando colide com n\^eutrons altamente energ\'eticos. Ademais, parte do $^{238}$U pode ser tamb\'em fission\'avel por esses n\^eutrons, enquanto outra parte \'e capaz de absorv\^e-los e se converter em $^{239}$Pu via subsequentes decaimentos $\beta$, \[ n + \,^{238}\text{U} \to \,^{239}\text{U} \overset{\beta}{\to} \,^{239}\text{Np} \overset{\beta}{\to} \,^{239}\text{Pu}.\] 
Assim, FNRs s\~ao reatores altamente eficientes porque parte do combust\'ivel (plut\^onio) \'e gerado colateralmente durante a rea\c{c}\~ao em cadeia a partir do is\'otopo mais comum de ur\^anio (sem necessidade, portanto, de enriquecimento). Mais ainda, n\^eutrons r\'apidos podem induzir fiss\~ao de actin\'ideos, que s\~ao subprodutos da fiss\~ao de elementos pesados e considerados parte do chamado ``lixo nuclear'' (a ser discutido logo abaixo, na se\c{c}\~ao~\ref{sec:ambiente}). Ou seja, em FNRs o ``lixo nuclear'' pode ser reciclado e utilizado como combust\'ivel. Infelizmente, FNRs ainda s\~ao uma tecnologia cara e pouco incentivada, visto que, dados os custos atuais de minera\c{c}\~ao e enriquecimento de ur\^anio, \'e mais barato operar com usinas ``tradicionais'', baseadas em n\^eutrons termais, i.e. lentos, pouco energ\'eticos. Para mais informa\c{c}\~oes vide refs.~\cite{WNA_FBR, NuclearEnergy21st}}.
   
\begin{figure}
	\centering
	\includegraphics[scale=.35]{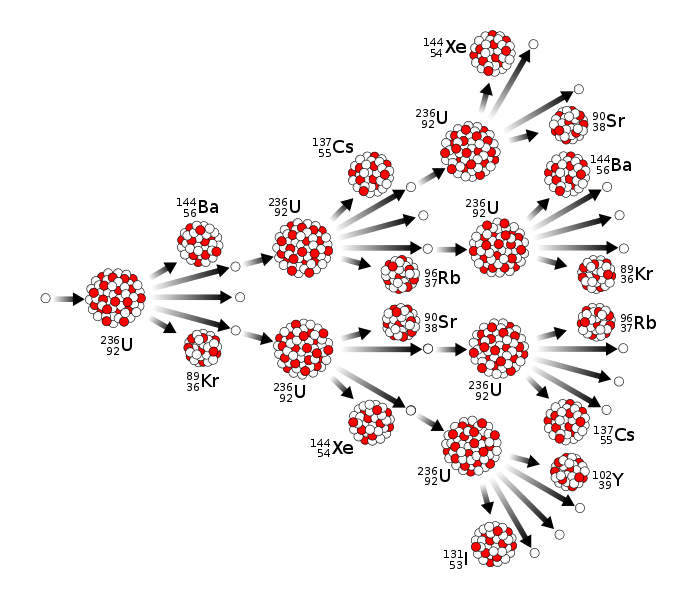}
	\caption{Rea\c{c}\~ao supercr\'itica de fiss\~ao nuclear em cadeia. Fonte: Wikimedia Commons, sob licen\c{c}a Creative Commons BY-SA 4.0~\cite{BY-SA40}.}
	\label{fig:chain}
\end{figure}

Para prop\'ositos b\'elicos, o objetivo \'e a libera\c{c}\~ao da maior quantidade de energia poss\'ivel. Nesse caso \'e preciso uma concentra\c{c}\~ao muito maior de $^{235}$U, acima de 85\%, para que a maioria dos n\^eutrons secund\'arios seja absorvida por outro nucl\'ideo de $^{235}$U e induza outra fiss\~ao, gerando um crescimento exponencial da energia liberada, como ilustrado na figura~\ref{fig:chain}. Uma tal rea\c{c}\~ao em cadeia \'e denominada \emph{supercr\'itica}.

\subsection{Fus\~ao nuclear e a origem dos elementos}
\label{sec:fusao}

\begin{figure}
	\centering
	\includegraphics[scale=.46]{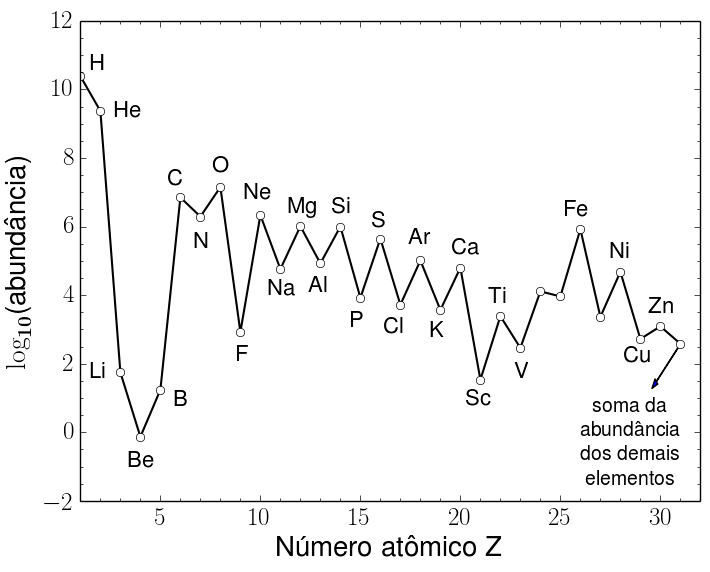}
	\caption{Abund\^ancia relativa de elementos no sistema solar, calculada como o n\'umero de \'atomos do elemento em quest\~ao para cada $10^6$ \'atomos de Si. Dados extra\'idos da ref.~\cite{Lodders}.}
	\label{fig:abundances}
\end{figure}

Uma aula sobre fus\~ao nuclear pode ser motivada pela seguinte quest\~ao-problema: como foram formados os elementos qu\'imicos que nos rodeiam? O(a) docente pode iniciar mostrando \`a turma o gr\'afico das abund\^ancias dos elementos no sistema solar, figura~\ref{fig:abundances}, e pedir que os alunos discutam o comportamento da curva. Quais s\~ao suas caracter\'isticas mais marcantes? Algum comportamento da curva lhes parece intuitivo? Algum parece contra-intuitivo? Acham estranho que haja um predom\'inio de hidrog\^enio e h\'elio no Universo? Podem oferecer uma explica\c{c}\~ao ao comportamento oscilat\'orio da curva, i.e. ao fato de que elementos com n\'umero at\^omico par s\~ao $\approx 10$ vezes mais abundantes que os elementos vizinhos, com $Z$ \'impar?

Ao instigar os alunos com essa problematiza\c{c}\~ao, deve-se esperar respostas que partam de uma perspectiva \emph{a-hist\'orica}, sugerindo que o Universo sempre teve a constitui\c{c}\~ao que observamos atualmente. O principal prop\'osito dessa aula deve ser desconstruir essa perspectiva, demonstrando que tamb\'em na F\'isica h\'a historicidade, e que tampouco nas ci\^encias naturais existe um \emph{status quo} eterno e absoluto, mas que para entendermos a Natureza \'e preciso compreender os \emph{processos} que resultaram em sua configura\c{c}\~ao atual. Mais ainda, a discuss\~ao culmina no vislumbre da hist\'oria humana como apenas um cap\'itulo de uma longa hist\'oria universal, em que simples part\'iculas elementares se organizam em estruturas cada vez mais complexas --- n\'ucleons, nucl\'ideos, \'atomos, mol\'eculas, nuvens gasosas e protoestruturas rochosas, estrelas e planetas, mol\'eculas org\^anicas autorreplicantes, a vida complexa ---, at\'e chegar na g\^enese do ser humano e na hist\'oria de nossas culturas. Com isso, o(a) estudante pode contemplar a interconex\~ao entre todas as sub\'areas em que se divide o conhecimento humano --- todas as ``mat\'erias'' ou ``disciplinas'' escolares ---, ao se enfatizar como cada uma lida com um cap\'itulo espec\'ifico de uma hist\'oria universal que as conecta e que tudo abrange. 

\subsubsection{Nucleoss\'intese primordial}
\label{sec:primordial}

Para entender a g\^enese dos elementos sob a perspectiva hist\'orica supracitada, basta introduzir um conceito muito simples, mas que \'e a base de toda a cosmologia contempor\^anea: o fato observ\'avel de que nosso Universo est\'a em expans\~ao~\cite{Dodelson, Liddle}. Uma forte evid\^encia em favor dessa asser\c{c}\~ao \'e a observa\c{c}\~ao de que as gal\'axias distantes est\~ao se afastando de n\'os. Se o Universo est\'a se expandindo, isso implica que no passado toda sua mat\'eria e energia estavam contidas em uma regi\~ao muito menor do que ocupam atualmente, o que significa que as temperaturas deviam ser alt\'issimas --- basta lembrar-se que um g\'as contra\'ido adiabaticamente se aquece. Ou seja, o Universo primordial era muito quente, e \`a medida que evolui e se expande, ele se resfria. 

Altas temperaturas significam alto poder dissociativo, ou seja, quanto maior a temperatura, mais dif\'icil \'e manter estruturas complexas, pois elas tendem a se dissociar em partes mais elementares. Exemplos s\~ao a dissocia\c{c}\~ao
da estrutura cristalina de um s\'olido quando do derretimento do material, e das liga\c{c}\~oes inter-moleculares de um l\'iquido durante a vaporiza\c{c}\~ao, bem como a ioniza\c{c}\~ao de um g\'as, que, aquecido a temperaturas suficientemente elevadas, tem os el\'etrons at\^omicos dissociados dos n\'ucleos, tornando-se um \emph{plasma}. Assim, o Universo inicia sua hist\'oria com mat\'eria e energia em sua forma mais simplificada, mais rudimentar, como simples part\'iculas elementares, e somente \`a medida que ele se expande e resfria \'e que estruturas mais complexas come\c{c}am a se formar.queo processo de \emph{fus\~ao nuclear} \'e energeticamente favorecido, pois a fus\~ao de dois nucl\'ideos mais leves resulta em um mais est\'avel --- i.e. com maior energia de liga\c{c}\~ao ---, e h\'a libera\c{c}\~ao de energia nesses processos.

Mas por que os nucl\'ideos se formam como estruturas complexas est\'aveis durante essa hist\'oria? A resposta vem novamente da figura~\ref{fig:binding_energy}, especificamente do
comportamento crescente da curva para nucl\'ideos leves. Esse comportamento indica que o processo denominado de \emph{fus\~ao nuclear} \'e energeticamente favorecido, em que a fus\~ao de dois nucl\'ideos mais leves resulta em um mais est\'avel ? i.e. com maior energia de
liga\c{c}\~ao ?, ocorrendo libera\c{c}\~ao de energia nesse processo. Ou, reciprocamente, a dissocia\c{c}\~ao dessas estruturas n\~ao ocorre espontaneamente, mas requer que se forne\c{c}a energia em quantidade que j\'a n\~ao est\'a dispon\'ivel quando a temperatura do Universo torna-se suficientemente baixa.

Consideremos, ent\~ao, o Universo nos primeiros segundos de sua hist\'oria, quando j\'a havia pr\'otons e n\^eutrons\footnote{Al\'em de estarem presentes tamb\'em outras part\'iculas elementares, como el\'etrons, f\'otons, neutrinos, entre outras.}, mas ainda n\~ao se haviam formado estruturas mais complexas --- ou seja, n\~ao havia \'atomos ou sequer n\'ucleos at\^omicos, e os pr\'otons e n\^eutrons encontravam-se livres\footnote{J\'a vimos anteriormente que a energia de liga\c{c}\~ao nuclear \'e muito maior que a at\^omica. Por isso, requer-se temperaturas muito maiores para dissociar um nucl\'ideo do que um \'atomo. Ou seja, na hist\'oria de resfriamento do Universo, os nucl\'ideos formam-se primeiro, j\'a nos primeiros minutos dessa hist\'oria, e s\'o ap\'os $\sim 380$ mil anos a temperatura atinge n\'iveis suficientemente baixos para que os el\'etrons capturados por esses nucl\'ideos n\~ao sejam mais dissociados, formando assim \'atomos neutros.}. \'E claro que, nessa situa\c{c}\~ao, suas energias de liga\c{c}\~ao s\~ao nulas, simplesmente por n\~ao estarem ligados a nada. Mas \`a medida que se aproximam o suficiente, entrando no alcance de atua\c{c}\~ao da intera\c{c}\~ao nuclear, uma for\c{c}a os atrai de modo a formarem um estado ligado: um nucl\'ideo de $^2$H, tamb\'em denominado \emph{d\^euteron}.

Como o processo de forma\c{c}\~ao do d\^euteron \'e espont\^aneo, h\'a libera\c{c}\~ao de energia durante a rea\c{c}\~ao. Ou, dito de outra forma: \'e mais energeticamente custoso manter pr\'otons e n\^eutrons isolados do que ligados. A energia liberada nada mais \'e do que a energia de liga\c{c}\~ao do d\^euteron\footnote{Afigura~\ref{fig:binding_energy} mostra que a energia de liga\c{c}\~ao do $^2$H por n\'ucleon \'e aproximadamente 1.1~MeV. Como o d\^euteron possui dois n\'ucleons, a energia de liga\c{c}\~ao total \'e duas vezes esse valor.}, e o processo \'e descrito por
\begin{equation}
	\text{p + n} \to\ ^2\text{H + 2.2~MeV,}
	\label{eq:pn_fusion}
\end{equation}
cuja ilustra\c{c}\~ao seria an\'aloga \`a figura~\ref{fig:desintegrar}, por\'em vista da direita para a esquerda.

\begin{figure}
	\centering
	\includegraphics[scale=.35]{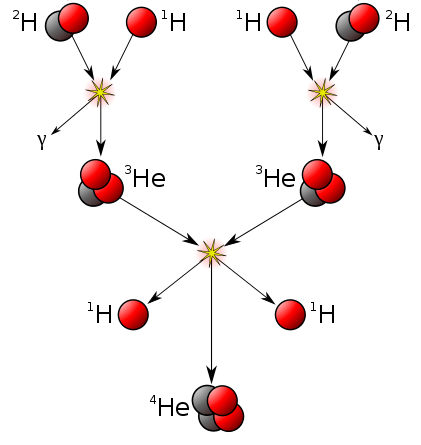}
	\caption{Exemplo de processo de fus\~ao de d\^euterons e pr\'otons para forma\c{c}\~ao de $^4$He. Fonte: Wikimedia Commons/Dom\'inio P\'ublico.}
	\label{fig:pd_fusion}
\end{figure}

Essa rea\c{c}\~ao ocorreu abundantemente quando o Universo tinha apenas poucos minutos de idade, e constitui o primeiro est\'agio do processo de forma\c{c}\~ao de nucl\'ideos mais complexos a partir dos pr\'otons e n\^eutrons que se encontravam inicialmente livres --- processo que recebe o nome de \emph{nucleoss\'intese primordial}. Esses processos primordiais de fus\~ao nuclear continuam at\'e culminarem na produ\c{c}\~ao de $^4$He, como exemplificado na cadeia de rea\c{c}\~oes da figura~\ref{fig:pd_fusion}. Esse nucl\'ideo \'e t\~ao mais est\'avel que seus vizinhos (i.e. os processos de fus\~ao que lhe d\~ao origem s\~ao t\~ao energeticamente favor\'aveis) que praticamente todos os d\^euterons, assim que formados, tendem a continuar se fundindo\footnote{Dito de outra forma: todos os n\^eutrons que se encontravam livres nessa \'epoca do Universo primordial terminam ligados em n\'ucleos de h\'elio.} at\'e formar $^4$He. 

Nesse ponto, o processo de fus\~ao encontra um gargalo, por n\~ao existirem elementos est\'aveis de n\'umeros de massa 5 ou 8 --- por exemplo, uma eventual fus\~ao de dois $^4$He produz $^8$Be, que logo decai novamente em 2~$^4$He. A cadeia de fus\~ao s\'o pode prosseguir, portanto, passando pela produ\c{c}\~ao de $^{12}$C --- este, sim, altamente est\'avel ---, mas isso demanda um encontro quase simult\^aneo de 3~$^4$He, o que \'e altamente improv\'avel exceto em situa\c{c}\~oes de alt\'issima densidade de mat\'eria, que n\~ao \'e o caso durante essa \'epoca cosmol\'ogica. 

Um outro fator importante a se considerar, nesse processo, \'e a elevada energia cin\'etica que as part\'iculas devem possuir\footnote{A energia cin\'etica m\'edia das part\'iculas em um plasma ou g\'as est\'a intimamente relacionada \`a temperatura de tal fluido via $\langle E \rangle \sim k_B T$, onde $k_B\approx 8.617\times 10^{-5}~\text{eV}\cdot\text{K}^{-1}$ \'e a constante de Boltzmann, uma constante universal da Natureza. Ou seja, pode-se refrasear a condi\c{c}\~ao acima afirmando-se que a fus\~ao \'e um processo que usualmente requer altas temperaturas.} para serem capazes de ``superar'' a intensa repuls\~ao coulombiana
e se aproximarem a uma dist\^ancia $\lesssim 1$~fm, de modo a possibilitar a atua\c{c}\~ao da intera\c{c}\~ao nuclear para promover a fus\~ao. A energia potencial de duas cargas elementares $e$ a uma dist\^ancia $r\sim 1$~fm \'e\footnote{Um resultado dessa ordem de magnitude \'e esperado, porque: (i) da figura~\ref{fig:binding_energy} sabemos que essa \'e a ordem de grandeza da energia de liga\c{c}\~ao nuclear, e (ii) sabemos que a repuls\~ao coulombiana entre n\'ucleons \'e capaz de competir com a atra\c{c}\~ao dos n\'ucleons pela for\c{c}a nuclear, devido \`a inflex\~ao na curva de energia de liga\c{c}\~ao para nucl\'ideos pesados (e tamb\'em por causa da diferen\c{c}a observ\'avel entre as energias de liga\c{c}\~oes de pares de \emph{nucl\'ideos espelhados}, i.e. que possuem mesmo n\'umero de massa mas diferentes n\'umeros de pr\'otons, como o $^3$H e o $^3$He).} $k e^2/r \sim 1~\text{MeV}$, portanto a energia m\'edia das part\'iculas n\~ao pode ser muito inferior a esse valor\footnote{Note, entretanto, que devido \`a possibilidade de tunelamento qu\^antico, os nucl\'ideos podem fusionar mesmo tendo energia ligeiramente menor do que a altura da barreira.}. \`A \'epoca da nucleoss\'intese essa energia m\'edia era $\sim 0.1$~MeV~\cite{Dodelson}, o que corresponde a temperaturas da ordem de 1 bilh\~ao de kelvins (!), muito acima da temperatura m\'edia do Universo atualmente, $\sim 3$~K.

Considerando-se todos os fatores acima, estima-se que, ap\'os conclu\'ida essa \'epoca de nucleoss\'intese primordial, cerca de 75\% da massa do Universo encontrava-se na forma de pr\'otons livres (i.e. $^1$H) e 25\% era $^4$He, com 0.01\% de deut\'erio ($^2$H) e $^3$He, e apenas tra\c{c}os de $^7$Li~\cite{Krane, Dodelson}. O excelente acordo entre essas previs\~oes te\'oricas e as abund\^ancias observadas no meio interestelar constitui uma das mais robustas evid\^encias de que o Universo j\'a passou por uma fase extremamente quente, como prev\^e a teoria do Big Bang.

\subsubsection{Nucleoss\'intese estelar} 
\label{sec:estelar}

Mas como foram formados os elementos mais pesados? Os pr\'e-requisitos listados acima, de alt\'issimas densidades e temperaturas, s\~ao satisfeitos no interior de estrelas supermassivas, como as chamadas \emph{gigantes vermelhas}. Para entender um pouco esse processo, vale discutir brevemente sobre o ciclo de vida dessas estrelas.

O in\'icio da vida de uma estrela se d\'a quando um aglomerado de gases primordiais, constitu\'ido essencialmente de hidrog\^enio e h\'elio, em abund\^ancia n\~ao muito diferente daquelas resultantes da nucleoss\'intese primordial, colapsa devido \`a sua pr\'opria atra\c{c}\~ao gravitacional. \`A medida que o g\'as \'e comprimido, sua temperatura aumenta, at\'e que, eventualmente, torna-se alta o suficiente para ativar a fus\~ao de hidrog\^enio em deut\'erio\footnote{Note que a forma\c{c}\~ao de deut\'erio no interior de estrelas n\~ao adv\'em da captura de n\^eutrons livres, como descrito acima para a nucleoss\'intese primordial --- pois n\^eutrons s\~ao escassos no meio estelar ---, mas da fus\~ao de dois pr\'otons seguida de um decaimento do tipo $\beta$, i.e. \[ p + p \to \,^2\text{H} + e^+ + \nu_e. \] O processo de decaimento de um pr\'oton em um n\^eutron (um decaimento $\beta$) \'e ditado pela intera\c{c}\~ao fraca, e n\~ao pela for\c{c}a nuclear. Por isso, n\~ao entraremos em maiores detalhes neste artigo (ver se\c{c}\~ao~\ref{sec:decaimentos}), mas em trabalhos futuros sobre outras partes desta sequ\^encia did\'atica.},  e, em \'ultima inst\^ancia, h\'elio, segundo a figura~\ref{fig:pd_fusion}. Essas rea\c{c}\~oes predominam no primeiro est\'agio de vida estelar --- como \'e o caso do Sol ---, e delas decorre libera\c{c}\~ao de energia na forma de radia\c{c}\~ao (assim como toda fus\~ao de elementos com $A\lesssim 56$). \'E da fus\~ao nuclear que as estrelas obt\^em sua energia --- \'e por isso que elas t\^em ``brilho pr\'oprio''. Mais ainda, a press\~ao dessa radia\c{c}\~ao contrabalanceia o colapso gravitacional da estrela e a mant\'em em equil\'ibrio.

\begin{figure}
	\centering
	\includegraphics[scale=.2]{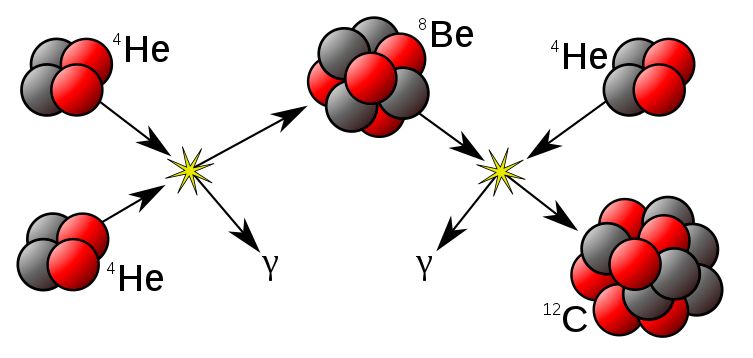}
	\caption{Dois nucl\'ideos de $^4$He se fundem e formam $^8$Be, que \'e inst\'avel e decai ap\'os cerca de $10^{-16}$~s. No entanto, no interior de estrelas muito massivas, a densidade de mat\'eria pode ser t\~ao alta que, antes de decorrido esse tempo, um outro nucl\'ideo de $^4$He se aproxima e se funde ao $^8$Be, produzindo $^{12}$C. Fonte: Wikimedia Commons, sob licen\c{c}a Creative Commons BY-SA 4.0~\cite{BY-SA40}.}
	\label{fig:fusion_C}
\end{figure}

O segundo est\'agio da vida estelar se inicia com o esgotamento do hidrog\^enio no n\'ucleo. Os processos de fus\~ao s\~ao temporariamente interrompidos, e, sem a press\~ao de radia\c{c}\~ao, o n\'ucleo volta a colapsar gravitacionalmente. Isso faz com que suas temperatura e densidade aumentem novamente, at\'e atingir um patamar que ativa a fus\~ao de h\'elio. Quando a densidade de h\'elio-4 \'e suficientemente alta, ocorre fus\~ao em $^8$Be e, antes que esse decaia, um outro nucl\'ideo de $^4$He se aproxima e se funde ao ber\'ilio, produzindo $^{12}$C, conforme ilustrado na figura~\ref{fig:fusion_C}. Uma vez superado o gargalo causado pela instabilidade do ber\'ilio, outras absor\c{c}\~oes de $^4$He podem ocorrer, formando uma s\'erie de elementos mais pesados como
\[ ^{12}\text{C}  \overset{^4\text{He}~}{\longrightarrow} 
   \,^{16}\text{O} \overset{^4\text{He}~}{\longrightarrow} 
   \,^{20}\text{Ne} \overset{^4\text{He}~}{\longrightarrow} 
   \,^{24}\text{Mg}\ldots
\]
Entretanto, \`a medida que se progride nessa cadeia, h\'a um aumento do n\'umero de pr\'otons nos nucl\'ideos envolvidos e, consequentemente, cresce a repuls\~ao coulombiana, de modo que s\~ao necess\'arias temperaturas cada vez maiores (ou seja, estrelas mais massivas) para continuar a fus\~ao. 

H\'a, ainda, outros processos nucleares, como a absor\c{c}\~ao de pr\'otons por nucl\'ideos pesados, decaimentos $\beta$, e emiss\~oes de pr\'otons e n\^eutrons, que d\~ao origem a outros is\'otopos e elementos de n\'umero at\^omico \'impar, como o nitrog\^enio. No entanto, s\~ao processos secund\'arios relativamente \`a cadeia de fus\~ao descrita acima, e portanto espera-se que a abund\^ancia de elementos com n\'umero at\^omico par seja relativamente maior do que para $Z$ \'impar.

Essa predi\c{c}\~ao \'e confirmada pela figura~\ref{fig:abundances}. A abund\^ancia de elementos com n\'umero at\^omico par \'e pelo menos uma ordem de magnitude maior do que para $Z$ \'impar. Note, tamb\'em, a abund\^ancia excepcionalmente grande de ferro (e tamb\'em de n\'iquel), devido ao fato de serem elementos finais nas cadeias de fus\~ao, ap\'os os quais a fus\~ao deixa de ser exot\'ermica, i.e. deixa de liberar energia 
 (vide figura~\ref{fig:binding_energy}). 

Quando, ap\'os essa sequ\^encia de fus\~oes, o n\'ucleo de uma estrela supermassiva passa a ser constitu\'ido majoritariamente de ferro e n\'iquel (i.e. elementos com $A\sim 56$, que est\~ao no \'apice da curva da figura~\ref{fig:binding_energy}), os processos de fus\~ao s\~ao interrompidos, por n\~ao serem mais energeticamente favor\'aveis, e portanto n\~ao h\'a mais energia liberada em radia\c{c}\~ao. O n\'ucleo colapsa sob sua pr\'opria gravidade, a densidade aumenta drasticamente, at\'e formar um buraco negro --- se a massa for suficientemente grande para isso --- ou uma explos\~ao de supernova, em que a maior parte do material estelar \'e ejetado, restando somente uma estrela de n\^eutrons\footnote{Durante o colapso do n\'ucleo a densidade fica t\~ao grande que h\'a v\'arias absor\c{c}\~oes de el\'etrons por parte dos pr\'otons, que s\~ao convertidos em n\^eutrons. Muitos desses n\^eutrons s\~ao ejetados durante a explos\~ao, e outros permanecem na regi\~ao do n\'ucleo que sobra, formando um objeto astrof\'isico denominado \emph{estrela de n\^eutrons}.}. A energia liberada nessa imensa explos\~ao \'e parcialmente utilizada para realizar fus\~oes posteriores (que s\~ao rea\c{c}\~oes endot\'ermicas), resultando na forma\c{c}\~ao de elementos mais pesados que $A\gtrsim 56$ atrav\'es da absor\c{c}\~ao de v\'arios n\^eutrons, que posteriormente sofrem decaimento do tipo $\beta$ e s\~ao convertidos em pr\'otons, originando elementos de elevado n\'umero at\^omico. No entanto, supernovas s\~ao objetos bastante mais raros no Universo atual do que estrelas massivas ativas, e a abund\^ancia de elementos muito pesados \'e, por isso, comparativamente muito menor. Na figura~\ref{fig:abundances} consta apenas a soma de todas essas abund\^ancias.

A explos\~ao de uma supernova
\'e tamb\'em o mecanismo pelo qual os elementos pesados, inicialmente produzidos no interior de estrelas, s\~ao lan\c{c}ados ao espa\c{c}o. A poeira dessa explos\~ao eventualmente se condensa, atra\'ida pela sua pr\'opria for\c{c}a gravitacional, e pode vir a formar um novo sistema estelar, como o nosso. Tudo o que \'e formado de elementos mais pesados do que o h\'elio foi gerado no interior de uma estrela supermassiva e lan\c{c}ado ao espa\c{c}o ap\'os uma tal explos\~ao de supernova. \'E por isso que se pode afirmar que \textbf{somos, todos, poeira de estrelas}~\cite{CarlSagan}.

\subsubsection{Extra\c{c}\~ao de energia por fus\~ao nuclear}
\label{sec:fusao_energia}

Por fim, nota-se que, como a curva da energia de liga\c{c}\~ao (figura~\ref{fig:binding_energy}) \'e muito mais \'ingreme para fus\~ao do que fiss\~ao, a efici\^encia da extra\c{c}\~ao de energia por fus\~ao (i.e. a quantidade de energia extra\'ida por quantidade de combust\'ivel utilizado) \'e muito maior. De fato, cada fiss\~ao de $^{235}$U libera, em m\'edia, 200~MeV de energia~\cite{Krane}, o que d\'a $\sim 8.2\times 10^{10}$~J por grama de combust\'ivel\footnote{O c\'alculo \'e simples, bastando converter 1~eV $\approx 1.6\times 10^{-19}$~J e lembrar que 1 mol de $^{235}$U possui $\approx 6.02\times 10^{23}$ \'atomos e totaliza uma massa de aproximadamente 235~g.}, ao passo que uma fus\~ao deut\'erio-tr\'itio,
\begin{equation}
	^2\text{H} \,+ \,^3\text{H} \to \,^4\text{He} + \text{n} + 17.6~\text{MeV},
\end{equation}
libera $\sim 3.4\times 10^{11}$~J/g, cerca de 4 vezes mais do que a fiss\~ao para cada grama de combust\'ivel. Entretanto, \'e muito mais dif\'icil produzir uma rea\c{c}\~ao controlada e autossustent\'avel de fus\~ao, porque o combust\'ivel deve ser mantido a temperaturas e densidades alt\'issimas, para que os nucl\'ideos possam superar a barreira de repuls\~ao coulombiana, como j\'a discutido anteriormente. A manuten\c{c}\~ao dessas condi\c{c}\~oes \'e custosa e tecnicamente complexa, o que faz com que a extra\c{c}\~ao controlada de energia por meio de fus\~ao ainda seja um problema em aberto na F\'isica Nuclear.

Tamb\'em \'e poss\'ivel produzir armamentos baseados na fus\~ao nuclear. Um tal dispositivo \'e denominado bomba de hidrog\^enio (bomba-H) ou arma termonuclear. Nesse caso, as altas temperaturas requeridas para realiza\c{c}\~ao da fus\~ao s\~ao fornecidas ao combust\'ivel por uma pr\'evia detona\c{c}\~ao de uma bomba de fiss\~ao.

\subsection{Energia nuclear: uma aula de CTSA}
\label{sec:CTSA}

Toda a discuss\~ao anterior culmina em uma oportunidade de se realizar um amplo debate com os(as) estudantes sobre os impactos socioambientais da energia nuclear. 

\subsubsection{Armas nucleares}

Em colabora\c{c}\~ao com o(a) professor(a) de hist\'oria de seu centro de ensino, o(a) docente pode discutir o contexto e as consequ\^encias do lan\c{c}amento das bombas nucleares sobre as cidades de Hiroshima e Nagasaki em 1945 --- at\'e hoje os \'unicos casos de uso militar de armamentos nucleares na hist\'oria ---, bem como a corrida armamentista subsequente que levou \`a prolifera\c{c}\~ao desses artefatos, e os atuais acordos e negocia\c{c}\~oes almejando a n\~ao-prolifera\c{c}\~ao e o desmantelamento dos m\'isseis existentes. Imagens ilustrando o poder destrutivo desse tipo de armamentos, incluindo compara\c{c}\~oes das cidades japonesas atingidas, antes e depois da destrui\c{c}\~ao, podem ser encontradas em~\cite{BBC, NuclearDarkness, Allthatisinteresting}.
	
\subsubsection{Ci\^encia e \'Etica}

N\~ao \'e incomum, na hist\'oria humana, que resultados de pesquisas cient\'ificas sejam utilizados para servir fins belicistas ou at\'e mesmo prop\'ositos genocidas. E, embora o senso-comum contempor\^aneo n\~ao hesite em trivialmente atribuir toda a responsabilidade aos pol\'iticos, fato \'e que nenhum desses projetos poderia ser bem sucedido se n\~ao fosse pela participa\c{c}\~ao de cientistas extremamente competentes. Por outro lado, essas mesmas linhas de pesquisa frequentemente originam aplica\c{c}\~oes indiscutivelmente ben\'eficas \`a humanidade, como aprimoramento de t\'ecnicas medicinais e farmacol\'ogicas. Nesse contexto, o(a) docente pode levantar uma salutar discuss\~ao interdisciplinar sobre quest\~oes como: como a ci\^encia e a sociedade civil poderiam/deveriam lidar com potenciais conflitos \'eticos advindos de uma determinada pesquisa cient\'ifica? Como deve se portar o cientista diante desses supostos dilemas? H\'a situa\c{c}\~oes em que \'e desej\'avel impor limites \`a nossa capacidade de explora\c{c}\~ao cient\'ifica? Ou seja, existem casos em que a ignor\^ancia possa ser mais ben\'efica do que o conhecimento? Poderia ser essa ignor\^ancia um meio eficiente e desej\'avel de se evitar cat\'astrofes humanit\'arias? 
	
Esse debate n\~ao se limita \`a F\'isica. Atualmente h\'a amplas discuss\~oes no meio cient\'ifico, e na sociedade civil em geral, sobre dilemas \'eticos envolvendo a engenharia gen\'etica, e os limites que se poderia ou se deveria imp\^or ao desenvolvimento dessas t\'ecnicas de manipula\c{c}\~ao g\^enica. 
	
Portanto essa atividade ofereceria uma excelente oportunidade de di\'alogo e interdisciplinaridade envolvendo tamb\'em os(as) docentes de Filosofia, Hist\'oria, Sociologia e Biologia do centro de ensino em que atua o(a) docente de F\'isica.
	
\subsubsection{Matrizes energ\'eticas e impactos ambientais}
\label{sec:ambiente}

Em di\'alogo com a Geografia, pode-se levantar a discuss\~ao sobre as vantagens e os problemas associados \`a energia nuclear, especialmente quando comparados aos impactos gerados por outras fontes de energia.
	
	\begin{itemize}
		\item Primeiramente, \'e relevante e interessante comparar a taxa de gera\c{c}\~ao de energia por cada grama de combust\'ivel consumido em usinas nucleares e na queima de combust\'iveis f\'osseis. J\'a estimamos anteriormente, na subse\c{c}\~ao~\ref{sec:fusao_energia}, que a fiss\~ao de 1~g de $^{235}$U produz cerca de $8\times 10^{10}$~J. No entanto, uma estimativa realista da efici\^encia de uma usina nuclear deve levar em conta fatores como o grau de enriquecimento do ur\^anio, especificidades do funcionamento do reator, bem como a perda de energia por calor, resultando finalmente em valores entre $5-500\times 10^{8}$~J/g de energia el\'etrica por grama de combust\'ivel consumido~\cite{NuclearEnergy21st, WNA_Heat}. Por sua vez, o poder calor\'ifico dos combust\'iveis f\'osseis (carv\~ao, g\'as, \'oleo, etc.) \'e da ordem de $4\times 10^4$~J/g~\cite{NuclearEnergy21st, WNA_Heat}. Ou seja, uma usina nuclear \'e de 10 mil a 1 milh\~ao de vezes mais eficiente na convers\~ao de combust\'ivel em energia --- corroborando a expectativa j\'a apontada ao fim da subse\c{c}\~ao~\ref{sec:binding}.
		
		Isso significa que se requer menos usinas nucleares para produzir a mesma quantidade de energia atualmente advinda de combust\'iveis f\'osseis, al\'em de haver uma redu\c{c}\~ao de gastos (e de consequentes impactos ambientais) relativos ao transporte do combust\'ivel desde a mina at\'e a usina.
		\item A rea\c{c}\~ao de fiss\~ao resulta em libera\c{c}\~ao de energia na forma de radia\c{c}\~ao e calor. Nas especifica\c{c}\~oes t\'ecnicas de um reator, a pot\^encia produzida nessa forma \'e medida em MWt, ou megawatts-t\'ermicos. Para converter essa energia em eletricidade, usa-se um \emph{material refrigerador} --- geralmente, mas n\~ao exclusivamente, \'agua. A \'agua circula em torno do reator, absorve o calor e se vaporiza. O vapor ent\~ao gira uma turbina que, por indu\c{c}\~ao eletromagn\'etica, produz uma corrente. Ou seja, o processo de produ\c{c}\~ao de eletricidade, em uma usina nuclear, \'e o mesmo que em hidrel\'etricas, termel\'etricas ou usinas e\'olicas: a rota\c{c}\~ao de uma turbina. O que varia \'e a \emph{fonte prim\'aria} da energia. A pot\^encia el\'etrica produzida pela usina \'e medida em MWe, ou megawatts-el\'etricos. 
		
		O refrigerador tamb\'em desempenha o importante papel de manter o n\'ucleo do reator a temperaturas moderadas, de modo que os materiais envolvidos n\~ao derretam, o que interromperia o funcionamento adequado da usina e provocaria s\'erios acidentes.
		
		Outros componentes de um reator j\'a foram mencionados anteriormente. O \emph{material moderador} --- normalmente \'agua ou grafite --- tem a fun\c{c}\~ao de absorver parte da energia cin\'etica dos n\^eutrons resultantes das fiss\~oes, tornando-os n\^eutrons lentos, o que aumenta a probabilidade de promoverem a fiss\~ao do $^{235}$U. E as \emph{barras de controle} s\~ao constitu\'idas de elementos com alta probabilidade de absor\c{c}\~ao dos n\^eutrons, controlando a taxa de fiss\~oes ocorrendo no reator e evitando assim um aumento exponencial e descontrolado da energia liberada (i.e. uma rea\c{c}\~ao supercr\'itica como na figura~\ref{fig:chain}), o que causaria poss\'iveis derretimentos do n\'ucleo do reator ou explos\~oes que destruiriam a usina e potencialmente lan\c{c}aria material radioativo ao ambiente\footnote{Note que, devido ao baixo n\'ivel de enriquecimento do ur\^anio usado em reatores, \'e imposs\'ivel ocorrer uma explos\~ao similar ao de uma bomba nuclear, mesmo nos acidentes em que a rea\c{c}\~ao escapa de controle. As explos\~oes de reatores, nesses casos, decorrem da elevada press\~ao de vapor que \'e produzida em quantidade maior do que a usina foi desenhada para suportar.}. Mais detalhes sobre os elementos constituintes e o funcionamento de reatores nucleares podem ser encontrados na ref.~\cite{WNA_Reactors}.
		\item A energia nuclear \'e vista pela opini\~ao p\'ublica como perigosa e danosa ao meio ambiente e ao ser humano. Entretanto, v\'arias pesquisas indicam que essa percep\c{c}\~ao n\~ao coaduna com os fatos. 
		
		Em primeiro lugar, nota-se que a produ\c{c}\~ao de energia nuclear incorre em n\'iveis baix\'issimos de emiss\~ao de gases de efeito estufa, 100 vezes menos do que a queima de carv\~ao e g\'as, e em quantidade compar\'avel \`as emiss\~oes por fontes e\'olica, hidr\'aulicas e solar~\cite{WNA_Environment}. Essa dr\'astica discrep\^ancia \'e ainda mais significativa ao notarmos que cerca de 60\% das emiss\~oes atuais de CO$_2$ adv\'em do setor energ\'etico~\cite{OWD_CO2}. Portanto, aumentar a fatia nuclear na matriz energ\'etica mundial contribuiria significativamente para a supera\c{c}\~ao do desequil\'ibrio clim\'atico ocasionado pela emiss\~ao massiva de gases de efeito estufa, sem comprometer a produ\c{c}\~ao energ\'etica em n\'iveis capazes de satisfazer \`a crescente demanda. De fato, v\'arios estudos comprovam essa correla\c{c}\~ao entre o aumento do uso de fontes nucleares de energia e a consequente \emph{redu\c{c}\~ao} dos n\'iveis de polui\c{c}\~ao~\cite{Menyah1, Menyah2}.
		
		Existem, tamb\'em, estudos comparando o n\'umero de mortes provocadas pela produ\c{c}\~ao de energia por diversas fontes, levando em conta tanto os acidentes durante o processo de produ\c{c}\~ao de energia (incluindo minera\c{c}\~ao e o trabalho nas usinas) quanto a polui\c{c}\~ao emitida. Os resultados mostram que, para cada terawatt-hora de energia produzida, a queima de combust\'iveis f\'osseis causa cerca de 78 perdas de vidas humanas, enquanto a energia nuclear est\'a associada a 0.07 mortes~\cite{OWD_Safety}.
		
		Quanto \`a preocupa\c{c}\~ao de contamina\c{c}\~ao ambiental, \'e importante mencionar que usinas nucleares n\~ao s\~ao a \'unica fonte energ\'etica que produz res\'iduos radioativos. As rochas extra\'idas na minera\c{c}\~ao do carv\~ao cont\^em, tamb\'em, pequenas concentra\c{c}\~oes de elementos radioativos. Durante a queima desse min\'erio em usinas termel\'etricas, esses res\'iduos radioativos s\~ao lan\c{c}ados ao ar junto com as cinzas, e contaminam o solo onde se depositam, sendo posteriormente absorvidos por plantas e ingeridos na cadeia alimentar. A dosagem radioativa recebida por habitantes nos arredores de uma usina termel\'etrica \'e usualmente \emph{maior} do que nas proximidades de uma usina nuclear --- embora em ambos os casos os n\'umeros costumam estar abaixo dos limites de seguran\c{c}a impostos pelas ag\^encias reguladoras~\cite{SciAm:CoalRadiation}.
		
		Em geral, no quesito seguran\c{c}a, um ponto favor\'avel \`a energia nuclear \'e justamente o fato de haver forte conscientiza\c{c}\~ao quanto aos riscos envolvidos, o que faz com que n\~ao faltem estudos e investimentos voltados a garantir a maior seguran\c{c}a em todos os processos de produ\c{c}\~ao, desde a extra\c{c}\~ao do mineral at\'e a reciclagem ou descarte dos res\'iduos p\'os-produ\c{c}\~ao.
		\item O maior problema associado \`as tecnologias atuais para produ\c{c}\~ao de energia nuclear \'e a co-produ\c{c}\~ao de res\'iduos altamente t\'oxicos no processo de fiss\~ao --- o chamado ``lixo nuclear''. Um componente desse lixo s\~ao os nucl\'ideos resultantes da fiss\~ao do combust\'ivel, como alguns exemplificados na figura~\ref{fig:chain}, que s\~ao altamente radioativos (e assim permanecem por muitos s\'eculos ou at\'e mil\^enios), apresentando alto risco de contamina\c{c}\~ao a seres vivos e ao meio ambiente, caso n\~ao sejam descartados adequadamente. Ademais, os muitos n\^eutrons resultantes da rea\c{c}\~ao em cadeia, ao incidirem nos materiais que circundam o reator, podem tamb\'em torn\'a-los fonte de radioatividade, que precisam ser tratados com igual cautela. N\~ao \'e exagero dizer que, at\'e o momento, n\~ao existe uma solu\c{c}\~ao totalmente satisfat\'oria para lidar com esse material t\'oxico. Uma parte poderia ser reprocessada e reutilizada na usina, ou transmutada, por meio de bombardeamentos radioativos, em outros elementos de menor toxicidade~\cite{WNA_Waste}. Quanto ao restante, uma pr\'atica comum \'e simplesmente enterr\'a-lo em dep\'ositos nas profundezas do subsolo, fora da biosfera, em uma regi\~ao com baixo risco de contamina\c{c}\~ao de aqu\'iferos, e l\'a deix\'a-los at\'e que decaiam completamente, mantendo constante monitoramento da situa\c{c}\~ao. Entretanto, em caso de aumento da contribui\c{c}\~ao da energia nuclear \`a matriz energ\'etica mundial, haveria aumento dr\'astico na produ\c{c}\~ao de lixo radioativo, o que requereria dep\'ositos cada vez maiores. Para uma discuss\~ao mais detalhada, vide ref.~\cite{Beck}.
		\item H\'a, ainda, outro fator importante na discuss\~ao sobre  impactos ambientais da produ\c{c}\~ao de energia: o quanto de \'agua precisa ser desviada ou consumida no processo. No caso de usinas nucleares, assim como nas usinas de queima de carv\~ao e biomassa, grandes quantidades de \'agua precisam circular pela usina para resfriar o reator e evitar superaquecimento. Parte desses recursos h\'idricos \'e perdida por evapora\c{c}\~ao, e outra parte retorna \`a fonte inicial (lagos ou mares) a temperaturas mais elevadas, o que tamb\'em impacta negativamente o ecossistema local. Um estudo realizado nos Estados Unidos\footnote{Os n\'umeros exatos podem variar muito, a depender da regi\~ao geogr\'afica em que se encontra a usina em quest\~ao.} mostra que, nesse quesito: (i) as hidrel\'etricas t\^em tipicamente o maior impacto, pela pr\'opria natureza de seu funcionamento: uma imensa quantidade de recursos h\'idricos \'e desviada para constituir o reservat\'orio, e h\'a grande perda devido \`a evapora\c{c}\~ao; (ii) os impactos das usinas nucleares s\~ao compar\'aveis \`as de combust\'iveis f\'osseis e biomassa, mas elas desviam/consomem muito mais recursos h\'idricos do que fontes renov\'aveis de energia como as usinas e\'olicas e solares (fotovoltaicas)~\cite{Macknick:2012}.		
		\item Outra ressalva comum do p\'ublico \`a produ\c{c}\~ao de energia nuclear adv\'em do receio de acidentes catastr\'oficos, com imensa contamina\c{c}\~ao do solo e da \'agua, e consequente exposi\c{c}\~ao da popula\c{c}\~ao a alt\'issimas doses de radia\c{c}\~ao, como o caso dos acidentes de Chernobyl e Fukushima. O problema adv\'em da possibilidade de um reator tornar-se supercr\'itico --- e liberar uma quantidade descontrolada de energia --- caso haja falha em algum mecanismo de controle e de seguran\c{c}a. Mais especificamente, a efic\'acia de um reator depende de um balan\c{c}o t\^enue entre a taxa de absor\c{c}\~ao e de modera\c{c}\~ao dos n\^eutrons, para que a quantidade de n\^eutrons termais no reator n\~ao seja nem demasiadamente baixa --- o que interromperia a rea\c{c}\~ao e a gera\c{c}\~ao cont\'inua de energia --- nem muito elevada --- o que causaria uma rea\c{c}\~ao supercr\'itica como na figura~\ref{fig:chain}. 
		
		Em Chernobyl, uma falha no desenho do reator, aliado a uma s\'erie de incidentes operacionais, fez com que o reator se tornasse supercr\'itico, produzindo mais vapor do que a constru\c{c}\~ao suportava, causando uma explos\~ao do n\'ucleo e expondo toneladas de material radioativo \`a atmosfera~\cite{WNA_Chernobyl}.
		
		O acidente de Fukushima deveu-se a um \emph{tsunami} que destruiu os sistemas de refrigera\c{c}\~ao dos reatores, causando superaquecimento e consequente derretimento do combust\'ivel radioativo, contaminando a \'agua e o solo da regi\~ao~\cite{WNA_Fukushima}. \'E curioso notar que a usina comportou-se de maneira extremamente robusta frente ao terremoto que precedeu o \emph{tsunami}, tendo os reatores sido desativados por sistemas de seguran\c{c}a diante da detec\c{c}\~ao dos tremores. Entretanto, um reator nuclear continua liberando calor por algum tempo ap\'os seu desligamento, e infelizmente os engenheiros negligenciaram o risco de \emph{tsunamis} subsequentes aos tremores de terra, que acabaram causando a cat\'astrofe. Ainda assim, a resili\^encia da usina diante do terremoto mostra como \'e poss\'ivel reduzir as possibilidades de acidentes dr\'asticos quando os riscos s\~ao devidamente acessados.
		
		A respeito de acidentes em usinas nucleares, \'e importante ressaltar que a explos\~ao resultante da rea\c{c}\~ao supercr\'itica em reatores \emph{n\~ao} possui poder de destrui\c{c}\~ao compar\'avel ao de um armamento nuclear, porque a concentra\c{c}\~ao de material fission\'avel \'e muito inferior nos reatores do que nos armamentos (i.e. o n\'ivel de enriquecimento do combust\'ivel \'e baixo, como dito na se\c{c}\~ao~\ref{sec:fissao}). O maior problema, em acidentes nucleares, n\~ao \'e a explos\~ao em si, mas o vazamento de material radioativo que a explos\~ao provoca.
		\item Os problemas citados acima, associados \`a produ\c{c}\~ao de energia por fiss\~ao nuclear, n\~ao ocorrem na extra\c{c}\~ao de energia por fus\~ao. Por um lado, os nucl\'ideos envolvidos na fus\~ao s\~ao leves e os subprodutos s\~ao muito menos radioativos do que os resultantes da fiss\~ao (ou seja, decaem muito mais rapidamente, ent\~ao o tempo de armazenamento e cautela \'e menor). Ademais, o risco de acidentes similares aos supracitados inexiste: ao contr\'ario da fiss\~ao, em que o desafio \'e \emph{controlar} a rea\c{c}\~ao em cadeia para que ela n\~ao se torne supercr\'itica, na fus\~ao o desafio \'e \emph{manter} a rea\c{c}\~ao ocorrendo. A manuten\c{c}\~ao de uma rea\c{c}\~ao de fus\~ao autossustent\'avel requer um ajuste t\~ao fino de condi\c{c}\~oes de temperatura e densidade do plasma que, havendo qualquer eventualidade que ocasione a perda de controle sobre qualquer etapa da cadeia produtiva (como um terremoto nas proximidades da usina), o processo inteiro seria automaticamente interrompido. 
		
		Em outras palavras, a fus\~ao nuclear tem o potencial de constituir uma fonte de energia muito mais limpa do que a queima de carv\~ao e a fiss\~ao nuclear. Infelizmente, justamente por ser t\~ao dif\'icil manter a fus\~ao autossustent\'avel, a extra\c{c}\~ao de energia por essa fonte \'e extremamente desafiadora e ainda um problema sem solu\c{c}\~ao. Entretanto, \'e essencial que haja cont\'inuos investimentos em pesquisa nessa \'area, especialmente considerando a urg\^encia de reduzir a queima de combust\'iveis f\'osseis e outros emissores de gases de efeito estufa, visando evitar a iminente cat\'astrofe ambiental a que rumamos devido ao aquecimento global.
	\end{itemize}

	Maiores informa\c{c}\~oes sobre diversos aspectos envolvidos na produ\c{c}\~ao de energia nuclear podem ser obtidas em~\cite{WNA}. Excelentes materiais did\'aticos a respeito da produ\c{c}\~ao de energia nuclear e da medicina nuclear, voltadas a estudantes de n\'ivel m\'edio, produzidos pela Comiss\~ao Nacional de Energia Nuclear (CNEN), podem ser encontradas em~\cite{CNEN1, CNEN2}.
	
	Como proposta de atividade, o(a) docente pode sugerir aos estudantes que pesquisem sobre o consumo de energia mensal ou anual de sua cidade/estado, e calculem a quantidade de $^{235}$U que deveria ser fissionado para satisfazer essa demanda. Os dados de consumo energ\'etico devem estar dispon\'iveis nos sites oficiais do governo do estado (ou da Secretaria de estado respons\'avel pela ger\^encia de infraestrutura e meio ambiente), ou no site da Ag\^encia Nacional de Energia El\'etrica (ANEEL)~\cite{ANEEL} e da Empresa de Pesquisa Energ\'etica (EPE)~\cite{EPE}. 
	
	Por se tratar de um tema pol\^emico em nossa sociedade, sem respostas fixas e triviais, essa tem\'atica oferece uma excelente oportunidade para motivar uma din\^amica de debates entre todos em sala. Aqui, o(a) docente pode e deve usar sua criatividade para fomentar atividades polivalentes, que visem aprimorar v\'arias compet\^encias do(a) estudante para al\'em do racioc\'inio f\'isico. Por exemplo, dividindo-se a turma em grupos, o(a) docente pode sugerir que cada um desses apresente suas perspectivas sobre os pr\'os e contras da energia nuclear --- ou, mais abrangentemente, sobre os problemas relacionados \`a matriz energ\'etica nacional e mundial --- na forma de um debate jornal\'istico, uma mesa-redonda, ou como uma pe\c{c}a teatral, um sarau de poesias e m\'usicas, ou um ``duelo'' de repentistas ou de \emph{rappers}, ou qualquer outra manifesta\c{c}\~ao art\'istico-cultural com que os(as) estudantes se identifiquem, fomentando n\~ao somente a alfabetiza\c{c}\~ao cient\'ifica do(a) estudante, mas tamb\'em o desabrochar de suas aptid\~oes art\'isticas atrav\'es da cria\c{c}\~ao e exposi\c{c}\~ao a diversas manifesta\c{c}\~oes art\'istico-culturais. 

\subsection{Decaimentos nucleares}
\label{sec:decaimentos}

\begin{figure}[h!]
    \centering
    \includegraphics[width=.4\textwidth]{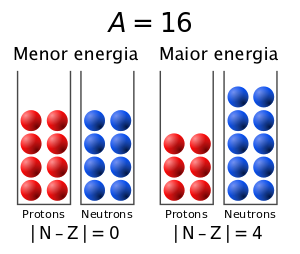}
    \caption{Distribui\c{c}\~ao esquem\'atica de pr\'otons e n\^eutrons em n\'iveis de energia. A configura\c{c}\~ao de menor energia total (ou seja, de maior energia de liga\c{c}\~ao) \'e aquela em que o n\'umero de pr\'otons e n\^eutrons \'e igual. Fonte: Wikimedia Commons, sob licen\c{c}a CC-BY-SA 3.0~\cite{BY-SA30}.}
    \label{fig:levels}
\end{figure}

Exceto pelo $^1\text{H}$, que consiste apenas em um pr\'oton, todos os nucl\'ideos observados na Natureza s\~ao constitu\'idos tanto por pr\'otons quanto por n\^eutrons. A presen\c{c}a dos n\^eutrons contribui para a estabilidade do nucl\'ideo, aumentando a atra\c{c}\~ao nuclear sem acrescentar nenhuma repuls\~ao eletrost\'atica. Mas, j\'a que os n\^eutrons n\~ao se repelem eletrostaticamente, por que n\~ao observamos nucl\'ideos formados apenas por n\^eutrons? O motivo, um pouco mais sutil, \'e o \emph{princ\'ipio da exclus\~ao de Pauli}, que imp\~oe que dois n\^eutrons n\~ao podem possuir simultaneamente as mesmas configura\c{c}\~oes. Isso significa que nem todos n\^eutrons podem ocupar o estado de menor energia no n\'ucleo, de modo que o excesso deve ocupar estados de energias cada vez maiores. O mesmo vale para pr\'otons, o que serve como ainda outro motivo (al\'em da repuls\~ao coulombiana) para coibir a exist\^encia de um nucl\'ideo com pr\'otons excessivos. Mas, se metade dos constituintes do n\'ucleo forem pr\'otons, e a outra metade forem n\^eutrons, a distribui\c{c}\~ao de energia ficaria mais igualit\'aria entre eles, e a energia total da configura\c{c}\~ao seria comparativamente menor (o que corresponde a uma maior energia de liga\c{c}\~ao, e portanto a um nucl\'ideo mais est\'avel). Esse efeito\footnote{O efeito \'e exatamente an\'alogo ao que ocorre para el\'etrons na eletrosfera at\^omica. Cada orbital comporta apenas dois el\'etrons, devido ao princ\'ipio da exclus\~ao, desde que estejam com \emph{spins} desemparelhados. Assim, quanto mais el\'etrons o \'atomo possuir, mais externos ser\~ao os orbitais por eles ocupados, o que significa que eles s\~ao mais facilmente ioniz\'aveis (i.e. possuem menor energia de liga\c{c}\~ao). Assim como a eletrosfera, o n\'ucleo tamb\'em possui camadas energ\'eticas, e cada uma comporta dois n\^eutrons e dois pr\'otons, desde que seus \emph{spins} estejam desemparelhados. Em nucl\'ideos mais pesados que o $^4\text{He}$, portanto com mais do que 2 pr\'otons e 2 n\^eutrons, o excesso deve ocupar n\'iveis de energia maiores, como mostrado na figura~\ref{fig:levels}.} est\'a ilustrado na figura~\ref{fig:levels}.

Se apenas esse efeito, devido ao princ\'ipio da exclus\~ao, fosse relevante no equil\'ibrio energ\'etico nuclear, os nucl\'ideos est\'aveis teriam igual n\'umero de pr\'otons e n\^eutrons. Mas j\'a sabemos que existem outros fatores relevantes nessa din\^amica, como a repuls\~ao coulombiana, que tende a favorecer a presen\c{c}a de n\^eutrons ao inv\'es de pr\'otons.

\begin{figure}
    \centering
    \includegraphics[width=.46\textwidth]{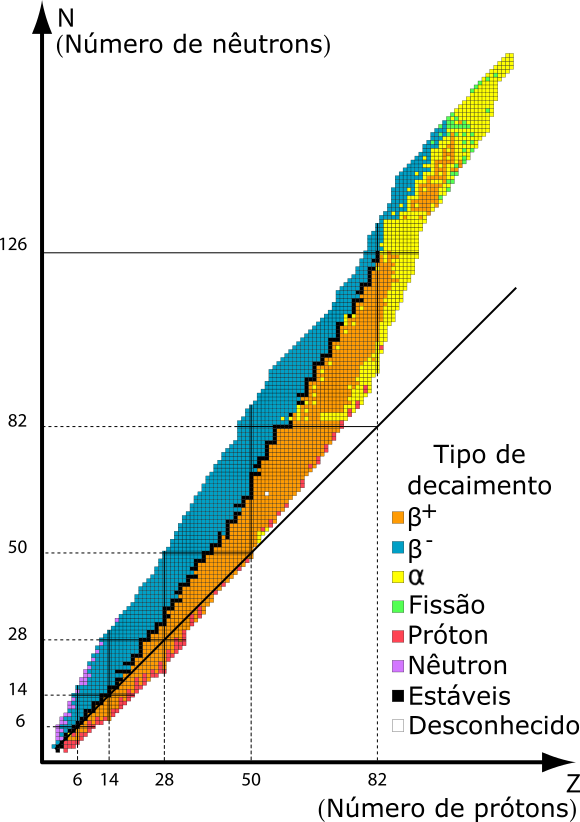}
    \caption{Distribui\c{c}\~ao de nucl\'ideos por n\'umero de n\^eutrons ($N$) e n\'umero de pr\'otons ($Z$), coloridos de acordo com o principal tipo de decaimento observado para cada um deles. Os pontos pretos representam nucl\'ideos est\'aveis, que n\~ao decaem. Nucl\'ideos acima desses pontos
    t\^em um excesso de n\^eutrons e decaem por decaimento $\beta^-$ (convertendo um n\^eutron em um pr\'oton, pontos azuis), ou pela emiss\~ao de um n\^eutron (pontos lil\'as). Abaixo dos pontos
    de estabilidade ocorre o oposto: os nucl\'ideos t\^em um excesso de pr\'otons e tendem a decair por $\beta^+$ (convertendo um pr\'oton em n\^eutron, pontos laranjas) ou, em casos extremos, por emiss\~ao de um pr\'oton (pontos vermelhos). Para nucl\'ideos mais pesados (sobretudo os que cont\^em mais que $\simeq 150$ n\'ucleons) o excesso de pr\'otons pode ser aliviado tamb\'em com decaimento $\alpha$, que frequentemente \'e seguido de um decaimento $\beta^+$. Note que n\~ao h\'a nucl\'ideos est\'aveis com mais que 82 pr\'otons e 126 n\^eutrons, que corresponde ao elemento \emph{chumbo} ($^{208}\text{Pb}$). A maior parte dos elementos nessa regi\~ao decai por emiss\~ao de part\'iculas $\alpha$, e alguns at\'e sofrem fiss\~ao espont\^anea --- ambos processos reduzem o n\'umero de pr\'otons e n\^eutrons, aliviando a repuls\~ao coulombiana e diminuindo o raio do nucl\'ideo, aumentando a intensidade m\'edia da for\c{c}a nuclear atrativa entre n\'ucleons, pois passam a estar mutuamente mais pr\'oximos. As linhas verticais e horizontais correspondem aos chamados ``n\'umeros m\'agicos''. Nucl\'ideos que possuem n\'umero de pr\'otons e/ou n\^eutrons igual(is) a 6, 14, 28, 50, 82 ou 126 s\~ao excepcionalmente est\'aveis, por possu\'irem camadas nucleares completas (da mesma forma que os gases nobres s\~ao quimicamente est\'aveis por possu\'irem orbitais eletr\^onicos completamente preenchidos). V\'arios comportamentos ilustrados na figura s\~ao explicados por esses n\'umeros m\'agicos. Vide, por ex., discuss\~ao no ap\^endice~\ref{sec:atividades}. Fonte: Wikimedia commons, sob licen\c{c}a CC-BY-SA 3.0~\cite{BY-SA30}. Um gr\'afico interativo e atualizado pode ser encontrado em~\cite{IAEA:decays}.
    }
    \label{fig:decays}
\end{figure}

O resultado l\'iquido desses efeitos competitivos \'e que nucl\'ideos est\'aveis tendem a ter um ligeiro excesso de n\^eutrons sobre o n\'umero de pr\'otons. Isso pode ser visto na figura~\ref{fig:decays}, que mostra os nucl\'ideos existentes na Natureza em termos de seu n\'umero de pr\'otons e n\^eutrons. A curva preta corresponde aos nucl\'ideos \emph{est\'aveis}, que n\~ao decaem. Nucl\'ideos est\'aveis e leves, com $A\lesssim 16$, t\^em mesmo n\'umero de pr\'otons e n\^eutrons, mas os mais pesados t\^em uma ligeira prefer\^encia por um excesso de n\^eutrons, pois eles contrabalanceiam a repuls\~ao eletrost\'atica. 

Os pontos coloridos da figura~\ref{fig:decays}, fora da curva preta, s\~ao nucl\'ideos que possuem mais pr\'otons ou mais n\^eutrons do que demandado pela condi\c{c}\~ao de estabilidade. Nesse caso ocorrer\'a um processo de transmuta\c{c}\~ao, em que esse excesso ser\'a aliviado de alguma forma, com tend\^encia a transformar aquele nucl\'ideo em uma configura\c{c}\~ao est\'avel. A esse processo de transmuta\c{c}\~ao d\'a-se o nome de \emph{decaimento radioativo}.

Existem diversos tipos de decaimento, dentre os quais os principais s\~ao chamados de decaimentos $\alpha$ (alfa), $\beta$ (beta) e $\gamma$ (gama). Esses nomes foram dados no in\'icio das pesquisas radioativas, quando n\~ao se sabia nada sobre esses decaimentos a n\~ao ser que eram qualitativamente diferentes --- por isso escolheram as tr\^es primeiras letras do alfabeto grego para nomin\'a-los. Hoje sabemos que cada tipo de decaimento corresponde a um tipo diferente de part\'icula que \'e emitida.

\subsubsection{Decaimento $\alpha$}

\begin{figure}
	\centering
	\includegraphics[scale=.3]{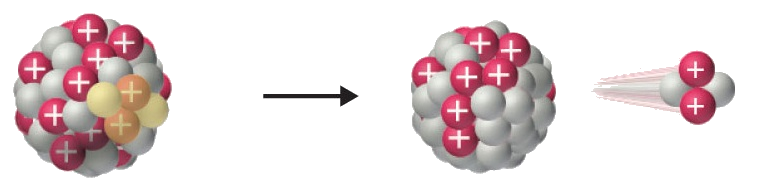}
	\caption{Decaimento do tipo $\alpha$, em que um nucl\'ideo de $^4$He (2 pr\'otons e 2 n\^eutrons) \'e emitido pelo nucl\'ideo m\~ae devido \`a repuls\~ao coulombiana. Fonte: Chemistry LibreTexts~\cite{ChemLibre}, sob licen\c{c}a Creative Commons BY-NC-SA 3.0~\cite{CCBYSA30}.}
	\label{fig:decay_alpha}
\end{figure}

O decaimento $\alpha$ ocorre para nucl\'ideos muito pesados, sobretudo para aqueles que t\^em mais de $\approx 150$ n\'ucleons, e consiste na emiss\~ao de um n\'ucleo de h\'elio-4, como ilustrado na figura~\ref{fig:decay_alpha}. Nesse contexto, o nucl\'ideo de $^4\text{He}$ \'e tamb\'em chamado de \emph{part\'icula $\alpha$}.

Trata-se de uma maneira eficiente de fazer um nucl\'ideo pesado reduzir a repuls\~ao coulombiana
--- por emitir dois pr\'otons --- e tamb\'em reduzir o seu tamanho, fazendo com que todos n\'ucleons fiquem, em m\'edia, mais pr\'oximos uns dos outros, aumentando a energia de liga\c{c}\~ao m\'edia (pois a for\c{c}a nuclear \'e mais intensa a menores dist\^ancias). O decaimento $\alpha$ se explica, portanto, pela repuls\~ao eletrost\'atica entre part\'iculas no interior do nucl\'ideo --- uma explica\c{c}\~ao bastante acess\'ivel a estudantes de n\'ivel m\'edio\footnote{Mais precisamente, o processo envolve o tunelamento qu\^antico de uma part\'icula $\alpha$ atrav\'es da barreira produzida pela atra\c{c}\~ao nuclear e a repuls\~ao coulombiana. Ou seja, nesse modelo a part\'icula $\alpha$ \'e vista como pr\'e-existente e confinada no interior do nucl\'ideo, rebatendo em suas ``paredes'' at\'e eventualmente tunelar. Esses detalhes esclarecem o car\'ater qu\^antico do fen\^omeno --- n\~ao se trata de uma repuls\~ao cl\'assica, at\'e porque o evento \'e probabil\'istico. No entanto ---  e esse \'e o ponto central dessa s\'erie de artigos --- a simples compreens\~ao do mecanismo central por tr\'as do decaimento $\alpha$, como uma disputa entre as intera\c{c}\~oes nuclear e eletrost\'atica, j\'a fundamenta uma compreens\~ao s\'olida do fen\^omeno, al\'em de fomentar o uso da intui\c{c}\~ao f\'isica por parte do(a) discente, demonstrando a aplicabilidade de conceitos simples, como a repuls\~ao eletrost\'atica, na compreens\~ao de fen\^omenos da F\'isica Contempor\^anea.}.

A maior estabilidade do nucl\'ideo-filha pode ser vista tamb\'em da figura~\ref{fig:binding_energy}. Os nucl\'ideos que decaem por $\alpha$ est\~ao \`a direita do $^{56}\text{Fe}$, onde a curva  \'e descendente. Os resultados desses decaimentos, i.e. os nucl\'ideos-filha, est\~ao sempre \`a esquerda dos nucl\'ideos-m\~ae nesta curva descendente e, portanto, t\^em maior energia de liga\c{c}\~ao por n\'ucleon, o que significa maior estabilidade.

Na figura~\ref{fig:decays}, os pontos amarelos representam nucl\'ideos que decaem majoritariamente por emiss\~ao de part\'iculas $\alpha$. Uma 
 {proposta de atividade}, nesse contexto, \'e usar essa figura como exer\'icio de interpreta\c{c}\~ao gr\'afica, bem como para desenvolver racioc\'inio f\'isico a respeito dos decaimentos --- vide ap\^endice~\ref{sec:atividades}.

Por fim, a partir do d\'eficit de massa entre os produtos do decaimento (nucl\'ideo-filha + part\'icula $\alpha$) e reagente (nucl\'ideo-m\~ae) \'e poss\'ivel calcular a energia liberada na rea\c{c}\~ao. Por exemplo, para o decaimento do pol\^onio-214, que Rutherford usou como fonte de suas part\'iculas $\alpha$ no experimento descrito na se\c{c}\~ao~\ref{sec:Rutherford}, tem-se
\begin{equation}
    ^{214}\text{Po} \to \,^{210}\text{Pb}+\alpha.
\end{equation}
Usando os valores tabelados para as massas envolvidas, encontra-se $E_\text{liberada}\approx 7.83$~MeV. Nem toda essa energia vai para a part\'icula $\alpha$ pois, por conserva\c{c}\~ao de momento linear, o nucl\'ideo-filha tamb\'em deve sofrer um recuo. \'E um simples exerc\'icio de mec\^anica\footnote{No referencial de repouso do nucl\'ideo-m\~ae, o momento inicial \'e zero, portanto o momento final deve tamb\'em se anular, ou seja,\[ m_\text{filha} v_\text{filha}=m_\alpha v_\alpha.\] Por outro lado, por conserva\c{c}\~ao de energia de repouso + cin\'etica,
\[ m_\text{m\~ae} c^2 = m_\text{filha} c^2 + \frac{m_\text{filha} v_\text{filha}^2}{2} + m_\alpha c^2 + \frac{m_\alpha v_\alpha^2}{2},\] de onde se chega \`a equa\c{c}\~ao~(\ref{eq:Ealpha}).} mostrar que a fra\c{c}\~ao de energia que vai para a part\'icula $\alpha$ \'e
\begin{equation}
    E_\alpha = \left(\frac{1}{1+\frac{m_\alpha}{m_\text{filha}}}\right)E_\text{liberada},
    \label{eq:Ealpha}
\end{equation}
que nesse caso d\'a $E_\alpha\approx 7.7$~MeV, valor utilizado na se\c{c}\~ao~\ref{sec:Rutherford}.

\subsubsection{Decaimentos $\beta$}
\label{sec:beta}

Os tipos mais comuns de decaimentos s\~ao os chamados $\beta^-$ e $\beta^+$. No primeiro caso, h\'a a convers\~ao de um n\^eutron em um pr\'oton, com emiss\~ao de um el\'etron (e tamb\'em um antineutrino\footnote{Neutrinos e antineutrinos s\~ao part\'iculas neutras e com massa muito pequena, desprez\'ivel para a maior parte dos efeitos. Essas part\'iculas s\~ao t\~ao elusivas que sua descoberta ocorreu d\'ecadas ap\'os os primeiros estudos sobre radioatividade. A hist\'oria e as caracter\'isticas dessas part\'iculas s\~ao importantes na discuss\~ao sobre a intera\c{c}\~ao fraca, que ser\'a tema de artigo posterior. Por isso, n\~ao falaremos muito sobre elas aqui.}),
\begin{equation}
    n\to p + e^- + \overline{\nu}_e
    \qquad
    (\text{decaimento}~\beta^-).
\end{equation}
Esses decaimentos ocorrem em nucl\'ideos que t\^em excesso de n\^eutrons, comparado ao valor \'otimo para estabilidade (acima dos pontos pretos na figura~\ref{fig:decays}).

Ao discutir esses decaimentos, o(a) professor(a) pode explorar o princ\'ipio da \emph{conserva\c{c}\~ao da carga el\'etrica}. Como o n\^eutron n\~ao possui carga, sua transmuta\c{c}\~ao em um pr\'oton (que tem carga positiva) requer a simult\^anea emiss\~ao de uma part\'icula com carga negativa, que, no caso, \'e o el\'etron.

\begin{figure}[h!]
    \centering
    \includegraphics[scale=.29]{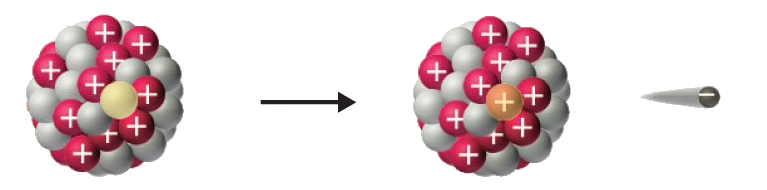}
    \caption{Decaimento $\beta^-$, em que um n\^eutron se converte em um pr\'oton, emitindo um el\'etron (e tamb\'em um neutrino, que n\~ao est\'a mostrado na figura). Fonte: Chemistry LibreTexts~\cite{ChemLibre}, sob licen\c{c}a Creative Commons BY-NC-SA 3.0~\cite{CCBYSA30}.}
    \label{fig:beta}
\end{figure}

O outro tipo de decaimento $\beta$ \'e o oposto: um pr\'oton decai em um n\^eutron, com emiss\~ao de uma part\'icula com carga igual \`a do pr\'oton (ou seja, igual \`a do el\'etron em m\'odulo, mas de sinal positivo). Ou seja,
\begin{equation}
    p\to n + e^+ +\nu_e
    \qquad
    \text{(decaimento}~\beta^+).
\end{equation}
Esse decaimento ocorre a fim de aliviar o excesso de pr\'otons. Nota-se, novamente, a conserva\c{c}\~ao da carga el\'etrica.

A part\'icula $e^+$ emitida nesse decaimento \'e id\^entica ao el\'etron em todos aspectos, exceto pelo sinal de sua carga, que \'e positivo (o m\'odulo da carga e a massa de ambas s\~ao iguais). Ou seja, trata-se de um ``el\'etron com carga oposta'', tamb\'em chamado de \emph{antiel\'etron} ou \emph{p\'ositron}. Diz-se que o p\'ositron \'e a antipart\'icula associada ao el\'etron (e, reciprocamente, o el\'etron \'e a antipart\'icula do p\'ositron). Uma discuss\~ao sobre a interpreta\c{c}\~ao dessas antipart\'iculas e suas aplica\c{c}\~oes pr\'aticas pode ser encontrada na primeira parte desta s\'erie de artigos~\cite{CarvalhoDorsch:2021lvd}.

Tamb\'em \'e pertinente comentar que o el\'etron (ou o p\'ositron) resultantes do decaimento \emph{n\~ao} estavam no ``interior'' dos n\^eutrons/pr\'otons que decaem. \'E comum se pensar que o decaimento \'e a emiss\~ao de uma part\'icula j\'a pr\'e-existente no reagente, e essa interpreta\c{c}\~ao de fato se aplica ao decaimento $\alpha$, mas \emph{n\~ao} para o $\beta$. Ou seja, \emph{n\~ao} \'e correto pensar que o n\^eutron \'e formado de um pr\'oton + um el\'etron, e que o decaimento $\beta^-$ apenas desintegra esse composto. At\'e porque isso traria problemas se tent\'assemos interpretar o decaimento $\beta^+$ da mesma forma\footnote{Afinal, seria o n\^eutron composto de um pr\'oton + el\'etron, ou seria o pr\'oton composto de um n\^eutron + p\'ositron? A resposta n\~ao \'e nenhuma das duas op\c{c}\~oes: os el\'etrons e p\'ositrons n\~ao est\~ao previamente presentes dentro desses n\'ucleons.}. O pr\'oton \emph{n\~ao} cont\'em um p\'ositron, nem o n\^eutron cont\'em previamente um el\'etron. O que ocorre \'e que esses el\'etrons e p\'ositrons s\~ao \emph{criados} no processo do decaimento. Conforme j\'a discutido na primeira parte desta s\'erie~\cite{CarvalhoDorsch:2021lvd}, part\'iculas podem ser criadas ou destru\'idas, pois a famosa f\'ormula de Einstein, $E=mc^2$, garante que massa pode ser convertida em energia, e vice-versa. Esse processo de cria\c{c}\~ao e destrui\c{c}\~ao de part\'iculas no decaimento $\beta$ ser\'a assunto de discuss\~oes mais detalhadas em um artigo futuro desta s\'erie.

\subsubsection{Decaimento $\gamma$}

O decaimento $\gamma$ se distingue dos discutidos anteriormente na medida em que o nucl\'ideo-m\~ae \emph{n\~ao} sofre transmuta\c{c}\~ao, ou seja, n\~ao se alteram os n\'umeros de pr\'otons e de n\^eutrons.

O que ocorre, nesse caso, \'e que um nucl\'ideo em um estado excitado decai para um estado de energia inferior, e a diferen\c{c}a de energia \'e liberada na forma de radia\c{c}\~ao eletromagn\'etica. O processo \'e exatamente an\'alogo \`a emiss\~ao de f\'otons por el\'etrons que transicionam entre diferentes camadas eletr\^onicas. A \'unica diferen\c{c}a \'e que, agora, s\~ao pr\'otons ou n\^eutrons transicionando entre camadas nucleares.

\begin{figure}
    \includegraphics[scale=.3]{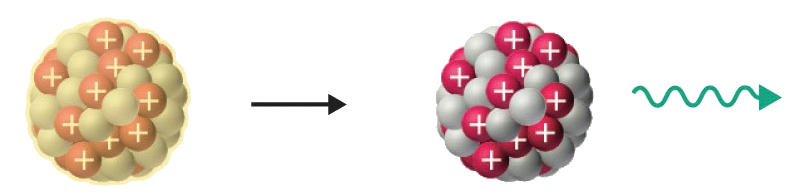}
    \caption{Decaimento $\gamma$, que corresponde \`a emiss\~ao de um f\'oton (ou seja, radia\c{c}\~ao eletromagn\'etica, vide ref.~\cite{CarvalhoDorsch:2021lvd}) por um nucl\'ideo excitado. \`A radia\c{c}\~ao emitida d\'a-se o nome de \emph{raios gama}. Fonte: Chemistry LibreTexts~\cite{ChemLibre}, sob licen\c{c}a Creative Commons BY-NC-SA 3.0~\cite{CCBYSA30}.}
    \label{fig:gamma}
\end{figure}

A diferen\c{c}a de energia entre camadas eletr\^onicas no \'atomo \'e tipicamente da ordem de poucos el\'etron-volts, e portanto a energia do f\'oton emitido em transi\c{c}\~oes at\^omicas \'e dessa ordem de magnitude. Mas j\'a vimos, na subse\c{c}\~ao~\ref{sec:binding}, que as energias associadas \`a intera\c{c}\~ao nuclear s\~ao da ordem de MeV, ou seja, a radia\c{c}\~ao emitida em decaimentos nucleares \'e um milh\~ao de vezes mais energ\'etica do que em transi\c{c}\~oes eletr\^onicas no \'atomo. A essa radia\c{c}\~ao emitida por transi\c{c}\~oes entre camadas nucleares d\'a-se o nome de \emph{raios gama} ou radia\c{c}\~ao gama. O fato de essa radia\c{c}\~ao ser extremamente energ\'etica \'e o motivo de ela ser t\~ao danosa a organismos vivos.

\begin{figure}[h!]
    \includegraphics[scale=1.3, page=1]{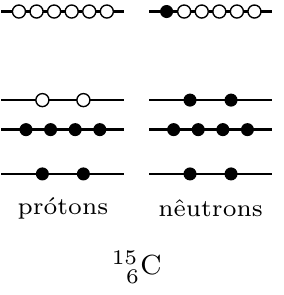}
    \qquad
    \includegraphics[scale=1.3, page=2]{tikz_figs2}
    \caption{Distribui\c{c}\~oes de pr\'otons e n\^eutrons em suas respectivas camadas para (esquerda) carbono-15 e (direita) nitrog\^enio-15. As camadas inferiores s\~ao menos energ\'eticas que as superiores. Quando o carbono-15 decai por $\beta^-$ em nitrog\^enio-15, um dos pr\'otons passa a ocupar a quarta camada (pois essa era a energia do n\^eutron que decaiu), sendo que ainda h\'a espa\c{c}os dispon\'iveis na terceira. Portanto, em seguida ao decaimento $\beta$ inicial, haver\'a um decaimento $\gamma$ para levar o nitrog\^enio-15 ao seu estado fundamental.}
    \label{fig:levels_gamma}
\end{figure}

Decaimentos $\gamma$ geralmente ocorrem ap\'os o nucl\'ideo ter passado anteriormente por um decaimento $\alpha$ ou $\beta$. Por exemplo, a figura~\ref{fig:levels_gamma} (esquerda) ilustra a distribui\c{c}\~ao de pr\'otons e n\^eutrons nas respectivas camadas para o carbono-15. Esse nucl\'ideo \'e inst\'avel por possuir excesso de n\^eutrons, e decai por $\beta^-$, convertendo o n\^eutron da quarta camada em um pr\'oton, e se transmutando em um nitrog\^enio-15. O nucl\'ideo-filha possui um pr\'oton na quarta camada (j\'a que essa era a energia do n\^eutron original), mas a terceira camada est\'a vazia, como mostra a figura~\ref{fig:levels_gamma} (direita). E as camadas mais superiores s\~ao mais energ\'eticas que as inferiores. Portanto, a fim de minimizar a energia da configura\c{c}\~ao, esse pr\'oton vai decair para a terceira camada, e a diferen\c{c}a de energia entre as camadas ser\'a liberada na forma de um f\'oton.

\subsubsection{Outros tipos de decaimentos}

Os decaimentos $\alpha$, $\beta$ e $\gamma$ descritos anteriormente foram os primeiros a serem identificados e s\~ao os mais comuns observados na Natureza. Mas, para nucl\'ideos que possuem um enorme excesso de n\^eutrons ou de pr\'otons, existem outros tipos de decaimentos poss\'iveis, como se v\^e da figura~\ref{fig:decays}. \'E poss\'ivel que o excesso de n\^eutrons (respectivamente pr\'otons) seja aliviado pela emiss\~ao de uma dessas part\'iculas, ao inv\'es de se convert\^e-la em um pr\'oton (respectivamente n\^eutron), como ocorre no decaimento $\beta$. Al\'em disso, para nucl\'ideos muito pesados, \'e poss\'ivel que sofram fiss\~ao espont\^anea e se subdividam em dois nucl\'ideos menores. Ou seja, esse caso seria an\'alogo ao ilustrado na figura~\ref{fig:fission}, por\'em sem a necessidade de se incidir um n\^eutron para induzir a fiss\~ao. Trata-se de um fen\^omeno muito raro. \'E importante mencionar que o ur\^anio e o plut\^onio, dois dos elementos mais usados como combust\'iveis nucleares, n\~ao sofrem fiss\~ao espont\^anea. Nesse caso a fiss\~ao precisa ser induzida por um n\^eutron incidente, ou seja, \'e preciso um gatilho que d\^e in\'icio \`a rea\c{c}\~ao. \'E isso o que permite que tenhamos controle sobre essas usinas, pois quando quisermos diminuir a produ\c{c}\~ao de energia basta capturarmos os n\^eutrons presentes no reator (usando um material chamado \emph{absorvedor}, presente nas chamadas \emph{barras de controle} do reator ---  ver se\c{c}\~ao~\ref{sec:CTSA}).
    
\subsection{Aplica\c{c}\~oes de decaimentos radioativos}

Os decaimentos radioativos descritos acima possuem uma s\'erie de aplica\c{c}\~oes tecnol\'ogicas presentes em nosso quotidiano, que possibilitam uma abordagem dessa tem\'atica em um contexto CTSA. Faremos aqui alguns apontamentos dessas aplica\c{c}\~oes, que obviamente n\~ao encerrar\~ao o tema.

\subsubsection{Data\c{c}\~ao radioativa}

O decaimento radioativo \'e um processo qu\^antico, portanto probabil\'istico. Isso significa que \'e \emph{imposs\'ivel} prever o momento exato em que um nucl\'ideo individual decair\'a. Mas a \emph{taxa de probabilidade} de ele decair (i.e. a probabilidade por unidade de tempo) \'e mensur\'avel, e, com isso, \'e poss\'ivel determinar quanto tempo levar\'a, em m\'edia, para que \emph{metade} dos nucl\'ideos em uma amostra\footnote{Lembrando que uma amostra com $1$ mol da subst\^ancia tem cerca de $10^{23}$ nucl\'ideos, que \'e um n\'umero imenso.} tenha deca\'ido. A situa\c{c}\~ao \'e an\'aloga ao do lan\c{c}amento de uma moeda: se lan\c{c}armos uma a cada segundo, \'e imposs\'ivel saber qual ser\'a o resultado de cada lan\c{c}amento, mas sabemos que precisaremos esperar aproximadamente 1000 segundos para obtermos 500 caras ou 500 coroas (e essa estimativa \'e tanto mais precisa, quanto maior o n\'umero de lan\c{c}amentos envolvidos).

Esse tempo para que metade da amostra tenha deca\'ido \'e chamado de \emph{tempo de meia-vida}, e cada elemento radioativo possui uma meia-vida caracter\'istica. Por exemplo, o tempo de meia-vida do carbono-14 \'e $5730$ anos. Isso significa que, se temos inicialmente uma amostra de $1$~g de $^{14}$C, ap\'os 5730 anos essa amostra conter\'a apenas $0.5$~g desse elemento, e ap\'os outros 5730 anos conter\'a apenas 0.25~g, e assim sucessivamente.

De todo carbono presente na atmosfera terrestre, uma pequena fra\c{c}\~ao\footnote{H\'a aproximadamente 1 \'atomo de radiocarbono para cada 1 trilh\~ao de \'atomos de carbono na atmosfera.} est\'a na forma de carbono-14, tamb\'em chamado de \emph{radiocarbono}. Enquanto um organismo est\'a vivo, respirando e se alimentando (seja realizando fotoss\'intese, produzindo a\c{c}\'ucares a partir de g\'as carb\^onico absorvido da atmosfera, seja fazendo parte de uma cadeia alimentar em cuja base est\~ao esses organismos realizadores de fotoss\'intese), os \'atomos de carbono de seu corpo s\~ao constantemente ``reciclados'', substitu\'idos por novos oriundos da atmosfera, e parte desses \'atomos s\~ao do is\'otopo carbono-14. Isso significa que a concentra\c{c}\~ao de carbono-14 em um organismo vivo \'e a mesma que na atmosfera. Quando o organismo morre,  o carbono-14 deixa de ser reposto, e sua concentra\c{c}\~ao de radiocarbono apenas diminui devido ao decaimento. Assim, se medirmos a concentra\c{c}\~ao\footnote{Na verdade n\~ao se mede diretamente a concentra\c{c}\~ao de radiocarbono, e sim a quantidade de decaimentos por cada grama de todo o carbono presente na amostra. Como a quantidade de decaimentos \'e obviamente proporcional \`a quantidade de \'atomos presentes, as duas medi\c{c}\~oes s\~ao equivalentes. Em um organismo vivo, h\'a cerca de $0.226$ decaimentos por segundo e por cada grama de carbono da amostra. Ent\~ao, se uma amostra tem 0.113 decaimentos por segundo e por grama de carbono, isso significa que o organismo morreu h\'a 5730 anos.} de radiocarbono em uma amostra de um organismo que j\'a foi vivo, podemos determinar quanto tempo transcorreu desde sua morte.

Como se v\^e, uma aula sobre data\c{c}\~ao por radiocarbono \'e uma excelente oportunidade para trazer a \emph{interdisciplinaridade} \`a sala de aula, discutindo ciclo de carbono na biosfera, fotoss\'intese e cadeia alimentar. 

\'E importante ressaltar que a data\c{c}\~ao por radiocarbono s\'o pode ser usada para datar mat\'eria org\^anica --- por exemplo, ossos, restos de plantas, ou at\'e artefatos arqueol\'ogicos, como cer\^amicas enriquecidas por mat\'eria org\^anica. Ademais, n\~ao \'e poss\'ivel datar com radiocarbono materiais com mais do que $\sim 50$ mil anos de idade. Isso porque, ap\'os esse tempo, j\'a transcorreram aproximadamente 10 meias-vidas do radiocarbono, e a fra\c{c}\~ao desse elemento na amostra j\'a \'e $1/2^{10}\sim 0.1\%$ da concentra\c{c}\~ao original. \'E claro que detectores mais precisos permitem extrapolar um pouco esse limite superior, mas n\~ao muito mais do que isso.

Isso significa que f\'osseis de dinossauros \emph{n\~ao} podem ser datados por radiocarbono, (i) porque um f\'ossil de dinossauro n\~ao \'e mat\'eria org\^anica, mas material rochoso que se sedimentou sobre os ossos do animal (ou seja, um f\'ossil \'e uma rocha), e (ii) porque dinossauros foram extintos h\'a cerca de 65 milh\~oes de anos, portanto qualquer vest\'igio de carbono-14 no f\'ossil j\'a teria desaparecido completamente ap\'os esse per\'iodo.

\begin{figure}[h!]
    \centering
    \includegraphics[scale=.6]{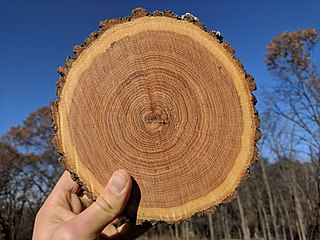}
    \caption{Corte transversal de um tronco de \'arvore mostrando seus an\'eis. A quantidade de an\'eis indica a idade da \'arvore. A colora\c{c}\~ao, a espessura e outras caracteriza\c{c}\~oes dos an\'eis indicam as condi\c{c}\~oes ambientais em que a \'arvore se desenvolveu. Fazendo um levantamento estat\'istico com diversas \'arvores, \'e poss\'ivel fazer uma associa\c{c}\~ao de quais anos calend\'aricos correspondem a quais sequ\^encias de an\'eis. Uma vez estabelecida essa rela\c{c}\~ao, qualquer artefato de madeira pode ser datado a partir da sequ\^encia de an\'eis da madeira. Fonte: Wikimedia commons, sob licen\c{c}a CC BY-SA 4.0~\cite{BY-SA40}.}
    \label{fig:treerings}
 \end{figure}

Isso n\~ao quer dizer que n\~ao se pode fazer data\c{c}\~ao de rochas por meio de decaimentos radioativos. Ao contr\'ario, essas data\c{c}\~oes s\~ao poss\'iveis, desde que se use o decaimento de outros elementos como ``r\'egua'', como o rub\'idio-87, que decai via
\begin{equation}
	\,^{87}\text{Rb} \to \,^{87}\text{Sr} + e^- + \bar{\nu}_e,
\end{equation} 
 com meia-vida $4.9\times 10^{10}$~anos. O m\'etodo \'e essencialmente o mesmo descrito acima, mas com algumas diferen\c{c}as. Nesse caso n\~ao se conhece a abund\^ancia inicial de rub\'idio-87 na amostra inicial, ent\~ao mede-se tamb\'em a abund\^ancia do elemento-filha desse decaimento, o estr\^oncio-87, e tamb\'em o outro is\'otopo estr\^oncio-86 (que \'e est\'avel). Com essas informa\c{c}\~oes \'e poss\'ivel determinar a idade de meteoros e, a partir da\'i, inferir a idade do Sistema Solar (e, portanto, do planeta Terra) como $4.5\times 10^9$ anos~\cite{White, Bowen}.

 \'E importante mencionar que a data\c{c}\~ao radioativa n\~ao \'e a \'unica forma de se datar um artefato, e nem sempre \'e a mais precisa. Na arqueologia, o mais comum \'e datar um objeto com base em outros itens encontrados em uma mesma escava\c{c}\~ao, como moedas, evid\^encias textuais (em l\'apides ou cer\^amicas), pr\'aticas culturais (por ex. a forma como o objeto foi fabricado ou o material utilizado), e outros ind\'icios que apontem para o contexto hist\'orico do objeto desenterrado. Ainda outra forma de data\c{c}\~ao muito precisa \'e a chamada \emph{dendrocronologia}\footnote{Da palavra grega $\delta\acute\epsilon\nu\delta\rho o\nu$ (``\'arvore''), ou seja, cronologia (atrav\'es) das \'arvores.}, t\~ao precisa a ponto de ser usada para calibrar a data\c{c}\~ao feita por m\'etodos radioativos descritos acima! Esse m\'etodo baseia-se no fato de que os troncos das \'arvores s\~ao formados por v\'arios an\'eis (vide figura~\ref{fig:treerings}), produzidos (em m\'edia) um por ano, e cuja espessura, colora\c{c}\~ao e outras caracter\'isticas dependem das condi\c{c}\~oes clim\'aticas e ambientais da regi\~ao onde a \'arvore se desenvolveu. Assim, fazendo um estudo de diversas \'arvores de uma regi\~ao em v\'arias \'epocas diferentes, podemos fazer uma tabela de quais anos calend\'aricos correspondem a quais caracter\'isticas de an\'eis. Tendo essa correspond\^encia em m\~aos, ao encontrarmos um peda\c{c}o de madeira com uma determinada sequ\^encia de an\'eis, podemos inferir a \'epoca em que a \'arvore viveu (e, consequentemente, inferir a \'epoca em que um certo artefato arqueol\'ogico de madeira foi produzido). A dendrocronologia \'e muito usada, por exemplo, para data\c{c}\~ao de obras de arte com base em suas molduras\footnote{Vale notar que, em caso de pinturas produzidas p\'os-Renascimento, i.e. nos \'ultimos 500 anos, os m\'etodos de data\c{c}\~ao radioativa n\~ao s\~ao t\~ao eficientes, pois transcorreu-se ainda pouco tempo desde a produ\c{c}\~ao do artefato (comparativamente \`a meia-vida do carbono-14) e n\~ao se detectaria ainda uma diferen\c{c}a significativa na abund\^ancia desse elemento relativamente ao que se espera para o organismo vivo.}.

 \'E interessante trazer a discuss\~ao do par\'agrafo acima para sala de aula, para evitar a dissemina\c{c}\~ao do mito de que os m\'etodos baseados em decaimentos radioativos s\~ao ``os melhores'' ou ``mais precisos''\footnote{Lembrando que a precis\~ao da data\c{c}\~ao por radiocarbono depende crucialmente de conhecermos a concentra\c{c}\~ao de carbono-14 na atmosfera na \'epoca em que o organismo viveu. Essa concentra\c{c}\~ao \'e aproximadamente constante, mas pode flutuar devido a eventos clim\'aticos ou at\'e antropom\'orficos. Por exemplo, \'e sabido que os in\'umeros testes com armamentos nucleares realizados desde a d\'ecada de 1950 aumentaram drasticamente a concentra\c{c}\~ao de radiocarbono na atmosfera. Portanto o valor que se mede hoje \emph{n\~ao} corresponde \`a concentra\c{c}\~ao de radiocarbono na \'epoca em que o organismo viveu h\'a s\'eculos. \'E preciso fazer uma corre\c{c}\~ao para levar em conta esses efeitos, e para isso usa-se outros m\'etodos de data\c{c}\~ao como refer\^encia, por exemplo a dendrocronologia.}. Ao contr\'ario, trata-se de um dentre muitos m\'etodos de data\c{c}\~ao que se complementam. Novamente o convite \`a \emph{interdisciplinaridade}
 faz-se presente, por se tratar de uma tem\'atica que envolve a f\'isica
 --- na parte dos decaimentos ---, a biologia --- na dendrocronologia ---, a hist\'oria --- na arqueologia --- e tamb\'em as artes ---  na medida em que essas t\'ecnicas s\~ao aplicadas para determinar a origem e a originalidade de pinturas e outras obras famosas.

 \subsubsection{Aplica\c{c}\~oes medicinais}

\begin{figure}[h!]
	\centering
	\includegraphics[width=.45\textwidth]{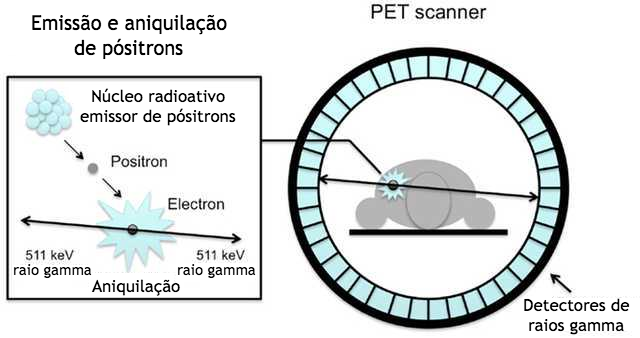}
	\caption{Esquema de funcionamento de um aparelho de Tomografia por Emiss\~ao de P\'ositrons (\emph{PET scan}). Fonte: adaptado de~\cite{physicsforums}.}
	\label{fig:PET}
\end{figure}

 Uma das aplica\c{c}\~oes medicinais de decaimentos radioativos j\'a foi discutida na primeira parte desta s\'erie de artigos: trata-se do uso de decaimentos $\beta^+$ para realiza\c{c}\~ao de exames como o \emph{PET scan}~\cite{CarvalhoDorsch:2021lvd}. Primeiramente, modifica-se uma mol\'ecula usualmente metabolizada pelo organismo (por exemplo, glucose) substituindo-se um de seus \'atomos por um elemento radioativo (por ex. o $^{18}$F) que decai por $\beta^+$, ou seja, emitindo p\'ositrons. Essa subst\^ancia \'e inserida intravenosamente no paciente, e o organismo, confundindo-a com a glucose usual, redireciona-a a \'org\~ao vitais, onde ser\'a metabolizada. Ali, o elemento radioativo decai\footnote{Para o caso do $^{18}$F a meia-vida \'e de aproximadamente $109$ minutos. Por isso o paciente deve chegar \`a cl\'inica e tomar a inje\c{c}\~ao com alguma anteced\^encia antes de realizar o exame.}, emitindo um p\'ositron. Esse p\'ositron se aniquilar\'a com um el\'etron do corpo do paciente, emitindo dois f\'otons  que ser\~ao observados pelo aparelho detector, como na figura~\ref{fig:PET}. Faz-se, assim, um mapeamento da atividade metab\'olica do organismo, podendo-se determinar a integridade do tecido cerebral, a exist\^encia de tumores, o funcionamento cardiovascular, dentre outras aplica\c{c}\~oes.

\begin{figure}
    \centering
    \includegraphics[width=.38\textwidth]{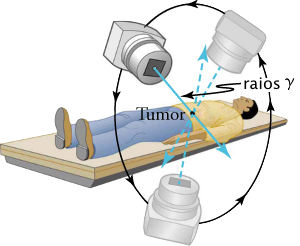}
    \caption{Bombardeamento de um tumor por raios $\gamma$ em sess\~ao de radioterapia. O \^angulo de incid\^encia da radia\c{c}\~ao \'e variado a fim de maximizar o dano ao tumor, minimizando o dano \`as regi\~oes saud\'aveis no seu entorno. Fonte: adaptado de~\cite{OpenStax}, distribu\'ido sob licen\c{c}a Creative Commons BY 4.0~\cite{BY-40}.}
    \label{fig:therapy}
 \end{figure}

Outra aplica\c{c}\~ao medicinal desses decaimentos radioativos \'e a chamada \emph{radioterapia} para combate a c\^anceres. As part\'iculas emitidas em decaimentos radioativos (as part\'iculas $\alpha$, os el\'etrons e p\'ositrons ou a radia\c{c}\~ao $\gamma$) s\~ao tipicamente muito energ\'eticas e, quando incidem sobre c\'elulas de organismos vivos, podem desintegrar o DNA e danificar sua capacidade reprodutiva. As c\'elulas cancer\'igenas s\~ao justamente aquelas que se reproduzem descontroladamente, a uma taxa muito maior que as c\'elulas normais do organismo. Assim, atacando a regi\~ao cancer\'igena com radia\c{c}\~ao e destruindo o DNA de suas c\'elulas, a reprodu\c{c}\~ao delas fica comprometida e o c\^ancer pode ser reduzido ou at\'e eliminado. Em geral, as c\'elulas saud\'aveis em torno do c\^ancer tamb\'em ser\~ao afetadas pela radia\c{c}\~ao, mas se variarmos o \^angulo de incid\^encia da radia\c{c}\~ao, mantendo a regi\~ao 
 afetada no centro (como ilustrado na figura~\ref{fig:therapy}) podemos minimizar o dano a regi\~oes saud\'aveis enquanto maximizamos o bombardeamento ao tumor.

 Outra forma de radioterapia \'e via implantes do tamanho de gr\~aos de arroz, contendo material radioativo, que s\~ao inseridos na regi\~ao do tumor. Esse material decai ao longo do tempo e danifica as c\'elulas tumorosas vizinhas ao implante, atuando como uma radioterapia prolongada e constante~\cite{OpenStax}.

Por fim, nota-se que a radioterapia \'e apropriada para o caso de tumores bem localizados em uma regi\~ao do corpo. H\'a, ainda, outros tratamentos contra c\^ancer, como a quimioterapia, que consiste na administra\c{c}\~ao de uma subst\^ancia que inibe a divis\~ao celular, fazendo com que as c\'elulas cancerosas morram sem se reproduzir. Nesse caso, n\~ao se trata de uma terapia com materiais radioativos, e, sim, com subst\^ancias qu\'imicas --- da\'i o nome do tratamento.

Como se v\^e, uma aula sobre aplica\c{c}\~oes medicinais da radioatividade abre uma janela para a \emph{explora\c{c}\~ao interdisciplinar} da f\'isica com a biologia, envolvendo tanto f\'isica nuclear quanto biologia celular, DNA e ciclo de reprodu\c{c}\~ao da c\'elula. Al\'em disso, trata-se de uma excelente oportunidade para, nesse contexto interdisciplinar, utilizar uma abordagem CTSA em sala de aula.

\subsubsection{Esteriliza\c{c}\~ao e conserva\c{c}\~ao de alimentos}

Como acabamos de mencionar, a radia\c{c}\~ao proveniente de decaimentos nucleares pode ser danosa a organismos vivos, a depender da energia da part\'icula emitida. Assim, outra utilidade pr\'atica desses decaimentos \'e na esteriliza\c{c}\~ao de equipamentos m\'edicos (quando se deseja eliminar microorganismos) e tamb\'em na conserva\c{c}\~ao de alimentos, eliminando-se bact\'erias, fungos, larvas, insetos e demais agentes biol\'ogicos indesej\'aveis. Al\'em disso, atrav\'es da irradia\c{c}\~ao pode-se postergar a matura\c{c}\~ao desses alimentos, fazendo com que durem mais tempo e possam ser transportados por maiores dist\^ancias sem que estraguem.

No Brasil, a irradia\c{c}\~ao de alimentos \'e uma t\'ecnica reconhecida e regulamentada pela ANVISA~\cite{ANVISA:Food}\footnote{ A regulamenta\c{c}\~ao atual vigora desde o ano de 2001, mas uma atualiza\c{c}\~ao est\'a prevista para 2023~\cite{ANVISA:Agenda}.}. De acordo com essa regula\c{c}\~ao, a radia\c{c}\~ao permitida para esse fim s\~ao os raios $\gamma$ provenientes de decaimentos do cobalto-60 e do c\'esio-137\,\footnote{O c\'esio-137 \'e um dos produtos colaterais de fiss\~oes que ocorrem na produ\c{c}\~ao de energia nuclear. Seu uso como fonte de radia\c{c}\~ao para preserva\c{c}\~ao de alimentos constitui um reaproveitamento do que seria considerado ``lixo nuclear''.}, ou raios-X (que tamb\'em s\~ao radia\c{c}\~ao eletromagn\'etica, como os raios $\gamma$, mas nesse caso de energia at\'e 5~MeV) ou at\'e mesmo el\'etrons acelerados a energias de no m\'aximo 10~MeV.

Assim como tudo o que envolve radioatividade, essa aplica\c{c}\~ao na ind\'ustria de alimentos gera preocupa\c{c}\~oes que merecem ser endere\c{c}adas com cuidado. 

A primeira preocupa\c{c}\~ao imagin\'avel \'e que o alimento possa se tornar radioativo. Isso \'e imposs\'ivel, porque a irradia\c{c}\~ao \'e feita com el\'etrons ou radia\c{c}\~ao eletromagn\'etica, que s\~ao incapazes de alterar os is\'otopos dos elementos presentes na comida\footnote{Se o bombardeamento fosse feito com n\^eutrons ou pr\'otons, essas part\'iculas poderiam ser absorvidas por algum n\'ucleo at\^omico presente no alimento, alterando sua estrutura e possivelmente tornando-o radioativo. Por exemplo, bombardeando-se o carbono-12 com n\^eutrons pode-se produzir o carbono-14, que \'e radioativo. Mas f\'otons n\~ao s\~ao capazes de converter um is\'otopo em outro.}$^,$\footnote{O(a) leitor(a) mais atento poderia pensar que o f\'oton bombardeado poderia excitar um n\'ucleo at\^omico presente no alimento, executando o processo inverso ao descrito na figura~\ref{fig:levels_gamma} (direita). Esse nucl\'ideo excitado tenderia a decair com nova emiss\~ao de $\gamma$, que, se ocorresse dentro de nosso corpo, poderia ser danoso \`as nossas c\'elulas. Ocorre que o f\'oton incidente s\'o pode ser absorvido se sua energia for \emph{exatamente} igual \`a diferen\c{c}a de energia entre dois n\'iveis nucleares mostrados na figura. Ent\~ao o f\'oton s\'o seria absorvido por um nucl\'ideo que tenha n\'iveis com a \emph{mesma} diferen\c{c}a de energia dos n\'iveis do cobalto-60 ou do c\'esio-137. Mas a energia dos n\'iveis nucleares servem como a ``impress\~ao digital'' do n\'ucleo, e cada elemento tem n\'iveis com espa\c{c}amentos caracter\'isticos. Ent\~ao a radia\c{c}\~ao $\gamma$ emitida pelo cobalto-60 n\~ao seria absorvida por um n\'ucleo de carbono, nitrog\^enio, oxig\^enio, ou outros elementos que encontramos na mat\'eria org\^anica.}. Por exemplo, \'e imposs\'ivel converter carbono-12 (est\'avel) em carbono-14 (radioativo) com bombardeamento de f\'otons ou el\'etrons.

Uma preocupa\c{c}\~ao mais pertinente diz respeito \`a poss\'ivel degrada\c{c}\~ao nutricional do alimento irradiado, bem como perdas de qualidades sensoriais, como sabor e aroma. Muitos estudos foram e continuam sendo realizados para investigar essa possibilidade. De fato, os estudos mostram que alguns nutrientes vitam\'inicos --- como a vitamina C, por exemplo --- s\~ao ligeiramente danificados pelo procedimento, a depender da temperatura em que eles s\~ao realizados\footnote{A redu\c{c}\~ao dos n\'iveis de vitamina s\~ao menores quando a irradia\c{c}\~ao \'e feita a frio.}, e da dosagem de radia\c{c}\~ao utilizada~\cite{OMS:1994, OMS:1999}. Contudo, assim como na discuss\~ao sobre energia nuclear, \'e preciso comparar esse resultado com o que ocorre nos processos de esteriliza\c{c}\~ao alternativos. Por exemplo, a pasteuriza\c{c}\~ao a altas temperaturas tamb\'em implica em perdas de nutrientes. Nessa perspectiva comparativa, as conclus\~oes dos estudos s\~ao de que o m\'etodo de esteriliza\c{c}\~ao por irradia\c{c}\~ao n\~ao \'e aplic\'avel a todos os alimentos --- at\'e porque muitos t\^em o sabor afetado de modo a torn\'a-los desagrad\'aveis ao paladar, como \'e o caso do leite ---, mas que em  outros casos as poss\'iveis perdas vitam\'inicas s\~ao compar\'aveis \`aquelas advindas dos outros m\'etodos mais usuais de esteriliza\c{c}\~ao.

Importante mencionar que esse m\'etodo \emph{n\~ao} deve ser aplicado para substituir as boas pr\'aticas de higiene na produ\c{c}\~ao e manuseio dos alimentos, e, sim, para complement\'a-las. Isso est\'a determinado muito claramente at\'e nas diretrizes da ANVISA~\cite{ANVISA:Food}. Ademais, essas diretrizes tamb\'em determinam que no r\'otulo do produto deve constar a frase ``ALIMENTO TRATADO POR PROCESSO DE IRRADIA\c{C}\~AO'', a fim de que o consumidor possa fazer uma escolha informada e consciente.

Justamente por ser uma tem\'atica que instiga o debate, e que tem impacto direto em nossa vida quotidiana --- por afetar nossa alimenta\c{c}\~ao ---, essa tem\'atica \'e ideal para uma aula CTSA em um contexto interdisciplinar envolvendo a f\'isica, a biologia --- na discuss\~ao do aspecto nutricional dos alimentos ---, a qu\'imica --- no debate sobre como a radia\c{c}\~ao pode destruir mol\'eculas ou catalizar rea\c{c}\~oes que produzam novas subst\^ancias --- e tamb\'em a geografia --- levantando a discuss\~ao sobre a import\^ancia de se ter t\'ecnicas de preserva\c{c}\~ao e prolongamento do tempo de vida de um alimento, e o impacto socioecon\^omico dessas tecnologias na ind\'ustria aliment\'icia. Abordar essas tem\'aticas em uma perspectiva de debate fomenta o desenvolvimento do racioc\'inio do(a) estudante e sua capacidade de elaborar uma opini\~ao eloquentemente. N\~ao h\'a certo ou errado, desde que todas opini\~oes sejam fundamentadas em dados e fatos. Pode-se, assim, incentivar a pesquisa por parte dos(as) discentes, para que se preparem \`a discuss\~ao. O debate pode tamb\'em servir para explorar a criatividade dos(as) estudantes, que podem apresentar suas falas em forma de um telejornal, pe\c{c}a teatral, m\'usica, ou outras formas que atraiam a aten\c{c}\~ao do(a) aluno(a).

\subsubsection{Tecnologias quotidianas}

\begin{figure*}
    \centering
    \includegraphics[trim=0 30 0 0 , clip, width=.35\textwidth]{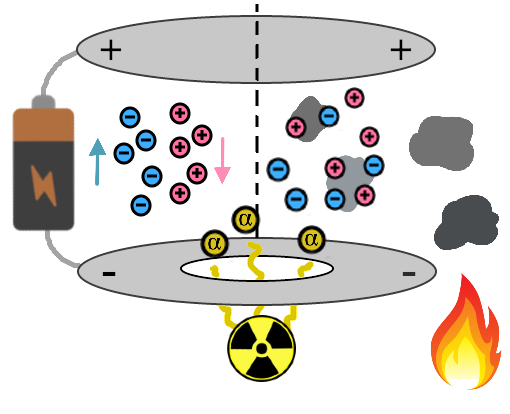}
    \qquad\qquad
    \includegraphics[scale=.25]{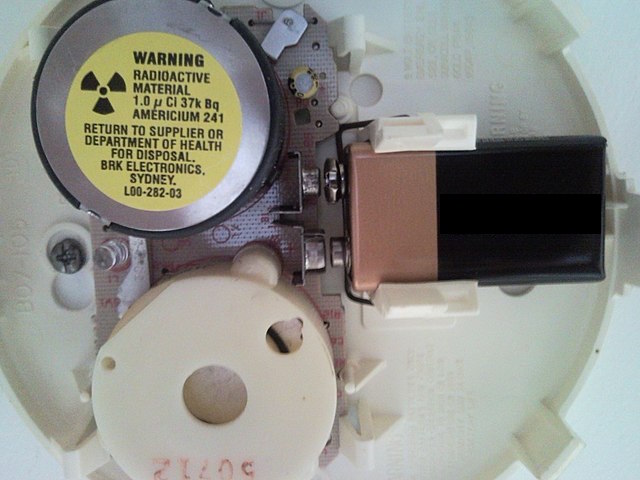}
    \caption{(Esquerda) Esquema do interior de uma c\^amara de ioniza\c{c}\~ao, parte integrante de um detector de fuma\c{c}a i\^onico. A c\^amara cont\'em um elemento emissor de $\alpha$ (tipicamente amer\'icio-241), que ioniza o ar no interior da c\^amara, provocando uma corrente entre as placas conectadas \`a bateria. Quando h\'a fuma\c{c}a, os \'ions s\~ao parcialmente reabsorvidos e o ar \'e novamente neutralizado, interrompendo a corrente e fazendo soar o alarme. (Direita) Interior de um detector de fuma\c{c}a i\^onico. A c\^amara de ioniza\c{c}\~ao est\'a marcada com a etiqueta amarela, que avisa sobre seu conte\'udo radioativo. Abaixo dela est\'a o alarme, e \`a direita a bateria que alimenta o circuito. Fonte: Wikimedia Commons, sob licen\c{c}a CC BY-SA 2.0~\cite{BY-SA20}.}
    \label{fig:smoke}
\end{figure*}

A radioatividade encontra aplica\c{c}\~oes tamb\'em em tecnologias quotidianas, presentes at\'e mesmo em muitas resid\^encias. Um exemplo s\~ao os detectores de fuma\c{c}a i\^onicos, ilustrados na figura~\ref{fig:smoke}. No interior desses detectores h\'a uma \emph{c\^amara de ioniza\c{c}\~ao}, que consiste de duas placas paralelas conectadas a uma bateria e exposta a um elemento radioativo emissor de part\'iculas $\alpha$, tipicamente o amer\'icio-241. Essas part\'iculas $\alpha$ ionizam o ar entre as placas (i.e. removem el\'etrons dos \'atomos), fazendo surgir uma corrente no circuito. As part\'iculas de fuma\c{c}a, quando entram na c\^amara, desionizam o ar, reabsorvendo as part\'iculas carregadas, interrompendo a corrente e fazendo soar o alarme.

\begin{figure}
    \includegraphics[scale=.3]{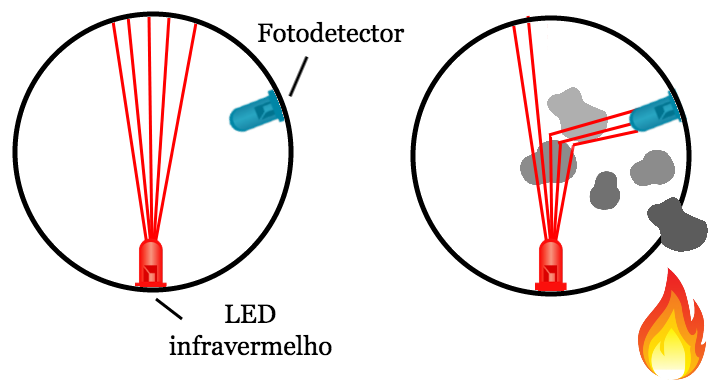}
    \caption{Esquema ilustrando o funcionamento de um detector de fuma\c{c}a \'otico. Em condi\c{c}\~oes normais, um LED emite radia\c{c}\~ao infravermelha que n\~ao atinge o fotodetector. Na presen\c{c}a de fuma\c{c}a, a luz \'e espalhada e desviada de sua trajet\'oria retil\'inea inicial, incidindo no detector e ativando o alarme.}
    \label{fig:smoke_photo}
\end{figure}

Cabe mencionar que este n\~ao \'e o \'unico tipo de detector de fuma\c{c}a dispon\'ivel no mercado. H\'a, ainda, os chamados detectores \'oticos, que funcionam \`a base do efeito fotoel\'etrico~\cite{CarvalhoDorsch:2021lvd}. Uma ilustra\c{c}\~ao do funcionamento desses detectores \'e dada na figura~\ref{fig:smoke_photo}.

\section{An\'alise das interven\c{c}\~oes}
\label{sec:analise}

\subsection{O contexto: o local e os sujeitos}
\label{subsec:contexto}

No ano de 2017, o N\'ucleo Cosmo-UFES da Universidade Federal do Esp\'irito Santo (UFES) iniciou o projeto de extens\~ao ``Universo na Escola''~\cite{UniversoEscola}, em que professores e pesquisadores s\~ao convidados a ministrar palestras sobre t\'opicos atuais de pesquisa em F\'isica a estudantes do ensino m\'edio da rede p\'ublica do estado do Esp\'irito Santo. O intuito do projeto \'e o de ampliar a forma\c{c}\~ao dos estudantes e despertar-lhes o interesse pelas ci\^encias em geral e, em particular, pela F\'isica. Ministradas fora do per\'iodo regular de aulas nas escolas participantes, as palestras se configuram como atividade complementar eletiva. 

Para este trabalho, os autores optaram por extrapolar o formato usual das palestras do contexto do projeto de extens\~ao. Para isso, elaboraram e ministraram uma sequ\^encia did\'atica sobre F\'isica de Part\'iculas contemplando dez encontros com dura\c{c}\~ao aproximada de uma hora cada, em hor\'arios extra-classe, durante os meses de setembro, outubro, novembro e dezembro de 2019, o que totalizou uma carga hor\'aria de dez horas de atividades de F\'isica de Part\'iculas desenvolvidas com os discentes. A sequ\^encia foi ministrada por um dos autores a uma turma mista com, no m\'aximo, quatorze estudantes das 1$^{\text{a}}$, 2$^{\text{a}}$ e 3$^{\text{a}}$ s\'eries do ensino m\'edio do Centro Estadual de Ensino M\'edio em Tempo Integral (CEEMTI) Prof$^{\text a}$ Maura Abaurre, no munic\'ipio de Vila Velha, no estado do Esp\'irito Santo. A organiza\c{c}\~ao local foi realizada pelo professor-coordenador da \'area de F\'isica na escola. Alguns participantes integravam um grupo com interesse pr\'evio em F\'isica, que j\'a era acompanhado e orientado durante o ano pelo professor-coordenador de F\'isica da escola. No entanto, no decorrer da sequ\^encia outros estudantes se juntaram ao grupo, frequentemente a convite dos demais que j\'a participavam das aulas. 

\subsection{Das aulas}
\label{subsec:aulas}

Na se\c{c}\~ao \ref{sec:momentos} apresentamos propostas de momentos did\'aticos para introduzir a tem\'atica de F\'isica Nuclear, organizados de maneira fluida em continuidade \`a primeira parte da sequ\^encia did\'atica, cujo foco foi a introdu\c{c}\~ao \`a F\'isica de Part\'iculas e a Eletrodin\^amica Qu\^antica~\cite{CarvalhoDorsch:2021lvd}. Da mesma forma que na primeira parte desta sequ\^encia did\'atica, os conceitos relevantes \`a tem\'atica eram constru\'idos em conjunto aos discentes a partir de conex\~oes com seus conhecimentos pr\'evios, de forma que o engajamento dos estudantes presentes nas aulas foi mantido ou aumentou consideravelmente, como veremos na se\c{c}\~ao~\ref{subsec:resultados}.

Na interven\c{c}\~ao realizada na CEEMTI Prof$^\text{a}$ Maura Abaurre, o conte\'udo apresentado neste trabalho, com foco na F\'isica Nuclear, serviu como base para a elabora\c{c}\~ao da segunda parte da sequ\^encia did\'atica sobre F\'isica de Part\'iculas. Por sua vez, tal parte foi dividida em duas aulas, a quarta e quinta aulas, ministradas nos dias 7 e 10 de outubro do ano de 2019, com dura\c{c}\~ao de 52 e 65 minutos, respectivamente. 

A primeira aula do bloco de aulas sobre F\'isica Nuclear, em an\'alise neste trabalho, foi a quarta aula da sequ\^encia did\'atica de dez aulas\footnote{Ressaltamos que as tr\^es primeiras aulas da sequ\^encia did\'atica versaram sobre uma introdu\c{c}\~ao \`a F\'isica de Part\'iculas e a intera\c{c}\~ao eletromagn\'etica, cujo material e an\'alise podem ser consultados em nosso primeiro artigo desta s\'erie~\cite{CarvalhoDorsch:2021lvd}.}. O primeiro momento da quarta aula foi de breve revis\~ao da terceira aula da sequ\^encia. Na terceira aula, t\'opicos introdut\'orios de Mec\^anica Qu\^antica foram abordados para a constru\c{c}\~ao dial\'ogica da dualidade onda-part\'icula, do f\'oton como mediador da intera\c{c}\~ao eletromagn\'etica e do conceito de antipart\'icula ou antimat\'eria a partir do p\'ositron, finalizando com um debate sobre aplica\c{c}\~oes na Medicina, como nos raios-X e na tomografia PET-scan. Mais detalhes sobre a terceira aula da sequ\^encia did\'atica podem ser encontrados em~\cite{CarvalhoDorsch:2021lvd}. Os principais t\'opicos abordados em sala no restante da quarta e da quinta aulas, com as respectivas subse\c{c}\~oes discutidas detalhadamente na se\c{c}\~ao~\ref{sec:momentos} acima, encontram-se listados nas tabelas~\ref{tab:Aula4Momentos} e \ref{tab:Aula5Momentos}.

\begin{table*}[h!]
	\centering
	\footnotesize
	\bgroup
	\def\arraystretch{1.1}
	\begin{tabular}{| >{\small} l |l | c |} 
		\hline
		\multicolumn{3}{|c|}{\small\textbf{Aula 4: F\'isica Nuclear -- Parte I}} \\
		\hline
		Se\c{c}\~ao & \multicolumn{2}{ c| }{\small Momento} \\
		\hline
		 & Breve revis\~ao da terceira aula& 1\\
		& As escalas dos efeitos qu\^anticos e relativ\'isticos & 2 \\
		\hline
		\ref{sec:Thomson} & Relembrando o \'atomo de Thomson & 3 \\
		\hline
		\ref{sec:Thomson_deflection} e \ref{sec:Rutherford_model} & O experimento de Rutherford-Geiger-Marsden: levantando hip\'oteses na ci\^encia & 4 \\
		\hline
		& Teste de hip\'oteses na ci\^encia & 5 \\
		\hline
		\ref{sec:ciencia} & Como ocorre o desenvolvimento da ci\^encia -- Parte I & 6 \\
		\hline
		\ref{sec:protons_neutrons} & Os pr\'otons e os n\^eutrons: o n\'ucleo at\^omico possui uma subestrutura & 7 \\
		\hline
		\ref{sec:estabilidade} & A intera\c{c}\~ao nuclear forte: por que o n\'ucleo n\~ao se desintegra? & 8 \\
		\hline
		\ref{sec:discussoes_nuclearforce} & Como ocorre o desenvolvimento da ci\^encia -- Parte II & 9 \\
		\hline
		\ref{sec:nuclearforce} & Comparando a intera\c{c}\~ao forte com as intera\c{c}\~oes gravitacional e eletromagn\'etica & 10 \\
		\hline
	\end{tabular}
	\egroup
	\caption{Os momentos da quarta aula da sequ\^encia.}
	\label{tab:Aula4Momentos}
\end{table*}

\begin{table*}[h!]
	\centering
 	\footnotesize
	\begin{tabular}{| >{\small} l |l | c|} 
		\hline
		\multicolumn{3}{|c|}{{\small \textbf{Aula 5: F\'isica Nuclear -- Parte II}}} \\
		\hline
		Se\c{c}\~ao & \multicolumn{2}{ c| }{\small Momento} \\
		\hline
		 &
		Breve revis\~ao da aula anterior & 1 \\
		\hline
		\ref{sec:reacoes} &
		Energia de liga\c{c}\~ao & 2\\
		\hline
		\ref{sec:fusao} & O processo de fus\~ao nuclear & 3\\
		\hline
		\ref{sec:primordial} e \ref{sec:estelar} & A energia do Sol e a forma\c{c}\~ao de elementos no Universo & 4\\
		\hline
		\multirow{4}{*}{\ref{sec:decaimentos}} &	O que s\~ao decaimentos nucleares? & 5\\
		\cline{2-3}
		 &	Decaimento alfa & 6\\
		\cline{2-3}
		 &	Decaimento beta & 7\\
		\cline{2-3}
		 &	Decaimento gama & 8 \\
		\hline
	    \ref{sec:fissao} &	O processo de fiss\~ao nuclear & 9 \\
		\hline
		\multirow{6}{*}{\ref{sec:CTSA}} & A radioatividade e seus perigos: Marie Curie e Chernobyl & 10\\
		\cline{2-3}
		 & A radioatividade e seus perigos: a bomba de Hiroshima & 11\\
		\cline{2-3}
		& A radioatividade e seus perigos: o desastre de Fukushima & 12\\
		\cline{2-3}
		& Mecanismos de controle em usinas nucleares & 13 \\
		\cline{2-3}
		& Vantagens da energia nuclear e quest\~oes pol\'iticas sobre enriquecimento de ur\^anio & 14\\
		\cline{2-3}
		 & O que fazer com o lixo radioativo? & 15\\
		\hline
		\ref{sec:fusao_energia} e \ref{sec:CTSA} & Energia nuclear a partir de fus\~ao e a bomba de hidrog\^enio & 16\\
		\hline
	\end{tabular}
	\caption{Os momentos da quinta aula da sequ\^encia.}
	\label{tab:Aula5Momentos}
\end{table*}

Em trabalho anterior~\cite{CarvalhoDorsch:2021lvd} ressaltamos o car\'ater dial\'ogico do professor com os estudantes em detrimento a uma postura meramente expositiva, a partir do qual v\'arios momentos surgiram de forma org\^anica das conversas, das interroga\c{c}\~oes dos pr\'oprios estudantes e de propostas de debates. Por exemplo, no in\'icio da quarta aula, ap\'os o momento 3, quando o professor relembrou o \'atomo de Thomson e comentou sobre como Rutherford teve a ideia de um experimento que pudesse testar a composi\c{c}\~ao do \'atomo de Thomson, e que foi implementada por seus assistentes Geiger e Marsden, um dos estudantes em sala de aula trouxe a informa\c{c}\~ao \`a turma, com autonomia, de que uma das part\'iculas alfa lan\c{c}adas ao \'atomo teria desviado muito mais do que o esperado. Assim, o professor aproveitou para aliar os conhecimentos pr\'evios dos estudantes sobre energia cin\'etica e potencial para obter, em conjuntos a esses mesmos estudantes, a velocidade final esperada das part\'iculas alfa caso o modelo de Thomson fosse uma descri\c{c}\~ao adequada para o \'atomo. Al\'em disso, o professor aproveitou para introduzir a curva esperada do n\'umero de part\'iculas alfa desviadas a diversos \^angulos para o \'atomo de Thomson, bem como os pontos experimentais obtidos pelos alunos de Rutherford, e abordou como a discrep\^ancia entre a curva esperada para o \'atomo de Thomson e a a curva dos pontos experimentais obtidos mostrou que algo n\~ao estava correto na descri\c{c}\~ao do \'atomo por Thomson, o que levou ao debate em sala de aula sobre um dos pilares da ci\^encia: o teste de hip\'oteses. Quando as previs\~oes de um modelo baseadas em determinadas hip\'oteses n\~ao condizem com as observa\c{c}\~oes, cientistas devem repensar as hip\'oteses iniciais e perspectivas com rela\c{c}\~ao \`a situa\c{c}\~ao investigada. Com o aux\'ilio do professor, o momento de discuss\~ao sobre o desenvolvimento da ci\^encia foi aprofundado: os estudantes constru\'iram a percep\c{c}\~ao de que os cientistas buscam, a todo momento, criar solu\c{c}\~oes para os problemas em aberto e, quando essas solu\c{c}\~oes aparecem e se mostram eficazes, coerentes com as observa\c{c}\~oes, surgem ainda outros novos problemas para os quais novas solu\c{c}\~oes devem ser buscadas. Tal momento foi guiado pela percep\c{c}\~ao de que a exist\^encia da intera\c{c}\~ao forte resolveu o problema da coes\~ao do n\'ucleo at\^omico, mas deixou um problema em aberto que seria abordado em aulas futuras: o fato de que n\~ao percebemos essa intera\c{c}\~ao no nosso dia a dia. 

Outros exemplos v\^em dos momentos 7 e 8 da aula 4, quando os estudantes foram estimulados a ir al\'em do \'atomo de Rutherford e levantaram hip\'oteses sobre a exist\^encia de subestruturas do n\'ucleo, que foram devidamente introduzidas pelo professor como os pr\'otons e os n\^eutrons. Em seguida, foi proposto um desafio aos estudantes: pensar como o n\'ucleo n\~ao se desintegra, uma vez que os n\^eutrons s\~ao neutros e os pr\'otons, de carga el\'etrica positiva, tendem a se repelir. Os estudantes expuseram seus pensamentos sobre a situa\c{c}\~ao, considerando el\'etrons que orbitam o n\'ucleo e as dist\^ancias envolvidas, at\'e um aluno levantar a hip\'otese de que o n\'ucleo seria formado por ainda outras part\'iculas, menores do que os pr\'otons e os n\^eutrons, e que dever\'iamos olhar para as cargas dessas part\'iculas. Dessa forma, o professor constr\'oi, junto aos alunos, a necessidade da exist\^encia de uma outra intera\c{c}\~ao, que eles ainda desconheciam, al\'em da eletromagn\'etica e da gravitacional, para vencer a repuls\~ao eletromagn\'etica e que seria fundamental para a atra\c{c}\~ao das part\'iculas do n\'ucleo a fim de mant\^e-lo coeso.

No \'ultimo momento da aula 4 houve uma compara\c{c}\~ao entre a nova intera\c{c}\~ao elaborada durante a aula -- a intera\c{c}\~ao forte -- com as intera\c{c}\~oes gravitacional e eletromagn\'etica, focos da primeira parte da sequ\^encia did\'atica~\cite{CarvalhoDorsch:2021lvd}.  Estimular tal pensamento relacional entre os t\'opicos abordados em sala de aula \'e sempre muito frut\'ifero, pois assim os estudantes percebem como a ci\^encia n\~ao \'e um apanhado de blocos tem\'aticos desconexos entre si mas, sim, que pode e deve explorar rela\c{c}\~oes das mais diversas, assim como com tudo o que eles v\^em e percebem do mundo, ampliando suas estruturas cognitivas e vis\~oes sobre o pr\'oprio mundo que os cerca. Al\'em disso, houve uma breve introdu\c{c}\~ao ao conceito dos p\'ions e a men\c{c}\~ao ao brasileiro envolvido na sua descoberta, C\'esar Lattes. Neste momento, partindo do pr\'oprio interesse dos alunos, houve uma discuss\~ao cr\'itica sobre as complexidades de se fazer ci\^encia no Brasil e sobre a hierarquia cient\'ifica acad\^emica e internacional a partir da informa\c{c}\~ao de que o Lattes n\~ao recebeu o pr\^emio Nobel por seu envolvimento na descoberta, o que demonstra interesse por parte dos pr\'oprios alunos em compreender as rela\c{c}\~oes sociais mundiais que envolvem a ci\^encia. 

Na quinta aula, destacamos os momentos sobre a fiss\~ao nuclear e a radioatividade como os mais dial\'ogicos da aula. Os estudantes trazem informa\c{c}\~oes da s\'erie de TV ``Chernobyl''~\cite{Chernobyl} para a aula, formulam diversas perguntas sobre os acontecimentos nela mostrados, e ainda trazem novas informa\c{c}\~oes sobre o acidente. Houve perguntas sobre a nuvem t\'oxica formada, que quase se espalhou por toda a Europa, sobre a forma\c{c}\~ao do c\^ancer nas v\'itimas do acidente e sobre como m\'aquinas ficam destru\'idas com a radia\c{c}\~ao ionizante. Foi um momento de bastante descontra\c{c}\~ao, mas que n\~ao ficou somente na surpresa com as informa\c{c}\~oes. Pelo contr\'ario, elas foram discutidas sistematicamente pelo professor para se aproximar o mais poss\'ivel de um real aprendizado dos conceitos envolvidos e n\~ao permanecer apenas na falsa percep\c{c}\~ao de aprendizado pelo fant\'astico~\cite{BACHELARD2005}. Ap\'os abordar com extens\~ao as desvantagens da energia nuclear, houve interesse dos alunos por uma discuss\~ao sobre suas vantagens e quest\~oes pol\'iticas que envolvem o seu uso, como o enriquecimento do ur\^anio. Foi abordado como as emiss\~oes de gases do efeito estufa, que aceleram o aquecimento global, s\~ao baix\'issimas no caso da energia nuclear. Outra vantagem \'e que h\'a muito menos impacto ambiental em termos de \'area que deve ser empregada para a constru\c{c}\~ao de uma usina nuclear, quando comparada, por exemplo, \`a regi\~ao utilizada para construir usinas hidrel\'etricas. Um dos estudantes, inclusive, trouxe com autonomia v\'arias informa\c{c}\~oes para essa discuss\~ao: o desmatamento consider\'avel na regi\~ao pr\'oxima aos rios e a retirada de residentes locais para a cria\c{c}\~ao dos reservat\'orios e das barragens das hidrel\'etricas, ao que o docente ainda mencionou o risco de rompimento dessas barragens. Os alunos perceberam a complexidade da tem\'atica sobre o uso de energia nuclear, refletindo para al\'em do b\'asico. No in\'icio da discuss\~ao foram abordadas apenas as desvantagens de seu uso, mas, aos poucos, a partir de intera\c{c}\~oes entre colegas e professor, a percep\c{c}\~ao final foi a de que h\'a diversas vantagens e pouca chance de ocorrer um acidente. Contudo, houve o consenso de que, caso ele ocorra, certamente ser\'a um acidente cr\'itico e, sem definir um posicionamento, conclu\'iram a favor da necessidade de mais investimentos em mecanismos de seguran\c{c}a para as usinas nucleares.  

Uma an\'alise preliminar das narrativas presentes em sala de aula, considerando as participa\c{c}\~oes dos estudantes, as trocas com o professor e colegas, e diversas outras observa\c{c}\~oes sobre todas as aulas da sequ\^encia did\'atica pode ser encontrada em~\cite{TCC_Thaisa}, e ser\'a publicada com mais detalhes em trabalho futuro dos autores.

\subsection{Metodologia e coleta de dados}
\label{subsec:metodologia}

Em nossa an\'alise adotamos um percurso metodol\'ogico de cunho qualitativo e car\'ater de estudo de caso. A pesquisa qualitativa considera todos os sujeitos envolvidos, os significados e os pontos de vista atribu\'idos \`as situa\c{c}\~oes a partir de dados coletados diretamente no ambiente natural de a\c{c}\~ao~\cite{LUEDKE1986}. A Secretaria de Educa\c{c}\~ao do Estado do Esp\'irito Santo e a diretoria da escola onde as interven\c{c}\~oes foram aplicadas autorizaram a observa\c{c}\~ao das interven\c{c}\~oes e anota\c{c}\~oes sobre as intera\c{c}\~oes entre os discentes, seus n\'iveis de interesse e de aten\c{c}\~ao, as perguntas realizadas, e demais rea\c{c}\~oes. Para isso, constru\'imos um di\'ario de campo considerando aspectos descritivos e reflexivos dos sujeitos envolvidos, dos objetos, do espa\c{c}o, das atividades e dos acontecimentos~\cite{LUEDKE1986, OLIVEIRA2014}, bem como reflex\~oes ap\'os as interven\c{c}\~oes propriamente ditas sobre as poss\'iveis rela\c{c}\~oes entre tais aspectos, sobre as atividades para casa e intera\c{c}\~oes imediatamente anteriores e posteriores \`as interven\c{c}\~oes. A metodologia de estudo de caso partiu da viv\^encia com o local onde as a\c{c}\~oes se desenvolveram, ou seja, em sala de aula, onde consideramos as perspectivas de significados dos sujeitos e a identifica\c{c}\~ao de rela\c{c}\~oes causais e padr\~oes em contextos complexos que n\~ao permitem a utiliza\c{c}\~ao de levantamentos e experimentos~\cite{MORAESTAZIRI2019}. Os dados coletados focam na compreens\~ao de conceitos-chave e suas rela\c{c}\~oes com tecnologia, sociedade e meio ambiente, avaliando o surgimento de indicadores de alfabetiza\c{c}\~ao cient\'ifica e de engajamento ao longo das aulas, bem como rela\c{c}\~oes entre esses indicadores. 

Para manter o sigilo e preservar suas identidades, optamos por adotar nomes fict\'icios para os estudantes, de cientistas que atuaram na \'area da F\'isica de Part\'iculas e/ou contribu\'iram para o seu desenvolvimento.

\subsection{Os indicadores de alfabetiza\c{c}\~ao cient\'ifica e de engajamento}
\label{subsec:indicadores}

A an\'alise da viabilidade e da efetividade da sequ\^encia did\'atica aplicada, bem como das potencialidades da inclus\~ao da tem\'atica de F\'isica de Part\'iculas em salas de aula do ensino m\'edio brasileiro, considerou a incid\^encia de indicadores de alfabetiza\c{c}\~ao cient\'ifica e de engajamento demonstrado pelos estudantes ao longo das aulas e suas atividades. Ressaltamos que as interven\c{c}\~oes incorporaram uma postura amplamente dial\'ogica do professor e acreditamos que sua combina\c{c}\~ao com uma tem\'atica mais contempor\^anea da F\'isica, a F\'isica de Part\'iculas, favoreceu a viabilidade e a efetividade de nossa proposta.

Os indicadores de alfabetiza\c{c}\~ao cient\'ifica utilizados se alicer\c{c}am na perspectiva freireana que destaca o(a) alfabetizado(a) cientificamente como aquele(a) que: compreende as rela\c{c}\~oes entre Ci\^encia, Tecnologia, Sociedade e Meio Ambiente; compreende a natureza da ci\^encia; compreende a \'etica que envolve o trabalho de um(a) cientista; e possui conhecimentos b\'asicos sobre as ci\^encias para atuar no mundo~\cite{FREIRE2000, FREIRE1989}. De acordo com~\cite{SasseronCarvalho2008}, em qualquer processo de ensino-aprendizagem voltado \`a alfabetiza\c{c}\~ao cient\'ifica identificam-se indicadores que demonstram o desenvolvimento de habilidades e conhecimentos associados ao trabalho de cientistas a partir de a\c{c}\~oes que: envolvem o trabalho com os dados obtidos em uma investiga\c{c}\~ao, que estruturam o pensamento cient\'ifico e que buscam o entendimento da situa\c{c}\~ao analisada. Cada conjunto de a\c{c}\~oes resume uma s\'erie de elementos do fazer cient\'ifico, como a seria\c{c}\~ao, a organiza\c{c}\~ao e a classifica\c{c}\~ao de informa\c{c}\~oes; o uso do racioc\'inio l\'ogico e/ou proporcional; o levantamento e o teste de hip\'oteses; a previs\~ao; entre outros. Tais elementos do fazer cient\'ifico foram considerados em nossa an\'alise como indicadores de alfabetiza\c{c}\~ao cient\'ifica e s\~ao resumidos na tabela~\ref{tab:Acoes}. Cada um desses indicadores \'e independente dos demais e eles podem, inclusive, aparecer concomitantemente. Para mais detalhes sobre o referencial te\'orico e considera\c{c}\~oes sobre os indicadores de alfabetiza\c{c}\~ao cient\'ifica adotados, recomendamos a leitura do primeiro artigo desta s\'erie sobre F\'isica de Part\'iculas no ensino m\'edio~\cite{CarvalhoDorsch:2021lvd} (ver tamb\'em~\cite{TCC_Thaisa}).
\begin{table*}[h!]
	\centering
	\scriptsize
	\def\arraystretch{1.75}
	\begin{tabular}{|m{5cm}| m{6cm}|} 
		\hline
		\textbf{A\c{c}\~ao do fazer cient\'ifico} &
		    \textbf{Indicadores de alfabetiza\c{c}\~ao cient\'ifica\newline
		    (Elementos do fazer cient\'ifico)}
		\\ \hline
		Trabalho com os dados obtidos\newline
		    em uma investiga\c{c}\~ao
		& 
		Seria\c{c}\~ao de informa\c{c}\~oes\newline 
		Organiza\c{c}\~ao de informa\c{c}\~oes\newline
		Classifica\c{c}\~ao de informa\c{c}\~oes		
		\\
		\hline
		Estrutura\c{c}\~ao do pensamento cient\'ifico
		& 
		Racioc\'inio l\'ogico\newline
		Racioc\'inio proporcional
		\\ \hline
		Entendimento da situa\c{c}\~ao analisada
		& 
		Levantamento de hip\'oteses\newline
		Teste de hip\'oteses\newline	
		Justificativa\newline		
		Previs\~ao\newline	 
		Explica\c{c}\~ao\\ \hline
	\end{tabular}
	\caption{Os indicadores de alfabetiza\c{c}\~ao cient\'ifica a partir de elementos do fazer cient\'ifico.}
	\label{tab:Acoes}
\end{table*}

Apesar da natureza multifacetada que reflete a complexa intera\c{c}\~ao social do estudante com o ambiente escolar, as situa\c{c}\~oes vivenciadas e os outros sujeitos desse espa\c{c}o, a literatura sobre engajamento escolar contempla tr\^es tipos de engajamento: o comportamental, o emocional e o cognitivo~\cite{FREDERICKSetal2004, COELHO2011, BORGESetal2005, SASSERONSOUZA2019, FARIAVAZ2019, Finn1993, Voelkl1997, STIPEK2002, CONNELLWELLBORN1991, BROPHY1987, AMES1992, DWECKLEGGETT1988, HARTER1981, CORNOMADINACH1983}. Todos se relacionam de forma din\^amica e n\~ao ocorrem em processos isolados. A partir desse referencial te\'orico, consideramos os indicadores de engajamento listados na tabela~\ref{tab:Engajamento} para avaliar a viabilidade, a efetividade e as potencialidades de inserir a tem\'atica de F\'isica de Part\'iculas em salas de aula do ensino m\'edio brasileiro. Para mais detalhes, novamente indicamos o primeiro artigo desta s\'erie sobre F\'isica de Part\'iculas no ensino m\'edio~\cite{CarvalhoDorsch:2021lvd} (ver tamb\'em~\cite{TCC_Thaisa}).
\bgroup
\def\arraystretch{1.4}
\begin{table*}[h!]
	\scriptsize
	\centering
	\begin{tabular}{|>{\centering}m{4cm}| >{\centering}m{4cm} | c|} 
		\hline
		\multicolumn{3}{|c|}{\textbf{Indicadores de engajamento}} \\
		\hline
		\textbf{Comportamental} &
		\textbf{Emocional} &
		\textbf{Cognitivo} \\
		\hline
		Participa\c{c}\~ao nas aulas &
		Emo\c{c}\~ao &
		Investimento no aprendizado \\
		\hline
		Participa\c{c}\~ao/execu\c{c}\~ao\newline
		   nas/das tarefas de sala &
		Identifica\c{c}\~ao com\newline 
		    a escola e/ou colegas &
		Autonomia \\
		\hline
		\multirow{3}{4cm}{\centering Participa\c{c}\~ao/execu\c{c}\~ao\newline
		    nas/das tarefas de casa} &
		Identifica\c{c}\~ao com o professor &
		Desejo de ir al\'em do b\'asico \\\cline{2-3}
		&
		Atribui\c{c}\~ao de valores \`a F\'isica\newline
		   e/ou \`as ci\^encias em geral &
		Uso de estrat\'egias \\		\hline
	\end{tabular}
	\caption{Os indicadores de engajamento.}
	\label{tab:Engajamento}
\end{table*}
\egroup

\subsection{Resultados}
\label{subsec:resultados}

Nesta se\c{c}\~ao apresentamos os resultados referentes \`a incid\^encia, durante a quarta e quinta aulas da sequ\^encia did\'atica, dos indicadores de alfabetiza\c{c}\~ao cient\'ifica e de engajamento, apresentados na se\c{c}\~ao~\ref{subsec:indicadores} acima. Uma publica\c{c}\~ao \`a parte aprofundar\'a na an\'alise da incid\^encia desses indicadores ao longo de toda a sequ\^encia. 

\begin{figure*}[h!]
	\centering
        \includegraphics[trim=0 265 0 0, clip, scale=.52]{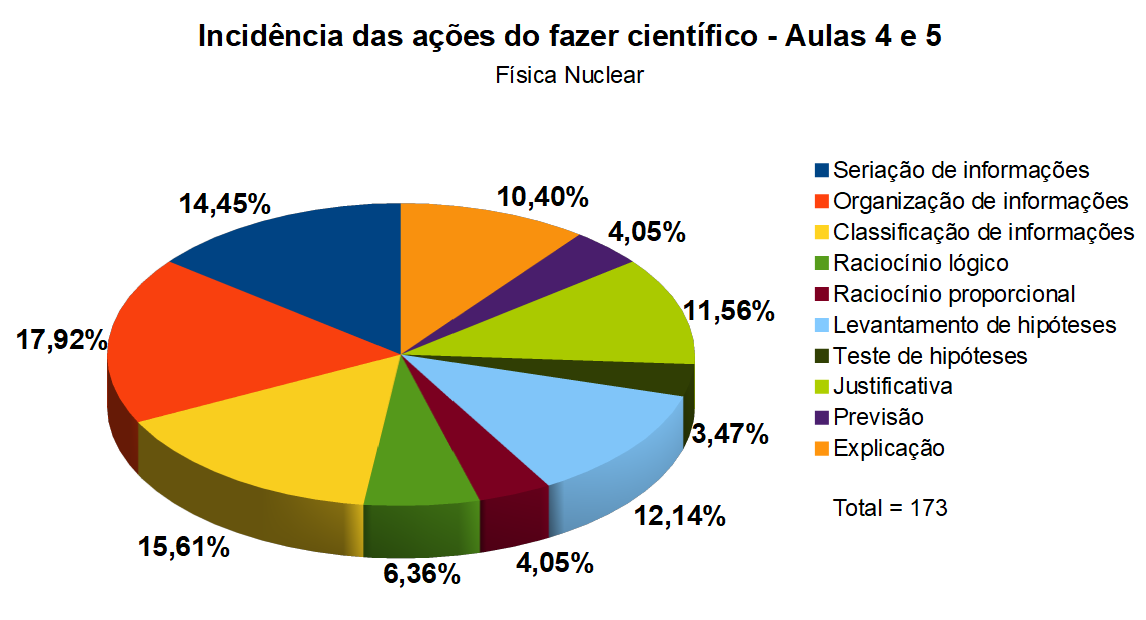}\\
	\includegraphics[trim=0 0 180 70, clip, scale=.52]{AC_45_Total}\qquad
        \includegraphics[trim=400 40 0 70, clip, scale=.7]{AC_45_Total}
	\caption{Incid\^encia dos indicadores de alfabetiza\c{c}\~ao cient\'ifica durante a quarta e quinta aulas sobre F\'isica Nuclear. A legenda est\'a ordenada no sentido anti-hor\'ario do gr\'afico a partir de ``seria\c{c}\~ao de informa\c{c}\~oes'' (azul escuro, 14,45\%).} 
	\label{fig:AC_45}
\end{figure*}

A incid\^encia dos indicadores de alfabetiza\c{c}\~ao cient\'ifica durante a quarta e quinta aula s\~ao apresentados na figura~\ref{fig:AC_45}.

Entre as a\c{c}\~oes que envolvem o trabalho com os dados obtidos em uma investiga\c{c}\~ao, houve uma incid\^encia similar de todos os tr\^es indicadores: seria\c{c}\~ao de informa\c{c}\~oes, organiza\c{c}\~ao de informa\c{c}\~oes e classifica\c{c}\~ao de informa\c{c}\~oes. No entanto, quando comparado ao resultado do primeiro bloco de aulas (as tr\^es primeiras aulas da sequ\^encia did\'atica~\cite{CarvalhoDorsch:2021lvd}), percebemos um aumento de, no m\'inimo, duas vezes nas incid\^encias de organiza\c{c}\~ao e classifica\c{c}\~ao de informa\c{c}\~oes, o que \'e natural pois, \`a medida que os alunos se deparam com novos conhecimentos, h\'a uma maior necessidade de arranj\'a-los em suas estruturas cognitivas e estabelecer rela\c{c}\~oes entre eles. 

Com rela\c{c}\~ao \`as a\c{c}\~oes que estruturam o pensamento cient\'ifico percebemos uma leve queda no racioc\'inio proporcional quando comparado ao resultado do primeiro bloco de aulas. Esse elemento do fazer cient\'ifico foi estimulado nos momentos 4 e 5 da quarta aula, sobre o experimento de Rutherford-Geiger-Marsden e durante a discuss\~ao sobre teste de hip\'oteses na ci\^encia (ver tabela~\ref{tab:Aula4Momentos}). No entanto, a incid\^encia foi menor do que nos momentos sobre as semelhan\c{c}as e diferen\c{c}as entre as leis de Coulomb e a da Gravita\c{c}\~ao, e sobre as equa\c{c}\~oes de Maxwell do primeiro bloco de aulas da sequ\^encia did\'atica~\cite{CarvalhoDorsch:2021lvd}. 

Com rela\c{c}\~ao \`as a\c{c}\~oes que buscam o entendimento da situa\c{c}\~ao analisada, observamos um aumento de quase 100\% na incid\^encia dos indicadores de justificativa e de explica\c{c}\~ao quando comparada \`a incid\^encia desses mesmos indicadores no primeiro bloco de aulas~\cite{CarvalhoDorsch:2021lvd}. Tal resultado foi devido, principalmente, \`a quinta aula pois, nela, os estudantes foram n\~ao s\'o estimulados a incorporar ideias durante todas as discuss\~oes em sala de aula, fortalecendo com argumenta\c{c}\~ao os aspectos abordados mas, tamb\'em, os estudantes agiram com autonomia para trazer novas informa\c{c}\~oes e fontes para as discuss\~oes. Inclusive, o indicador de explica\c{c}\~ao re\'une uma s\'erie de a\c{c}\~oes do fazer cient\'ifico e n\~ao foi mera coincid\^encia obter sua maior incid\^encia na quinta aula, justamente a que mais engajou os estudantes at\'e o presente momento da sequ\^encia.

\begin{figure}[h!]
	\centering
	\includegraphics[width=.485\textwidth]{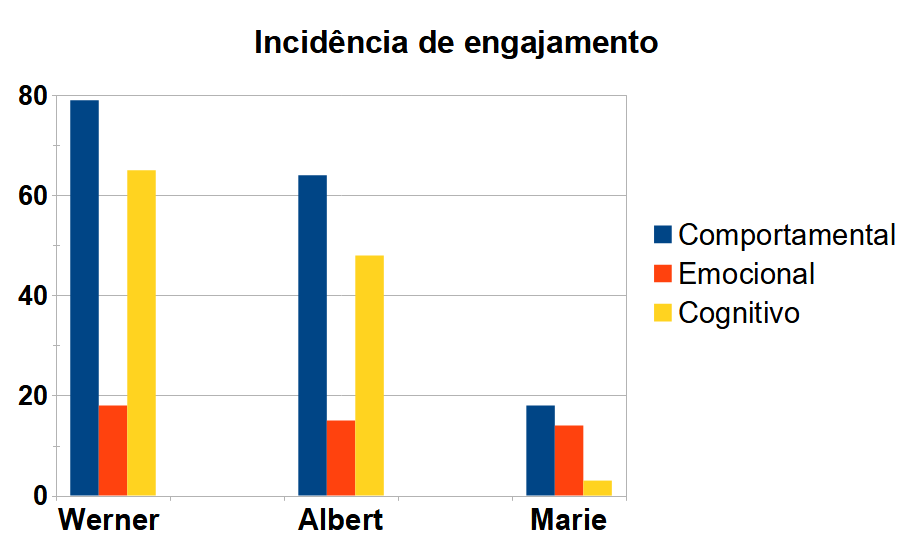}
	\caption{Incid\^encia dos indicadores de engajamento comportamental, emocional e cognitivo para os tr\^es estudantes mais presentes nas interven\c{c}\~oes da quarta e da quinta aulas da sequ\^encia did\'atica.}
	\label{fig:Resultados_Engajamento_Bloco2}
\end{figure}

Na figura~\ref{fig:Resultados_Engajamento_Bloco2} apresentamos os indicadores de engajamento presentes nas aulas consideradas neste trabalho para cada um dos tipos de engajamento e para tr\^es dos cinco alunos mais presentes nas aulas --- Werner, Albert, Marie, Peter e Emmy, que estiveram presentes em 10, 10, 8, 7 e 6 das 10 aulas da sequ\^encia, respectivamente. N\~ao apresentamos os resultados para os estudantes Peter e Emmy, como fizemos em trabalho anterior~\cite{CarvalhoDorsch:2021lvd} sobre as tr\^es primeiras aulas da sequ\^encia, pois eles n\~ao estiverem presentes justamente na quarta e quinta aulas, foco deste trabalho.

As maiores incid\^encias de engajamento comportamental e cognitivo partiram dos estudantes Werner e Albert, todas com expressividade, ainda mais quando notamos que houve apenas 10 e 16 momentos na quarta e na quinta aula, respectivamente. Al\'em disso, cabe ressaltar que o estudante Albert foi o menos participativo nas tr\^es primeiras aulas da sequ\^encia e, a partir da quarta aula, ap\'os se mostrar mais aberto e \`a vontade em sala, revelou um enorme salto na incid\^encia de engajamento comportamental. A partir da quinta aula, o mesmo estudante demonstrou ainda mais engajamento, iniciando pelo comportamental e se aprofundando no engajamento emocional e, principalmente, no engajamento cognitivo. De fato, enquanto nas tr\^es primeiras aulas da sequ\^encia Albert apresentou apenas 7, 14 e zero interven\c{c}\~oes demonstrando engajamento comportamental, emocional e cognitivo, respectivamente~\cite{CarvalhoDorsch:2021lvd}, esse mesmo estudante obteve, na quarta e quinta aulas, 64, 15 e 48 interven\c{c}\~oes de engajamento comportamental, emocional e cognitivo, respectivamente.  Considerando todas as aulas da sequ\^encia que ser\~ao abordadas em futuras publica\c{c}\~oes, Albert apresentou uma significativa evolu\c{c}\~ao nas incid\^encias de todos os tipos de engajamento. Junto a Werner, com engajamento significativo ao longo de todas as aulas da sequ\^encia, Albert foi um dos estudantes com maiores n\'iveis de engajamento da sequ\^encia. Esse crescimento progressivo nos \'indices de Albert \'e indicativo de que a sequ\^encia foi aplicada de maneira bem-sucedida com base na dialogicidade do docente aliada \`a tem\'atica contempor\^anea da F\'isica e mostrou-se capaz de trazer os estudantes para as aulas, at\'e mesmo aqueles mais inibidos ou inseguros nas primeiras aulas da sequ\^encia.

\section{Conclus\~oes}
\label{sec:conclusoes}

Nesta segunda parte da s\'erie ``F\'isica de Part\'iculas no Ensino M\'edio'' exploramos a tem\'atica de F\'isica Nuclear com \^enfase em uma perspectiva CTSA. Na sequ\^encia proposta, o assunto \'e motivado pela busca da estrutura fundamental da mat\'eria, mas logo se desenvolve rumo \`as aplica\c{c}\~oes tecnol\'ogicas que impactam a sociedade e o meio ambiente. Mais do que isso, foi mostrado como cada subt\'opico pode ser discutido em sala sob um vi\'es interdisciplinar, em di\'alogo com outros professores, fomentando a conjun\c{c}\~ao dos conhecimentos, em contraposi\c{c}\~ao \`a compartimenta\c{c}\~ao que frequentemente predomina nos curr\'iculos escolares.

Parte da sequ\^encia discutida neste artigo foi ministrada em duas aulas, de aproximadamente 1 hora cada, a uma turma mista de uma escola estadual da cidade de Vila Velha, estado do Esp\'irito Santo. Na ocasi\~ao, ficou patente o interesse agudo dos(as) estudantes pela tem\'atica, que se manifestou nos resultados da an\'alise de engajamento discutidos na se\c{c}\~ao~\ref{sec:analise}.

Apresentamos, tamb\'em, atividades que podem ser realizadas com a turma e que fomentam diversas habilidades em todas as compet\^encias norteadas pela Base Nacional Comum Curricular (BNCC) na \'area de Ci\^encias da Natureza e suas tecnologias no ensino m\'edio~\cite{BNCC}. Trabalha-se, assim, n\~ao apenas o conte\'udo desejado, mas tamb\'em outras capacidades, saberes e talentos dos(as) estudantes.

Assim como na primeira parte dessa s\'erie e artigos~\cite{CarvalhoDorsch:2021lvd}, o material contido aqui n\~ao precisa, necessariamente, ser aplicado na \'integra, e \'e at\'e desej\'avel que haja alguma sele\c{c}\~ao. O intuito n\~ao \'e fornecer um material formulaico, a ser seguido \`a risca pelo(a) professor(a), mas uma mat\'eria bruta a ser por ele(a) lapidada. O objetivo jamais deve ser o ``conteudismo'', mas o uso desse conte\'udo para motivar o engajamento e proporcionar alfabetiza\c{c}\~ao cient\'ifica e inclus\~ao do(a) discente no mundo contempor\^aneo. Esperamos que o presente trabalho tenha demonstrado ao(\`a) leitor(a) o potencial da f\'isica nuclear em servir a esse prop\'osito em sala de aula.

\section*{Agradecimentos}

Os autores agradecem a J\'ulio C\'esar Fabris, organizador do projeto ``Universo na Escola'', e a Thiago Pereira da Silva, professor de F\'isica da CEEMTI Prof$^{\text{a}}$ Maura Abaurre (Vila Velha -- ES) no per\'iodo das interven\c{c}\~oes, por viabilizarem o presente estudo. Agradecemos tamb\'em a Geide Rosa Coelho, por sugest\~oes e coment\'arios relevantes \`a an\'alise das interven\c{c}\~oes. O presente trabalho foi realizado com apoio da Coordena\c{c}\~ao de Aperfei\c{c}oamento de Pessoal de N\'ivel Superior - Brasil (Capes) - C\'odigo de Financiamento 001.

\appendix
\section{Ap\^endice: Propostas de Atividades}
\label{sec:atividades}

\subsection*{Se\c{c}\~ao~\ref{sec:reacoes}}

Um bom exerc\'icio, tanto para consolidar o conceito de energia de liga\c{c}\~ao nuclear, quanto para incentivar c\'alculos simples e com aplica\c{c}\~ao pr\'atica em discuss\~oes sobre energia nuclear, \'e usar a figura~\ref{fig:binding_energy} para determinar a quantidade de energia liberada em uma determinada rea\c{c}\~ao nuclear.

Por exemplo, conforme discutido na se\c{c}\~ao~\ref{sec:fusao_energia}, uma das principais rea\c{c}\~oes que poderiam ser usadas na gera\c{c}\~ao de energia nuclear por fus\~ao \'e a produ\c{c}\~ao de h\'elio-4 por fus\~ao de um deut\'erio
e um tr\'itio,
\begin{equation}
	^2\text{H} \,+ \,^3\text{H} \to \,^4\text{He} + \text{n} + E_\text{liberada}.
\end{equation}
A energia liberada \'e dada pela diferen\c{c}a entre a energia de repouso dos reagentes e do produto,
\begin{equation}
    E_\text{liberada} = (m_{^2\text{H}} + m_{^3\text{H}} - m_{^4\text{He}}  - m_n)\,c^2.
\end{equation}
Mas, pela equa\c{c}\~ao~(\ref{eq:Ebinding}), a massa do nucl\'ideo \'e a diferen\c{c}a da massa dos seus constituintes e de sua energia de liga\c{c}\~ao,
\begin{equation}\begin{split}
    m_{^2\text{H}}\,c^2 &= (m_p + m_n)\,c^2 - E(^2\text{H}),\\
    m_{^3\text{H}}\,c^2 &= (m_p + 2m_n)\,c^2 - E(^3\text{H}),\\
    m_{^4\text{He}}\,c^2 &= (2m_p + 2m_n)\,c^2 - E(^4\text{He}),
\end{split}\end{equation}
de modo que 
\begin{equation}
    E_\text{liberada} = E(^4\text{He}) - E(^2\text{H}) - E(^3\text{H}).
\end{equation}

Os valores das energias de liga\c{c}\~ao podem ser lidos da figura~\ref{fig:binding_energy}, lembrando que a figura d\'a a energia de liga\c{c}\~ao \emph{por n\'ucleon}. Dessa figura podemos ver que
\begin{equation}\begin{split}
    \frac{E(^4\text{He})}{4} &\approx 7.1~\text{MeV},\\
    \frac{E(^3\text{H})}{3} &\approx 2.8~\text{MeV},\\
    \frac{E(^2\text{H})}{2} &\approx 1.1~\text{MeV},
\end{split}\end{equation}
resultando em $E_\text{liberada}\approx 17.8~\text{MeV}$ para cada rea\c{c}\~ao, um valor muito pr\'oximo a $\approx 17.59~\text{MeV}$ calculado com valores tabelados para as energias de liga\c{c}\~ao.

Da mesma forma \'e poss\'ivel estimar, da figura~\ref{fig:binding_energy}, a quantidade de energia liberada em uma fiss\~ao de $^{235}\text{U}$. Para esse nucl\'ideo tem-se, da figura,
\begin{equation}
    \frac{E(^{235}\text{U})}{235}\approx 7.6~\text{MeV}.
\end{equation}
Essa fiss\~ao n\~ao ocorre sempre da mesma maneira, ou seja, o par de nucl\'ideos resultantes de uma fiss\~ao varia a cada vez que fiss\~ao for realizada (vide figura~\ref{fig:chain})\footnote{Como o processo \'e qu\^antico, existe uma \emph{probabilidade} de a fiss\~ao ocorrer de uma determinada maneira. N\~ao \'e poss\'ivel determinar, de antem\~ao, qual ser\'a o resultado da fiss\~ao. Mesmo que repliquemos o experimento exatamente sob as mesmas condi\c{c}\~oes, o resultado da fiss\~ao em cada caso vai variar.}. Mas podemos estimar a energia liberada em m\'edia, supondo que o resultado ser\~ao dois nucl\'ideos id\^enticos, com n\'umero de massa $\approx 118$. Da figura v\^e-se que a energia de liga\c{c}\~ao m\'edia para tal nucl\'ideo \'e $\approx 8.5~\text{MeV}$, ent\~ao $E_\text{liberada}\approx 235\times (8.5-7.6)~\text{MeV}\approx 211~\text{MeV}$.

\subsection*{Se\c{c}\~ao~\ref{sec:decaimentos}}

De acordo com a Base Nacional Comum Curricular (BNCC)~\cite{BNCC}, uma habilidade que se almeja desenvolver com os estudantes no \^ambito das ci\^encias exatas \'e a an\'alise e interpreta\c{c}\~ao de dados em gr\'aficos. Para esse prop\'osito, uma atividade com base na figura~\ref{fig:decays} mostra-se apropriada. A figura \'e simples, mas possui uma riqueza de informa\c{c}\~oes que o(a) docente pode explorar em in\'umeras atividades, que tamb\'em servem para consolidar o conceito e os tipos de decaimentos nucleares.

Algumas das quest\~oes que podem ser discutidas com base nesse gr\'afico s\~ao:
\begin{enumerate}
    \item \emph{Por que o ``vale de estabilidade'', mostrado nos pontos pretos, desvia da curva $Z=N$? E por que esse desvio \'e tanto maior, quanto maior for o n\'umero de massa dos nucl\'ideos (i.e. quanto mais \`a direita do gr\'afico o nucl\'ideo estiver)?}
    
    A resposta, como j\'a discutido no corpo principal do texto, \'e que nucl\'ideos mais pesados precisam de um ligeiro excesso de n\^eutrons sobre pr\'otons, para que a atra\c{c}\~ao nuclear compense a repuls\~ao coulombiana.
    \item \emph{Por que a maioria dos pontos acima do vale de estabilidade s\~ao azuis (decaem por $\beta^-$) e os pontos abaixo s\~ao, em sua maioria, laranjas (decaem por $\beta^+$)? }
    
    Pontos acima da curva de estabilidade possuem um excesso de n\^eutrons sobre pr\'otons. O decaimento $\beta^-$ alivia esse excesso ao converter um n\^eutron em um pr\'oton. O oposto ocorre abaixo da curva, em que $\beta^+$ converte um pr\'oton em um n\^eutron.
    \item \emph{Por que os pontos amarelos se tornam predominantes para nucl\'ideos muito pesados?}
    
    Nesses casos a estabilidade do nucl\'ideo \'e prejudicada tanto pela repuls\~ao coulombiana, quanto pelo tamanho do nucl\'ideo, que faz com que os n\'ucleons estejam mutuamente afastados (em m\'edia), o que reduz a energia de liga\c{c}\~ao m\'edia por n\'ucleon. O decaimento $\alpha$ resolve esses dois problemas simultaneamente, pois reduz o n\'umero de pr\'otons e o n\'umero de massa (e portanto o raio do nucl\'ideo).
    \item \emph{A maior parte dos decaimentos $\alpha$ ocorrem para nucl\'ideos mais pesados que o $^{126}\text{Pb}$ (o \'ultimo nucl\'ideo est\'avel), ou para nucl\'ideos muito \`a direita do vale de estabilidade (ou seja, com muito mais pr\'otons do que seria a quantidade \'otima para a estabilidade). Contudo, existem alguns pontos amarelos na figura~\ref{fig:decays} que est\~ao bastante pr\'oximos do vale de estabilidade, ou at\'e mesmo bem no meio desse vale. Por exemplo, o neod\'imio-144, $^{144}_{\ 60}\text{Nd}$, aparece na figura como um ponto amarelo bem em meio a pontos pretos que denotam outros is\'otopos est\'aveis do mesmo elemento. Por que isso ocorre?
    }
    
    Se o $^{144}\text{Nd}$ decair por emiss\~ao de $\alpha$, transformar-se-\'a no c\'erio-140, que possui 58 pr\'otons e 82 n\^eutrons. O n\'umero ``82'' \'e um dos chamados \emph{n\'umeros m\'agicos} na f\'isica nuclear (por isso aparecem na figura~\ref{fig:decays} como linhas verticais e horizontais), porque um nucl\'ideo que possui 82 pr\'otons (como o chumbo) ou 82 n\^eutrons (como o c\'erio-140 j\'a citado) \'e excepcionalmente est\'avel. Isso porque uma das camadas nucleares fica totalmente preenchida com esse n\'umero de n\'ucleons (analogamente ao que ocorre com os gases nobres, que s\~ao quimicamente est\'aveis porque possuem uma camada eletr\^onica totalmente preenchida). Assim, \'e mais vantajoso ao $^{144}\text{Nd}$ decair por $\alpha$ para chegar a esse estado mais est\'avel. Note que o mesmo n\~ao acontece para o $^{145}\text{Nd}$ ou $^{143}\text{Nd}$. Nesses casos o decaimento $\alpha$ \emph{n\~ao} levaria a uma configura\c{c}\~ao com n\'umero m\'agico de n\^eutrons, e por isso esses decaimentos n\~ao s\~ao favorecidos, e esses nucl\'ideos s\~ao est\'aveis, como se v\^e na figura.
    
    V\'arios outros pontos amarelos na vizinhan\c{c}a da curva de estabilidade levam a uma sequ\^encia de decaimentos que resulta, ao final, em um nucl\'ideo com n\'umero m\'agico de n\^eutrons. Para ver isso, basta notar que, se um ponto amarelo da figura decair por $\alpha$, o nucl\'ideo-filha estar\'a dois pontos \`a esquerda e dois pontos abaixo da posi\c{c}\~ao inicial. Seguindo-se a linha de decaimentos, v\^e-se que, para v\'arios deles, a sequ\^encia termina na curva de estabilidade ap\'os um \'unico decaimento (o que faz com que o decaimento $\alpha$ seja mais vantajoso que outra forma de decaimento), ou segue uma sequ\^encia de decaimentos que termina em um nucl\'ideo com 82 n\^eutrons. Vide, por exemplo, o caso
    \[
        ^{154}_{\ 66}\text{Dy}
        \stackrel{\alpha}{\longrightarrow}
        \,^{150}_{\ 64}\text{Gd}
        \stackrel{\alpha}{\longrightarrow}
        \,^{146}_{\ 62}\text{Sm}
        \stackrel{\alpha}{\longrightarrow}
        \,^{142}_{\ 60}\text{Nd}.
    \]
    O resultado final \'e est\'avel, com $82$ n\^eutrons.
\end{enumerate}


\thebibliography{99}

\bibitem{CarvalhoDorsch:2021lvd}
G.~C. Dorsch e T.~C. da C. Guio,
``{\it F\'\i{}sica de Part\'\i{}culas no ensino m\'edio Parte I: Eletrodin\^amica Qu\^antica}'',
Rev. Bras. Ens. Fis. \textbf{43} (2021), e20210083
[arXiv:2103.04946 [physics.ed-ph]].


\bibitem{CarlSagan}
    C.~Sagan, \emph{``As liga\c{c}\~oes c\'osmicas: uma perspectiva extraterrestre''}. Lisboa: Ed. Gradiva (2001).

\bibitem{Bybee1987} 
W. Bybee. {\it ``Science education and the science-technology-society (STS) theme''},
Science Education \textbf{71}, n. 5, pp. 667-683 (1987).

\bibitem{LopezCerezo1996}
J. L. L. L\'opez e J. A. Cerezo. {\it ``Educaci\'on CTS en acci\'on: ense\~nanza
secundaria y universidad''}. In: GARC\'IA, Marta. I. Gonz\'alez; CEREZO, Jos\'e A. L\'opez;
L\'OPEZ, Jos\'e L. Luj\'an. {\it ``Ciencia, tecnologia y sociedad: una introducci\'on al
estudio social de la ciencia y la tecnologia''}. Madrid: Editorial Tecnos S. A. (1996).

\bibitem{Krasilchik1987}
M. Krasilchik. {\it ``O professor e o curr\'iculo''}. S\~ao Paulo: EDUSP (1987).

\bibitem{SantosMortimer2001}
W. L. Pereira dos Santos e E. F. Mortimer, {\it ``Tomada de decis\~ao
para a\c{c}\~ao social respons\'avel no ensino de Ci\^encias''}, Ci\^encia e Educa\c{c}\~ao \textbf{7}, n. 1
pp. 95-111 (2001). 

\bibitem{SantosMortimer2002}
W. L. Pereira dos Santos e E. F. Mortimer, {\it ``Uma an\'alise de
pressupostos te\'oricos da abordagem C-T-S (Ci\^encia - Tecnologia - Sociedade) no contexto
da educa\c{c}\~ao brasileira''}. Ensaio: Pesquisa em Educa\c{c}\~ao em Ci\^encias, Belo Horizonte,
v. 2, n. 2, pp. 110-132 (2002).

\bibitem{SasseronCarvalho2011}
L. H. Sasseron e A. M. P. Carvalho, {\it ``Alfabetiza\c{c}\~ao Cient\'ifica:
uma revis\~ao bibliogr\'afica''}, Investiga\c{c}\~oes em Ensino de Ci\^encias \textbf{16},
n. 1, pp. 59-77 (2011).

\bibitem{Waks1990}
L. J. Waks, {\it ``Educaci\'on en ciencia, tecnolog\'ia y sociedad: origenes, desarrollos
internacionales y desaf\'ios actuales''}. In: MEDINA, Manuel; SANMART\'IN, Jos\'e (Eds.)
{\it ``Ciencia, tecnolog\'ia y sociedad: estudios interdisciplinares en la universidad,
en la educaci\'on y la gesti\'on p\'ublica''}. Barcelona: Anthropos, Leioa (Vizcaya):
Universidad del Pa\'is Vasco (1990).

\bibitem{BNCC}
    Minist\'erio da Educa\c{c}\~ao,
    {\it ``Base Nacional Comum Curricular''}.
    Dispon\'ivel em: \url{http://basenacionalcomum.mec.gov.br}.
    Acesso em 28 de fevereiro de 2023.
    
\bibitem{Thomson:1904bjw}
  J.~J.~Thomson,
  ``{\it On the structure of the atom: an investigation of the stability and periods of oscillation of a number of corpuscles arranged at equal intervals around the circumference of a circle; with application of the results to the theory of atomic structure}'',
  Philosophical Magazine, Series 6, {\bf 7} (1904) no.~39,  237.
    
\bibitem{Geiger:1910}
   H.~Geiger,
   \emph{``The scattering of $\alpha$-particles by matter''},
   Roy. Soc. Proc. A {\bf 83} (1910), 492
   
\bibitem{Geiger:1913}
   H.~Geiger e E.~Marsden,
   \emph{``The laws of deflexion of a particles through large angles''},
   Philosophical Magazine, Series 6, {\bf 25} (1913), 604
   
\bibitem{Hyperphysics}
   R.~Nave, {``\it The Thomson model of the atom''}, Hyperphysics. 
   Dispon\'ivel em: \url{http://hyperphysics.phy-astr.gsu.edu/hbase/Nuclear/rutsca3.html}. 
   Acesso em 28 de fevereiro de 2023.

\bibitem{BY-SA40}
   Creative Commons,
   {\it ``Licen\c{c}a Attribution-Share Alike 4.0 (BY-SA 4.0)''}. 
   Dispon\'ivel em: \url{https://creativecommons.org/licenses/by-sa/4.0/deed.en}. 
   Acesso em 28 de fevereiro de 2023.
   
\bibitem{Eisberg}
   R.~Eisberg e R.~Resnick, 
   \emph{``F\'isica Qu\^antica''},
   Ed. Campus.

\bibitem{Rutherford:1911}
   E.~Rutherford,
   \emph{``The scattering of $\alpha$ and $\beta$ particles by matter and the structure of the atom''},
   Philosophical Magazine, Series 6, {\bf 21} 669 (1911).
  
\bibitem{Rutherford:1919}
   E.~Rutherford,
   {\it ``Collision of $\alpha$ particles with light atoms. IV. An anomalous effect in nitrogen''},
  Philosophical Magazine, Series 6, {\bf 37} 581 (1919).
  
\bibitem{Rutherford:1920}
  E.~Rutherford, 
  {\it ``Bakerian Lecture. Nuclear Constitution of Atoms''},
  Roy. Soc. Proc. A, {\bf 97} 374 (1920).
   
\bibitem{Elcio}
   E.~S.~Lopes,
   {\it ``E o el\'etron? \'E onda ou \'e part\'icula? --- Uma proposta para promover a ocorr\^encia da alfabetiza\c{c}\~ao cient\'ifica de f\'isica moderna e contempor\^anea em estudantes do ensino m\'edio''},
    Disserta\c{c}\~ao de Mestrado em Ensino de F\'isica, Universidade de S\~ao Paulo (2013).

\bibitem{PPlato}
   PPlato/FLAP, {\it ``Flexible Learning Approach to Physics''}.
   Dispon\'ivel em: \url{http://www.met.reading.ac.uk/pplato2/h-flap/phys6\_4.html}. 
   Acesso em 28 de fevereiro de 2023.
   
\bibitem{Angeli}
   I.~Angeli e K.~Marinova,
   {\it ``Table of experimental nuclear ground state charge radii: An update''},
   Atomic Data and Nuclear Data Tables {\bf 99}, Issue 1, p. 69 (2013).

\bibitem{Ehrenberg}
   H.~Ehrenberg et al.,
   {\it ``High-Energy Electron Scattering and the Charge Distribution of Carbon-12 and Oxygen-16''},
   Phys.~Rev.~{\bf 113} p. 666 (1959).
   
\bibitem{Krane}
   K.~Krane,
   {\it ``Introductory Nuclear Physics''},
   Ed. John Wiley \& Sons (1988).
   
\bibitem{Tu:2004}
   L.-C.~Tu e J.~Luo,
   {\it ``Experimental tests of Coulomb's Law and the photon rest mass''},
   Metrologia {\bf 41}, Number 5 S136 (2004).
   
\bibitem{ChemLibre}
   Chemistry LibreTexts, 
   {\it ``General Chemistry, Ch. 24: Nuclear Chemistry''}.
   Dispon\'ivel em \url{https://chem.libretexts.org/Courses/Howard\_University/General\_Chemistry\%3A\_An\_Atoms\_First\_Approach/Unit\_8\%3A\_\_Materials/Chapter\_24\%3A\_Nuclear\_Chemistry}. Acesso em 28 de fevereiro de 2023.
   
\bibitem{CCBYSA30}
   Creative Commons, 
   {\it ``Licen\c{c}a Attribution-NonCommercial-ShareAlike 3.0 Unported (CC BY-NC-SA 3.0)''}.
   Dispon\'ivel em: \url{https://creativecommons.org/licenses/by-nc-sa/3.0}.
   Acesso em 28 de fevereiro de 2023.

\bibitem{WNA_FBR}
   World Nuclear Association,
   {\it ``Fast Neutrons Reactors''}.
   Dispon\'ivel em: \url{https://www.world-nuclear.org/information-library/current-and-future-generation/fast-neutron-reactors.aspx}.
   Acesso em 28 de fevereiro de 2023.

 \bibitem{NuclearEnergy21st}
    I.~Hore-Lacy,
    {\it ``Nuclear energy in the 21st century''}.
    London: Ed. Academic Press (2007).    
    
\bibitem{Dodelson}
   S.~Dodelson,
   {\it ``Modern Cosmology''},
   Ed. Elsevier (2003).

\bibitem{Liddle}
   A.~Liddle,
   {\it ``An Introduction to Modern Cosmology''}.
   Chichester: Ed. John Wiley \& Sons (2003).
   
\bibitem{Lodders}
   K.~Lodders,
   {\it ``Solar System Abundances and Condensation Temperatures of the Elements''},
   The Astrophysical Journal {\bf 591}, no.~2 (2003) pp.~1220--1247.

\bibitem{BBC}
   BBC News,
   {\it ``In pictures: Hiroshima, the first atomic bomb''}.
   Dispon\'ivel em: \url{https://www.bbc.com/news/in-pictures-33787169}.
   Acesso em 28 de fevereiro de 2023.
   
\bibitem{NuclearDarkness}
   Nuclear Darkness, Global Climate Change \& Nuclear Famine,
   {\it ``Hiroshima''}.
   Dispon\'ivel em: \url{http://www.nucleardarkness.org/hiroshima/}.
   Acesso em 28 de fevereiro de 2023.
 
 \bibitem{Allthatisinteresting}
    I.~Dickinson,
    {\it ``33 Photos of the Hiroshima aftermath that reveal the bombing's true devastation''} .
    Dispon\'ivel em: \url{https://allthatsinteresting.com/hiroshima-aftermath-pictures}.
    Acesso em 28 de fevereiro de 2023.

\bibitem{WNA_Heat}
   World Nuclear Association,
   {\it ``Heat Values of Various Fuels''}.
   Dispon\'ivel em: \url{https://www.world-nuclear.org/information-library/facts-and-figures/heat-values-of-various-fuels.aspx} .
   Acesso em 28 de fevereiro de 2023.
 
\bibitem{WNA_Reactors}
   World Nuclear Association,
   {\it ``Nuclear Power Reactors''}.
   Dispon\'ivel em: \url{https://www.world-nuclear.org/information-library/nuclear-fuel-cycle/nuclear-power-reactors/nuclear-power-reactors.aspx} .
   Acesso em 28 de fevereiro de 2023.
   
 \bibitem{WNA_Environment}
    World Nuclear Association,
    {\it ``CO$_2$ Implications of Electricity Generation''}.
    Dispon\'ivel em: \url{https://www.world-nuclear.org/information-library/energy-and-the-environment/co2-implications-of-electricity-generation.aspx}.
    Acesso em 28 de fevereiro de 2023.
 
 \bibitem{OWD_CO2}
    H.~Ritchie e M.~Roser,
    {\it ``CO$_2$ and greenhouse gas emissions''}. 
    \emph{In}: Our World in Data.
    Dispon\'ivel em: \url{https://ourworldindata.org/co2-and-other-greenhouse-gas-emissions}.
    Acesso em 28 de fevereiro de 2023.
    
 \bibitem{Menyah1}
    K.~Menyah e Y.~Wolde-Rufael,
    {\it ``CO$_2$ emissions, nuclear energy, renewable energy and economic growth in the US''},
    Energy Policy \textbf{38}, Issue 6, pp. 2911-2915 (2010).
    
 \bibitem{Menyah2}
    N.~Apergis, J.~E.~Payne, K.~Menyah e Y.~Wolde-Rufael,
    {\it ``On the causal dynamics between emissions, nuclear energy, renewable energy, and economic growth''},
    Ecological Economics \textbf{69}, Issue 11, pp. 2255-2260 (2010).
 
\bibitem{OWD_Safety}
   H.~Ritchie,
   {\it ``What are the safest sources of energy?''}.
   \emph{In:} Our World in Data.
   Dispon\'ivel em: \url{https://ourworldindata.org/safest-sources-of-energy}.
   Acesso em 28 de fevereiro de 2023.
   
\bibitem{SciAm:CoalRadiation}
    M. Hvistendahl,
    {\it ``Coal Ash Is More Radioactive Than Nuclear Waste''}.
    \emph{In:} Scientific American.
    Dispon\'ivel em: \url{https://www.scientificamerican.com/article/coal-ash-is-more-radioactive-than-nuclear-waste/} .
    Acesso em 28 de fevereiro de 2023.

    
\bibitem{WNA_Waste}
   World Nuclear Association,
   {\it ``Radioactive Waste - Myths and Realities''}.
   Dispon\'ivel em: \url{https://www.world-nuclear.org/information-library/nuclear-fuel-cycle/nuclear-wastes/radioactive-wastes-myths-and-realities.aspx}.
   Acesso em 28 de fevereiro de 2023.
   
\bibitem{Beck}
   P.~W.~Beck,
   {\it ``Nuclear energy in the 21st century: Examination of a contentious subject''},
   Annual Review of Energy and the Environment \textbf{24} pp. 113-137 (1999).

\bibitem{Macknick:2012}
    J.~Macknick, R.~Newmark, G.~Heath e K.~C.~Hallet,
    {\it ``Operational water consumption and withdrawal factors for electricity generating technologies: a review of existing literature''},
    Environmental Research Letters \textbf{7} 045802 (2012). 

\bibitem{WNA_Chernobyl}
    World Nuclear Association,
    {\it ``Chernobyl Accident 1986''}.
    Dispon\'ivel em: \url{https://www.world-nuclear.org/information-library/safety-and-security/safety-of-plants/chernobyl-accident.aspx} .
    Acesso em 28 de fevereiro de 2023.
 
\bibitem{WNA_Fukushima}
    World Nuclear Association,
    {\it ``Fukushima Daiichi Accident''}.
    Dispon\'ivel em: \url{https://www.world-nuclear.org/information-library/safety-and-security/safety-of-plants/fukushima-daiichi-accident.aspx} .
    Acesso em 28 de fevereiro de 2023.
    
\bibitem{WNA}
    World Nuclear Association,
    {\it ``Information Library''}.
    Dispon\'ivel em: \url{https://www.world-nuclear.org/information-library.aspx}.
    Acesso em 28 de fevereiro de 2023.
    
 \bibitem{CNEN1}
    E.~M.~Cardoso, 
    {\it ``Apostila educativa: A energia nuclear''},
    3$^\text{a}$ ed.
    Rio de Janeiro: Comiss\~ao Nacional de Energia Nuclear (CNEN) (2012).
    Dispon\'ivel em: \url{http://antigo.cnen.gov.br/images/cnen/documentos/educativo/apostila-educativa-aplicacoes.pdf}.
    Acesso em 28 de fevereiro de 2023.
    
 \bibitem{CNEN2}
    R.~P.~de Carvalho e S.~M.~V.~de Oliveira,
    {\it ``Aplica\c{c}\~oes da energia nuclear na sa\'ude''}.
    S\~ao Paulo: Sociedade Brasileira para o Progresso da Ci\^encia (SBPC) e International Atomic Energy Agency (IAEA) 
    (2017).
    Dispon\'ivel em: \url{https://www.gov.br/cnen/pt-br/material-divulgacao-videos-imagens-publicacoes/publicacoes-1/aplicacoesenergianuclearnasaude.pdf}.
    Acesso em 28 de fevereiro de 2023.
    
 \bibitem{ANEEL}
    Ag\^encia Nacional de Energia El\'etrica (ANEEL),
    {\it ``Relat\'orios de Consumo e Receita de Distribui\c{c}\~ao''}.
    Dispon\'ivel em: \url{https://antigo.aneel.gov.br/web/guest/relatorios-de-consumo-e-receita}.
    Acesso em 28 de fevereiro de 2023.
    
 \bibitem{EPE}
    Empresa de Pesquisa Energ\'etica (EPE),
    {\it ``Anu\'ario Estat\'istico de Energia El\'etrica''}.
    Dispon\'ivel em: \url{http://www.epe.gov.br/pt/publicacoes-dados-abertos/publicacoes/anuario-estatistico-de-energia-eletrica}.
    Acesso em 28 de fevereiro de 2023.

\bibitem{BY-SA30}
   Creative Commons,
   {\it ``Licen\c{c}a Attribution-Share Alike 3.0 (BY-SA 3.0)''}. 
   Dispon\'ivel em: \url{https://creativecommons.org/licenses/by-sa/3.0/deed.en}. 
   Acesso em 28 de fevereiro de 2023.

\bibitem{IAEA:decays}
    International Atomic Energy Agency, \emph{``Live Chart of Nuclides''}. Dispon\'ivel em: \url{https://www-nds.iaea.org/relnsd/vcharthtml/VChartHTML.html}. Acesso em 28 de fevereiro de 2023.

\bibitem{White}
	W.~M.~White, 
	\emph{``Geochemistry''}. Chichester: Ed. John Wiley \& Sons (2013).

\bibitem{Bowen}
	R.~Bowen,
	\emph{``Isotopes in the earth sciences''}.
	Londres: Ed.~Chapman \& Hall (1994).

\bibitem{physicsforums}
A. Klotz, \textit{``Physics Forums Insights''}. Dispon\'ivel em: \url{www.physicsforums.com/insights/basics-positron-emission-tomography-pet}. Acesso em 28 de fevereiro de 2023.

\bibitem{OpenStax}
OpenStax College, \emph{``College Physics''}.
Dispon\'ivel em: \url{https://openstax.org/books/college-physics/pages/32-introduction-to-applications-of-nuclear-physics}. Acesso em 28 de fevereiro de 2023.

\bibitem{BY-40}
   Creative Commons,
   {\it ``Licen\c{c}a Attribution 4.0 International (BY 4.0)''}. 
   Dispon\'ivel em: \url{https://creativecommons.org/licenses/by/4.0/}. 
   Acesso em 28 de fevereiro de 2023.
   
\bibitem{ANVISA:Food}
    Ag\^encia Nacional de Vigil\^ancia Sanit\'aria (ANVISA),
    \emph{``Resolu\c{c}\~ao da Diretoria Colegiada no.~21 de 26 de novembro de 2001''}.
    Dispon\'ivel em: \url{http://antigo.anvisa.gov.br/legislacao#/visualizar/26672}. 
    Acesso em 28 de fevereiro de 2023.

\bibitem{ANVISA:Agenda}
    Ag\^encia Nacional de Vigil\^ancia Sanit\'aria (ANVISA),
    \emph{``Agenda Regulat\'oria 2021/2023 (Projeto 3.12)''}.
    Dispon\'ivel em: \url{https://www.gov.br/anvisa/pt-br/assuntos/regulamentacao/agenda-regulatoria/agenda-2021-2023}.
    Acesso em 28 de fevereiro de 2023.

\bibitem{OMS:1994}
    Organiza\c{c}\~ao Mundial da Sa\'ude,
    \emph{``Safety and nutritional adequacy of irradiated food''}. Genebra, 1994. Dispon\'ivel em: \url{https://apps.who.int/iris/handle/10665/39463}. Acesso em 28 de fevereiro de 2023.
    
\bibitem{OMS:1999}
    Organiza\c{c}\~ao Mundial da Sa\'ude,
    \emph{``High-dose irradiation : wholesomeness of food irradiated with doses above 10 kGy : report of a Joint FAO/IAEA/WHO study group''}. Genebra, 1999.
    Dispon\'ivel em: \url{https://apps.who.int/iris/handle/10665/42203}. Acesso em 28 de fevereiro de 2023.
    
\bibitem{UniversoEscola}
   N\'ucleo Cosmo-UFES, {``\it Projeto Universo na Escola''}.
   Dispon\'ivel em: \url{http://www.cosmo-ufes.org/universo-na-escola.html}.
   Acesso em 28 de fevereiro de 2023.
   
\bibitem{BY-SA20}
    Creative Commons,
   {\it ``Licen\c{c}a Attribution-ShareAlike 2.0 Generic (BY-SA 2.0)''}. 
   Dispon\'ivel em: \url{https://creativecommons.org/licenses/by-sa/2.0/deed.en}. 
   Acesso em 28 de fevereiro de 2023.

\bibitem{Chernobyl}
Chernobyl [Seriado]. Dire\c{c}\~ao: Johan Renck. Produ\c{c}\~ao: Sanne Wohlenberg. Estados Unidos: HBO Home Entertainment (2019). 

\bibitem{BACHELARD2005}
G. Bachelard, {\it ``A forma\c{c}\~ao fo esp\'irito cient\'ifico: contribui\c{c}\~ao para uma psican\'alise do conhecimento''}. Tradu\c{c}\~ao Esteia dos Santos Abreu. Rio de Janeiro: Ed. Contraponto, 5$^{\text{a}}$ reimpress\~ao (2005).

\bibitem{TCC_Thaisa}
T. C. da C. Guio, \emph{``Uma sequ\^encia did\'atica para o ensino de F\'isica de Part\'iculas no ensino m\'edio: ind\'icios de alfabetiza\c{c}\~ao cient\'ifica e engajamento de estudantes''}. 164pp. Monografia (Licenciatura em F\'isica). Universidade Federal do Esp\'irito Santo (UFES), Vit\'oria, 2020. Dispon\'ivel em: \url{https://labec.ufes.br/sites/labec.ufes.br/files/field/anexo/labec_thaisaguio01corrigida.pdf}. Acesso em 28 de fevereiro de 2023.

\bibitem{LUEDKE1986}
M. L\"udke e M.~E.~D.~A Andr\'e, \textit{``Pesquisa em educa\c{c}\~ao: abordagens qualitativas''}. S\~ao Paulo: EPU, (1986). 

\bibitem{OLIVEIRA2014}
R. de C. M. de Oliveira, {\it ``(Entre)Linhas de uma pesquisa: o di\'ario de campo como dispositivo de (in)forma\c{c}\~ao na/da abordagem (auto)biogr\'afica''}, Revista Brasileira de Educa\c{c}\~ao de Jovens e Adultos \textbf{2}, n. 4, p. 69 (2014).

\bibitem{MORAESTAZIRI2019}
V. R. A. de Moraes e J. A. Taziri, {\it ``A motiva\c{c}\~ao e o engajamento de alunos em uma atividade na abordagem do ensino de ci\^encias por investiga\c{c}\~ao''}, Investiga\c{c}\~oes em Ensino de Ci\^encias \textbf{24}, n. 2, 72 (2019).

\bibitem{FREIRE2000}
P. Freire, \textit{``Educa\c{c}\~ao como pr\'atica da liberdade''}. S\~ao Paulo: Paz e Terra (2000).

\bibitem{FREIRE1989}
P. Freire, \textit{``A import\^ancia do ato de ler - em tr\^es artigos que se completam''}. S\~ao Paulo: Cortez (1989). 

\bibitem{SasseronCarvalho2008}
L. H. Sasseron e A. M. P. Carvalho, {\it ``Almejando a alfabetiza\c{c}\~ao cient\'ifica no ensino fundamental: a proposi\c{c}\~ao e a procura de indicadores do processo''}, Investiga\c{c}\~oes em Ensino de Ci\^encias \textbf{13}, 333 (2008).

\bibitem{FREDERICKSetal2004}
J. A. Fredericks, P. C. Blumenfeld e A. H. Paris, {\it ``School Engagement: Potential of the Concept, State of the Evidence''}, Review of Educacional Reseach \textbf{74}, p. 59 (2004).

\bibitem{COELHO2011}
G. R. Coelho, \textit{``A evolu\c{c}\~ao do entendimento dos estudantes em eletricidade: um estudo longitudinal''}. Tese de Doutorado, Universidade Federal de Minas Gerais (2011).

\bibitem{BORGESetal2005}
O. Borges, J. M. J\'ulio e G. R. Coelho, in \textit{``Atas do V ENPEC''}, Bauru (2005). 

\bibitem{SASSERONSOUZA2019}
L. H. Sasseron e T. N. de Souza, {\it ``O engajamento dos estudantes em aula de f\'isica: apresenta\c{c}\~ao e discuss\~ao de uma ferramenta de an\'alise.''}, Investiga\c{c}\~oes em Ensino de Ci\^encias \textbf{24}, n. 1, p. 139 (2019).

\bibitem{FARIAVAZ2019}
A. F. Faria e A. M. Vaz, {\it ``Engajamento de estudantes em investiga\c{c}\~ao escolar sobre circuitos el\'etricos simples''}, Ensaio: Pesquisa em Educa\c{c}\~ao em Ci\^encias \textbf{21}, e10545, 1 (2019).

\bibitem{Finn1993}
J. D. Finn, \textit{``School engagement and students at risk''}. National Center for Education Statistics, Washington (1993).

\bibitem{Voelkl1997}
K. E. Voelkl, {\it ``Identification with School''}, American Journal of Education \textbf{105}, 294 (1997).

\bibitem{STIPEK2002}
D. Stipek, in \textit{``Development of achievement motivation: a volume in Educational Psychology''}, editado por A. Wigfield and J. S. Eccles. San Diego: Academic Press (2002).

\bibitem{CONNELLWELLBORN1991}
J. P. Connell e J. G. Wellborn, in \textit{``Minnesota Symposium on Child Psychology 23 Self processes and development''}, editado por M. R. Gunnar and L. A. Sroufe. Chicago: University of Chicago Press (1991).

\bibitem{BROPHY1987}
J. E. Brophy, in \textit{``Advances in motivation and achievement: enhancing motivation''}, editado por M. L. Maehr and D. A. Kleiber. Greenwich: JAI Press, (1987).

\bibitem{AMES1992}
C. Ames, {\it ``Classrooms: Goals, structures, and student motivation''}, Journal of Educational Psychology \textbf{84}, 261 (1992).

\bibitem{DWECKLEGGETT1988}
C. S. Dweck e E. L. Legget, {\it ``A social-cognitive approach to motivation and personality''}, Psychological Review \textbf{95}, 256 (1988).

\bibitem{HARTER1981}
S. Harter, {\it ``A new self-report scale of intrinsic versus extrinsic orientation in the classroom: Motivational and informational components''}, Development Psychology \textbf{17}, 300 (1981).

\bibitem{CORNOMADINACH1983}
L. Corno e E. B. Madinach, {\it ``The role of cognitive engagement in classroom learning and motivation''}, Educational Psychologist \textbf{18}, 88 (1983).
 \end{document}